\documentclass[12pt,preprint]{aastex}

\begin{document}

\title{Constrained Cluster Parameters from Sunyaev-Zel'dovich Observations}

\author{Neelima Sehgal and Arthur Kosowsky}
\affil{Department of Physics and Astronomy, Rutgers University, 136
Frelinghuysen Road, Piscataway, NJ 08854-8019;
sehgal@physics.rutgers.edu}
\author{Gilbert Holder}
\affil{CIAR Scholar; Department of Physics, McGill University,
3600 rue University, Montreal, QC H3A 2T8 }

\begin{abstract}
Near-future SZ surveys such as ACT, SPT, APEX, and Planck will soon
find thousands of galaxy clusters.  Multi-frequency
arcminute-resolution SZ observations can, in principle, determine
each cluster's gas temperature ($T_{e}$), bulk velocity
($v_{\mathrm{pec}}$), and optical depth ($\tau$).  However, the
frequency bands and detector sensitivity employed by upcoming
surveys will generally not be sufficient to disentangle the
degeneracy between these three cluster parameters, even in the
absence of SZ signal contamination from point sources and imperfect
primary microwave background subtraction.  Assuming contaminants can
be removed, we find that near-future SZ surveys will be able to
constrain well two cluster gas parameters that are linear
combinations of $\tau T_{e}$, $\tau v_{z}$, and $\tau T_{e}^{2}$.
Because the SZ intensity shift is nearly a linear function of $\tau
T_{e}$, $\tau v_{z}$, and $\tau T_{e}^{2}$, a correspondence exists
between the two effective gas parameters that SZ surveys can
constrain and simple line-of-sight integrals through the three
dimensional cluster.  We illustrate the parameter constraints and
correspondence to line-of-sight integrals using three dimensional
Nbody + hydro cluster simulations and a Markov chain Monte Carlo
method.  We show that adding an independent temperature measurement
to upcoming SZ data breaks the parameter degeneracy and that the
cluster effective velocity thus constrained is approximately the
optical-depth-weighted velocity integrated along the cluster line of
sight.  A temperature prior with an error as large as 2 keV still
gives bulk velocity errors of 100 km/sec or less, even for a more
typical cluster with an electron temperature of 3 keV, for ACT-like
SZ observations in the absence of signal contamination. The Markov
chain constraints on $v_{\mathrm{pec}}$ and $\tau$ that we obtain
are more encouraging and most likely more accurate than those
obtained from Fisher matrices.
\end{abstract}


\section{INTRODUCTION}

Observations of the Sunyaev-Zel'dovich (SZ) effect \citep{sunyaev}
offer the hope of revealing much about the properties of galaxy
clusters and the evolution of large-scale structure.  Several
instruments are being built (ACT, SPT, APEX, Planck) that will
make use of the SZ effect to produce deep galaxy cluster surveys,
and upgrades to current experiments, such as SuZIE III, will
produce deep targeted observations of known galaxy clusters.
These surveys, in addition to measuring the number density of
clusters, can, in principle, reveal each galaxy cluster's peculiar
velocity, gas temperature, and optical depth, if the SZ
information is fully exploited.  The aim of this paper is to
quantify some of the difficulties with determining these individual
cluster parameters from future SZ measurements and to discuss what
cluster parameters these future surveys can constrain.

The SZ effect is a spectral distortion of the cosmic microwave
background caused by an intervening galaxy cluster.  The hot gas in
the intracluster medium inverse Compton scatters the microwave
photons creating this distortion.  For reviews of the SZ effect see
\citet{rephaeli}, \citet{sunyaev2}, \citet{birkinshaw}, and
\citet{carlstrom}. The dependence of this spectral distortion on the
gas temperature ($T$), radial peculiar velocity ($v$), and optical
depth ($\tau$), including relativistic corrections, has been
computed in several papers \citep{challinor, sazonov, itoh1, itoh2,
molnar, dolgov}. Since the amplitude of this distortion is also a
function of frequency, SZ measurements at three different observing
frequencies would ideally be enough to disentangle the three
unknowns ($T, v, \tau$) contributing to the SZ signal.  Some first
attempts to determine cluster temperatures using the relativistic SZ
effect were carried out by \citet{pointecouteau} and
\citet{hansen3}, the former simulating clusters observed with Planck
and the latter providing a temperature estimate of Abell 2163 with
quite large uncertainties. More recently, \citet{hansen2} has
developed code to extract cluster parameters using the SZ effect
which results in rather large errors if given prospective data from
upcoming surveys.  The difficulty, as pointed out in \citet{holder}
and \citet{aghanim}, is that there exist degeneracies among the
physically interesting parameters ($T, v, \tau$) that can best be
broken by choosing one frequency relatively low (around 30 GHz) and
placing the other two around 150 GHz and 300 GHz. In addition, the
observations need to be of arcminute-resolution to resolve
individual clusters.  A 30 GHz observing band is not a realistic
option for upcoming arcminute-resolution bolometer based instruments
(such as ACT, SPT, and APEX) since high-sensitivity bolometers
sharply lose sensitivity below 90 GHz and the single-dish diameter
required for arcminute-resolution observations at 30 GHz is
unrealistically large. In addition, large interferometers capable of
observing at 30 GHz with arcminute-resolution can only view small
areas of the sky at a time, making a large survey impractical. Thus
it is important to determine what information about individual
galaxy clusters these future SZ surveys will be able to constrain
given the reality that they will have approximately
arcminute-resolution observations at or above 90 GHz.

In this paper we make a preliminary investigation of information we
can potentially extract from SZ measurements.  In particular, we do
not include the effects of a variety of real world complications
including radio and infrared point sources, galactic dust, non-SZ
microwave fluctuations or instrumental systematic effects.  While
these are certainly important considerations for any data set, the
goal of this paper is to elucidate the maximum information about
individual clusters we can in principle obtain.  Rough estimates of
the impact of these various contributions are briefly discussed in
\S8. Detailed analysis including these effects is in progress.

In the next section we briefly summarize the SZ effect.  In \S3 we
investigate how varying observing frequencies and detector
sensitivity affects parameter degeneracies and parameter extraction.
In \S4 a Markov chain/Fisher matrix method is used to determine
which cluster parameters can be well constrained by future SZ
measurements, and in \S5 this method is applied to simulated
Nbody+gasdynamics clusters and the results presented. In \S6 we
discuss the near-linearity of the SZ intensity shift with respect to
$\tau T_{e}$, $\tau v_{z}$, and $\tau T_{e}^{2}$, and show the
resulting close correspondence between the constrained effective
parameters from 2D SZ images and line-of-sight integrals through the
3D cluster. In \S7 we show that an independent measurement of
$T_{e}$ breaks the parameter degeneracy and use a simple analytical
model to show that the velocity thus determined is approximately the
optical-depth-weighted velocity integrated along the cluster line of
sight.  We then use a Markov chain to calculate errors on cluster
velocities and optical depths given X-ray temperature priors.  In
\S8 we discuss sources of contamination to the SZ signal, and we
conclude with a summary of the above.

\section{SZ EFFECT}

When microwave photons pass through the hot gas in the
intracluster medium of a galaxy cluster, roughly $1\%$ of the
photons interact with the free electrons in the gas.  These
photons are inverse Compton scattered and energy is transferred
from the hot electrons to the cool photons, causing a slight
distortion of the microwave background spectrum. This
up-scattering of photons causes the intensity of photons with
frequencies below about 220 GHz to decrease while the intensity of
photons with higher frequencies increases.  This process is called
the thermal SZ effect, and it causes an effective temperature
shift relative to the mean microwave background temperature on the
order of one part in $10^{4}$. If the galaxy cluster has some bulk
velocity with respect to the microwave background rest frame, then
this will Doppler shift the scattered microwave photons and cause
an additional spectral distortion. This further shift in the
microwave spectrum is referred to as the kinematic SZ effect and
is typically an order of magnitude smaller than the thermal SZ
effect. The derivation of the combined SZ effect can be found in
\citet{sunyaev1} and \citet{sunyaev}, and more recent papers (e.g.
\citet{itoh1, itoh2}) have included relativistic corrections to
these derivations.

The expression for the SZ effect we use throughout this work is
from \citet{itoh2} and is given by
\begin{eqnarray}
\lefteqn{\frac{\triangle
I_{\nu}}{I_{0}}=\frac{X^{4}e^{X}}{(e^{X}-1)^{2}}\tau \theta_{e}
\bigl[Y_{0} + \theta_{e} Y_{1} + \theta_{e}^{2} Y_{2} +
\theta_{e}^{3} Y_{3} + \theta_{e}^{4} Y_{4} \bigr]}
                                         \nonumber\\
& & {}+\frac{X^{4}e^{X}}{(e^{X}-1)^{2}}\tau
(\frac{v_{\mathrm{tot}}}{c})^{2} \bigl[\frac{1}{3} Y_{0} +
\theta_{e} (\frac{5}{6} Y_{0} + \frac{2}{3} Y_{1}) \bigr]
                                         \nonumber\\
& & {}+\frac{X^{4}e^{X}}{(e^{X}-1)^{2}}\tau \frac{v}{c} \bigl[1 +
\theta_{e} C_{1} + \theta_{e}^{2} C_{2} \bigr]
\end{eqnarray}
where $I_{0}=2(k_{B} T_{CMB})^{3}/(hc)^{2}$, $X=h\nu/k_{B} T_{CMB}$,
$\theta_{e}=k_{B}T_{e}/m_{e}c^{2}$, $v_{\mathrm{tot}}$ is the
peculiar velocity, $v$ is the peculiar line-of-sight velocity, $\tau
= \sigma_{T}\int n_{e} dl$ is the optical depth, and the $Y$'s and
$C$'s are numbers that depend on frequency in a well-defined way.
Note that $\tau \theta_{e}$ is equivalent to the well known
Compton-y parameter which we will show is tightly constrained.  The
most dominant terms in this expression are proportional to $\tau
T_{e}$, $\tau T_{e}^{2}$, and $\tau v$. It is also important to note
that the above expression is independent of redshift. This makes the
SZ effect a powerful probe of the high-redshift universe because the
amplitude of the SZ signal does not weaken at high redshift (for
fixed $T$, $v$, $\tau$), unlike X-ray and optical signals.  Since
this microwave intensity shift for a given frequency is a
non-trivial function of the cluster's gas temperature, peculiar
velocity, and optical depth, one could hope that choosing at least
three well placed observing frequencies would allow these three
cluster parameters to be separated and measured. In practice, the
observing frequencies and sensitivities available to upcoming SZ
surveys result in degeneracies among these parameters. These
degeneracies are illustrated in the next section.

\section{PARAMETER EXTRACTION}
\subsection{Creating Likelihood Surfaces}

To understand the intrinsic limitations in determining a galaxy
cluster's gas temperature, peculiar velocity, and optical depth from
multi-frequency SZ measurements, we create likelihood surfaces for
these parameters and compare the 1-$\sigma$ regions for various
choices of observing frequencies and detector sensitivity. We do
this by first assuming some region of gas with uniform temperature
$T_{e}$, peculiar line-of-sight velocity $v$, and total optical
depth $\tau$. We neglect any transverse peculiar velocity since the
full SZ effect generates a temperature shift on the order of one
part in $10^{4}$, and the transverse velocity component of the SZ
effect contributes a temperature shift on the order of one part in
$10^{7}$.

Using the expression for the intensity shift given in eq. (1), we
calculate the change in intensity one would measure from our
fiducial gas region at three different observing frequencies.  We
perform these calculations for several frequency sets.  The first
observing frequency set we choose to be (30, 150, 300 GHz), which is
the optimal frequency set found by \citet{holder} and roughly that
found by \citet{aghanim}. We also choose the sets (90, 150, 300 GHz)
and (145, 225, 265 GHz), the latter being the frequencies planned
for ACT (see \citet{kosowsky}). After obtaining the intensity shift
for each of the three frequencies in each set, we change variables
to the ratios $x_{1}=\Delta I_{\nu_{1}}/\Delta I_{\nu_{2}}$ and
$x_{2}=\Delta I_{\nu_{3}}/\Delta I_{\nu_{2}}$, where $\nu_{1}$,
$\nu_{2}$, and $\nu_{3}$ are the three frequencies in a set. These
ratios are independent of $\tau$ (though their errors depend on
$\tau$), so we have two equations and two unknowns ($T_{e}$ and
$v$). We eliminated $\tau$ so that we could more easily construct
and view the likelihood surfaces of the remaining parameters.
Assuming first $1\mu K$ detector noise per arcminute beam and then
$10\mu K$ noise per beam, we step through the $T_{e} - v$ parameter
space and calculate the likelihood of the model described by
($T_{e}$, $v$), given the underlying fiducial model, in the usual
way. Thus, we create a 2-dimensional likelihood surface for the
parameters $T_{e}$ and $v$ for each observing frequency set at both
detector noise levels. A 1-$\sigma$ contour is then drawn for each
likelihood surface by connecting all the points with $\chi^{2} =
2.3$ (e.g. \citet{numerical}).

\subsection{Degeneracy Between Cluster Gas Parameters}

The 1-$\sigma$ contours we obtain from these likelihood surfaces
verify that the set of frequencies 30, 150, 300 GHz puts tighter
constraints on $T_{e}$ and $v$ than the other sets.  With $1\mu K$
detector noise per arcminute beam, the set 145, 225, 265 GHz
exhibits a clear degeneracy between the $T_{e}$ and $v$ parameters,
and with $10\mu K$ detector noise per arcminute beam all three
frequency sets show significant degeneracy. Figures 1a-1c show the
comparison of 1-$\sigma$ contours for 10 keV ($1.2\times 10^{8}$ K)
gas with a line-of-sight velocity of 200 km/sec, optical depth of
0.012, and $1\mu K$ detector noise. We choose gas parameters
corresponding to a hot cluster because it is the case which has the
highest signal-to-noise.  These parameter values roughly correspond
to the known cluster MS 0451 \citep{reese}. The gas temperature is
constrained to within 0.5 keV and its velocity to within 25 km/sec
at the 1-$\sigma$ level for 30, 150, and 300 GHz observing
frequencies (figure 1a). Shifting the lowest frequency to 90 GHz
increases these uncertainties by a factor of two (figure 1b). Using
observing frequencies at 145, 225, and 265 GHz results in
temperature uncertainties of 6 keV and velocity uncertainties of 220
km/sec (figure 1c). Figures 2a-2c show the 1-$\sigma$ likelihood
contours for 10 keV gas with a line-of-sight velocity equal to -200
km/sec, optical depth of 0.012, and $1\mu K$ detector noise.
Comparing these to the previous group of figures indicates that a
negative line-of-sight velocity somewhat increases the uncertainties
obtained for all frequency sets, as has been pointed out in
\citet{aghanim}. Adding a 30 GHz or 90 GHz observing frequency to
the 145, 225, 265 GHz frequency set with $1\mu K$ noise greatly
reduces the 1-$\sigma$ regions and makes the temperature and
velocity constraints as tight as for the 30, 150, 300 GHz and 90,
150, 300 GHz frequency sets respectively. These results confirm
previous results by \citet{holder} and \citet{aghanim}. It is clear
that measurements near the null of the SZ effect are not
particularly useful for SZ studies from a signal-to-noise
perspective, but such measurements will be useful for microwave
observations that aim to minimize SZ contamination and as a useful
diagnostic of point source contamination.

With $10\mu K$ detector noise per arcminute beam, figures 3a-3c show
the 1-$\sigma$ contours for the same 10 keV gas region with 200
km/sec line-of-sight velocity and an optical depth of 0.012. The gas
temperature is constrained to within 5 keV and the velocity to
within 200 km/sec at the 1-$\sigma$ level for the 30, 150, and 300
GHz set (figure 3a).  Increasing the lower frequency to 90 GHz
constrains the temperature to within 8 keV and the velocity to
within 250 km/sec (figure 3b).  The 145, 225, and 265 GHz frequency
set gives 1-$\sigma$ errors of 9 keV and 350 km/sec for temperature
and velocity respectively with $10\mu K$ detector noise (figure 3c).
Thus to constrain the cluster parameters well using SZ observations
alone (in the absence of SZ signal contamination from point sources
or residual primary microwave background), one needs both a low
frequency band (90 GHz or lower) and a detector sensitivity not much
higher than $1\mu K$ per arcminute beam.

We also compare the 1-$\sigma$ likelihood contours for gas with
different Compton-y parameters keeping the observing frequencies
fixed.  In figure 4a we see the 1-$\sigma$ contour for a 7 keV gas
region with an optical depth of 0.009, 200 km/sec line-of-sight
velocity, $1\mu K$ detector noise, and observing frequencies at 30,
150, and 300 GHz. Figure 4b shows the 1-$\sigma$ contour for a 3 keV
gas region with an optical depth of 0.004 and the same line-of-sight
velocity, detector noise, and observing frequencies. The 1-$\sigma$
contour for the gas region with lower Compton-y parameter shows a
larger parameter degeneracy for the same set of observing
frequencies and detector noise.  This is because a decrease in
Compton-y parameter lowers the signal-to-noise.  The signal-to-noise
is reduced by two effects as one moves to lower mass clusters: a
lower signal in all bands and reduced relative importance of
relativistic effects.  We have verified that the former is a more
important contributor to parameter degeneracies than the latter.
Table \ref{Table1} lists the above 1-$\sigma$ errors on $T_{e}$ and
$v$ for differing observing frequencies, detector noise, and
Compton-y parameters for convenient reference.

\section{CONSTRAINED PARAMETERS FROM FUTURE SZ SURVEYS}

Despite the inability of upcoming experiments like ACT, SPT, and
APEX to determine all of the cluster gas parameters, they will
provide tight constraints on certain combinations of parameters.
The next two sections explicitly demonstrate the parameter
combinations which will be determined with good precision by these
kinds of experiments.

\subsection{Markov Chain Analysis}

We create a Markov chain Monte Carlo (MCMC) using the parameters
$T_{e}$, $v$, and $\tau$ to find realistic error regions for all
three parameters.  This MCMC is made using the Metropolis-Hastings
algorithm which randomly steps through a parameter space and accepts
all points whose likelihood is greater than the previous point. If
the likelihood is less than the previous point, the current point is
accepted with a probability given by the ratio of the two
likelihoods.  A comprehensive review of MCMCs can be found in
\citet{gilks}; they were introduced into cosmology by \citet{chris},
\citet{chris2}, \citet{lewis}, and \citet{kosowsky2}. The region in
the 3-dimensional parameter space containing $68\%$ of all the
points accepted in the chain we define as the 1-$\sigma$ region.
This 1-$\sigma$ region is projected onto 2 dimensions in the figures
below. Figures 5a-5c show the 1-$\sigma$ regions generated by a MCMC
for 10 keV gas with 200 km/sec line-of-sight velocity and an optical
depth of 0.012. The frequency set 145, 225, and 265 GHz was used
with $1\mu K$ detector noise per beam to simulate ACT observations.
In the Markov chain, the parameter space was restricted to $T_{e}\in
(0, 2\times 10^{8}$ K (17 keV)), $\tau\in$(0, 0.02), and
$v\in$(-1500 km/sec, 1500 km/sec), and 5 million steps were used
with about $50\%$ acceptance rate in the chain.  In addition, each
step was taken in all three parameter directions simultaneously with
different step sizes in each parameter direction.  From the
1-$\sigma$ regions in figures 5a-5c, we can see a clear degeneracy
among all three parameters.

Figures 6a-6c show the same 1-$\sigma$ region as figures 5a-5c
except using the parameter directions $\tau T_{e}$, $\tau
T_{e}^{2}$, and $\tau v$.  We have verified that the MCMC in either
variable set gives the same results, although it is significantly
more efficient to use the second set of variables. We choose this
set of parameters since these are the dominant terms in the
intensity shift expression in eq. (1).  These figures demonstrate
that the cluster parameters lie in a nearly 1-dimensional subspace
of the 3-dimensional parameter space given by $\tau T_{e}$, $\tau
T_{e}^{2}$, and $\tau v$. We can find two axes within this parameter
space (both orthogonal to the degeneracy direction) along which the
cluster parameters are tightly constrained and one axis (parallel to
the degeneracy direction) along which the cluster parameters are
largely unconstrained.

\subsection{Fisher Matrix Determination of Constrained Parameters}

The three orthogonal directions in this parameter space that allow
us to tightly constrain the gas parameters in two directions with
one direction unconstrained correspond to the principal axes of the
1-$\sigma$ error ellipsoid which is projected in 2 dimensions in
figures 6a-6c.  To find these principal axes we calculate a Fisher
matrix at the most likely point found by the MCMC. The Fisher matrix
describes the curvature of the likelihood surface at a given point
in parameter space.  It is expressed by the formula
\begin{eqnarray} F_{\alpha\beta}=\left\langle-
\frac{d^{2}\ln\mathcal{L}}{dp_{\alpha}dp_{\beta}}\right\rangle.
\end{eqnarray}
Since we have $\mathcal{L} \propto e^{-\chi^{2}/2}$, the Fisher
matrix becomes $F_{\alpha\beta}=\frac{1}{2}\frac{d^{2}
\chi^{2}}{dp_{\alpha}dp_{\beta}}$.  Using
\begin{eqnarray}
\chi^{2} = \sum_{i=1}^{3}\Bigg(\frac{\Delta
I_{\nu_{i}}(\textbf{\emph{p}}) - \Delta
I_{\nu_{i}}^{o}}{e_{\nu_{i}}}\Bigg)^{2},
\end{eqnarray}
where $\textbf{\emph{p}}=($$\tau T_{e}$/($10^{6}$ K), $\tau v$/(1
km/sec), $\tau T_{e}^{2}$/($10^{14}$ K$^{2}$)), $\Delta
I_{\nu_{i}}^{o}$ is the observed $\Delta I_{\nu_{i}}$, and
$e_{\nu_{i}}$ is the error on $\Delta I_{\nu_{i}}^{o}$, and
averaging gives
\begin{eqnarray}
F_{\alpha\beta}=
\sum_{i=1}^{3}\frac{1}{(e_{\nu_{i}})^{2}}\frac{\partial\Delta
I_{\nu_{i}}(\textbf{\emph{p}})}{\partial
p_{\alpha}}\frac{\partial\Delta
I_{\nu_{i}}(\textbf{\emph{p}})}{\partial p_{\beta}}.
\end{eqnarray}
(See \citet{dodelson} for a good explanation of the Fisher matrix
and its applications.)  The eigenvectors of the Fisher matrix
evaluated at the minimum point of $\chi^{2}$ correspond to the
principal axes of the error ellipsoid.

This technique is essentially a version of principal component
analysis (PCA).  In PCA it is known that the principal components
one finds depend on how the variables are scaled. There is no
single correct scaling to choose, but the one that is widely
preferred in physical uses is scaling all the variables to order
unity, which has the advantage that all the variables are weighted
the same.  (\citet{jackson} provides a good overview of PCA.)  We
scale $\tau T_{e}$, $\tau v$, and $\tau T_{e}^{2}$ by $10^{6}$ K,
1 km/sec, and $10^{14}$ K$^{2}$ respectively, which are
characteristic cluster values for these variables.

The principal axes are linear combinations of the previous
parameter directions such that
\begin{eqnarray}
\textbf{\emph{a}}=\mathrm{C}(\textbf{\emph{p}}_{\ast})\textbf{\emph{p}},
\end{eqnarray}
where the rows of $\mathrm{C}(\textbf{\emph{p}}_{\ast})$ are the
eigenvectors of $\mathrm{F}$, $\textbf{\emph{a}}=(a, b, c)$ is a
point in the new parameter space, and $\textbf{\emph{p}}$ are the
old parameter directions described above. The vector
$\textbf{\emph{p}}_{\ast}$ describes $\textbf{\emph{p}}$ evaluated
at the maximum likelihood point. Note
$\mathrm{C}(\textbf{\emph{p}}_{\ast})$ (and thus
$\textbf{\emph{a}}$) will differ with $\textbf{\emph{p}}_{\ast}$,
but the variation with $\textbf{\emph{p}}_{\ast}$ is fairly weak for
realistic parameter regions. Roughly, the Fisher matrix eigenvectors
are $\textbf{\emph{e}}_{1}=(1, 0, 0)$, $\textbf{\emph{e}}_{2}=(0,
0.4, 0.9)$, and $\textbf{\emph{e}}_{3}=(0, 0.9, 0.4)$ normalized to
unity. Therefore the $a$ parameter is dominated by $\tau T_{e}$, and
the $b$ and $c$ parameters are primarily linear combinations of
$\tau v$ and $\tau T_{e}^{2}$. Thus, SZ measurements provide precise
measures of the $y$ parameter and one linear combination of the
kinematic SZ effect and the relativistic corrections.

Figures 7a-7c show that the gas parameters for this fiducial model
are constrained to within $1\%$ in the $a$ direction, $3\%$ in the
$b$ direction, and $70\%$ in the c direction. For a 4 keV gas region
with 200 km/sec line-of-sight velocity, an optical depth of 0.005,
and detector noise of $1\mu K$, the gas parameters are constrained
to within $4\%$ in the $a$ direction, $16\%$ in the $b$ direction,
and $200\%$ in the $c$ direction.  The gas parameters are
constrained to within $8\%$, $22\%$, and $160\%$ in the $a$, $b$,
and $c$ directions respectively for a 10 keV gas region with 200
km/sec line-of-sight velocity, an optical depth of 0.012, and $10\mu
K$ detector noise.

The $a$ and $b$ parameters, which are linear combinations of $\tau
T_{e}$, $\tau T_{e}^{2}$, and $\tau v$, are therefore well
constrained by an ACT-like SZ survey with observing frequencies
near 145, 225, and 265 GHz and $1\mu K$ detector noise.  This
technique can be applied to determine the constrained gas
parameters from any multi-frequency SZ observations that are
without arcminute-resolution observations at frequencies below 100
GHz or that have low frequency information but with a noise component
considerably larger than $1\mu K$.
By combining this gas information from SZ surveys with a data
set that can constrain just one of the parameters $T_{e}$, $\tau$,
$v$ (such as an X-ray survey of clusters that constrains $T_{e}$
for each cluster), all three cluster gas parameters can be well
determined.

\section{RESULTS USING SIMULATED CLUSTERS}

We now apply this technique to simulated SZ observations which we
generate using simulated galaxy clusters.  In \S5.1 we describe the
two simulated clusters we use, one about 9 keV and the other about 3
keV, and in \S5.2 we discuss how SZ simulations are created from
these.  The results of applying the technique in \S4 to simulated
ACT-like SZ maps of both clusters and to simulated Planck-like SZ
maps of the 9 keV cluster are contained in \S5.3.

\subsection{Cluster Simulations}

The clusters we use are two high-resolution 3D cluster simulations
that were made using the Adaptive Refinement Tree (ART)
Nbody+gasdynamics code \citep{kravtsov1, kravtsov2}.  These
clusters were simulated using a $\Lambda$CDM model with
$\Omega_{m}=0.3$, $\Omega_{b}=0.043$, $h=0.7$, and
$\sigma_{8}=0.9$. Each cluster has a redshift of z=0.43 and is
contained in a cube of side length 2 Mpc ($\simeq$ 6 arcminutes).
Each grid element within the larger cube has a side length of
0.0078 Mpc ($\simeq$ 0.02 arcminute).  The simulations track the
density of dark matter particles, the density of gas particles,
the gas temperature, and the 3-dimensional gas velocity for each
grid element.  One cluster is similar in size to the Coma cluster
and has a mass of $\simeq 10^{15} M_{\odot}$, an optical depth of
$\simeq 0.01$, and an average gas temperature of 9 keV. The other
cluster is similar in size to the Virgo cluster and has a mass of
$\simeq 2 \times 10^{14} M_{\odot}$, an optical depth of $\simeq
0.005$, and an average gas temperature of 3 keV. Both clusters
have characteristic bulk velocities of several hundred km/sec. The
morphology of the Virgo-size cluster indicates that it consists of
a recent merger of two smaller clusters. These cluster simulations
do not include the effects of gas cooling, stellar feedback,
magnetic fields, and thermal conduction.  For a more detailed
cluster description see \citet{daisuke1} and \citet{daisuke2}.

\subsection{SZ Map Generation}

To create a simulation of an SZ observation, we choose one of the
cluster simulations and a set of observing frequencies. The SZ
intensity shift that microwave photons would experience passing
through each grid element is calculated using eq. (1) and the gas
temperature, gas density, and gas velocity of each element. Every
$\Delta I$ is then integrated over a frequency band centered around
each observing frequency in the set.  We use a 3 GHz frequency
bandwidth, as opposed to a more realistic bandwidth of $\simeq$ 25
GHz, for numerical convenience.  However, the bandwidth size has a
negligible effect on cluster constraints, which we verified by
redoing some of our results with a 25 GHz bandwidth. Thus, for each
frequency band we end up with an SZ cube of $\Delta I$ values. This
cube is then projected along the line of sight into a
two-dimensional SZ distortion of the sky. We do not include the
primary microwave background in our simulations, assuming it is
perfectly subtracted, since it varies on scales large compared to
the cluster. The main effect of residual primary microwave
contamination will be as a source of noise for extracting estimates
of the peculiar velocity from the constrained parameters. Figures
8a-8c and 9a-9c show the 2D SZ images of the 9 keV and 3 keV
clusters after this projection process for the frequency bands
centered on 145, 225, and 265 GHz. After creating a 2D SZ image for
each observing frequency band, we smooth each image by convolving it
with a Gaussian beam, and increase the pixel size of our images to
sizes realistic for upcoming SZ surveys by averaging together
smaller pixels. Finally, Gaussian random noise of standard deviation
equal to our chosen detector sensitivity is added to each 2D pixel.

We make SZ simulations using ACT instrument specifications for both
the 3 keV and 9 keV clusters and using Planck specifications for the
9 keV cluster only.  The ACT-like SZ images of the 9 keV cluster
(assuming perfect microwave background and point source removal) are
shown in figures 10a-10c. For these images we use the frequency
bands centered on 145, 225, and 265 GHz.  We assume a beam size of 1
arcminute and choose a pixel size of 0.3' x 0.3'. Each 0.3 arcminute
pixel is given $3\mu K$ of Gaussian random detector noise. Our
results are not qualitatively different for moderately different
noise levels between channels. Figures 11a-11c show SZ simulations
of the 3 keV cluster with ACT specifications as above. For
comparison, we also made maps appropriate to the Planck experiment,
with a beam size of 4 arcminutes and detector noise of $16\mu K$ per
2' x 2' pixel, at the same frequencies as the ACT maps (which are
similar to the actual Planck bands centered at 143, 217, and 353
GHz).

\subsection{Parameter Constraints from Simulated 3D Clusters}

We now apply the Markov chain/Fisher matrix technique described in
\S4 to the simulated ACT-like and Planck-like SZ images of the
simulated clusters.  The only change is that in eq. (3) we assume
$3\mu K$ noise for ACT-like images and $16\mu K$ noise for
Planck-like images. The cluster parameters we constrain using this
method are really effective parameters that correspond to integrals
of the cluster parameters along a line of sight. The SZ intensity
(eq. [1]) is not linear, so there is no guarantee that the sum of SZ
signals of varying temperature and velocity can be fit to a single
temperature and velocity. We fit the resulting SZ intensity as a
function of frequency to a model with a single temperature and
velocity and call the constrained parameters
$\textbf{\emph{a}}_{\mathrm{eff}}$; we discuss what integrals these
effective parameters correspond to within the three dimensional
cluster in \S6.

Figures 12a-12c show the projected 1-$\sigma$ contours for the
$\textbf{\emph{a}}_{\mathrm{eff}}$ parameters for the central pixel
of ACT-like SZ images of the simulated 9 keV cluster.  We do this
analysis on a pixel-by-pixel basis to potentially obtain the most
information about cluster substructure. (See \citet{daisuke2} for
some discussion of substructure.)  In a low signal-to-noise
experiment or as a means to average out substructure, it could be
advantageous to add together pixels.  The 1-$\sigma$ errors on
$a_{\mathrm{eff}}$, $b_{\mathrm{eff}}$, and $c_{\mathrm{eff}}$ for
this central pixel are 0.06, 0.5, and 5 respectively. The range of
parameters $T \in (0, 2\times 10^{8}$ K), $\tau \in (0, 0.02)$, and
$v \in$ (-1500 km/sec, 1500 km/sec) correspond to ranges of about $a
\in (0, 4)$, $b \in (-5, 20)$, and $c \in (-30, 24)$. Clusters with
larger SZ signals (and thus larger $T_{e}$'s and $\tau$'s) tend to
have larger $a$, $b$, and $c$ parameters, and this is borne out in
the comparison of our results for the 9 keV and 3 keV clusters.
Since the $a$, $b$, and $c$ parameters have been scaled to roughly
order unity for characteristic cluster values, the absolute errors
are a meaningful reflection of how well the gas properties are
constrained.

Similar results are obtained from the ACT-like SZ simulation of the
3 keV cluster and the Planck-like SZ simulation of the 9 keV
cluster.  Figures 13a-13c show the projected 1-$\sigma$ contours for
the central pixel of the ACT-like SZ images of the 3 keV cluster.
These figures demonstrate $\sigma_{a_{\mathrm{eff}}} \simeq 0.02$,
$\sigma_{b_{\mathrm{eff}}} \simeq 0.2$, and
$\sigma_{c_{\mathrm{eff}}} \simeq 1.5$. Thus $a_{\mathrm{eff}}$,
$b_{\mathrm{eff}}$, and $c_{\mathrm{eff}}$ are constrained to a
small region of the available parameter space, with the first two
components especially well-constrained. Figures 14a-14c show the
projected 1-$\sigma$ contours for the central pixel of the
Planck-like SZ images of the 9 keV cluster. These figures show
$\sigma_{a_{\mathrm{eff}}} \simeq 0.06$, $\sigma_{b_{\mathrm{eff}}}
\simeq 0.7$, and $\sigma_{c_{\mathrm{eff}}} \simeq 3$, indicating
again that the measurements are providing strong constraints within
the available parameter space.

These results demonstrate that the $a_{\mathrm{eff}}$ and
$b_{\mathrm{eff}}$ cluster parameters are well constrained and
$c_{\mathrm{eff}}$ is moderately well constrained by SZ observations
typical of ACT and Planck, assuming perfect primary microwave
background and point source removal.

\section{CORRESPONDENCE BETWEEN CONSTRAINED EFFECTIVE PARAMETERS AND LINE-OF-SIGHT INTEGRALS}

The cluster gas parameters we have constrained in the previous
sections using a projected two dimensional SZ image correspond to
integrals along the line of sight of the three dimensional cluster.
Previous studies of cluster projection effects have been done
indicating that the temperature in principle obtainable from SZ
measurements is really a Compton-averaged quantity (i.e. each
line-of-sight integral of $T_{e}$ is weighted by the Compton
parameter) \citep{hansen,knox}. Here we show that the line-of-sight
integrals corresponding to the $\textbf{\emph{a}}_{\mathrm{eff}}$
parameters are even more straightforward.

The reason for the simple correspondence is that the SZ intensity
shift given by eq. (1) is nearly linear with respect to $\tau T$,
$\tau v$, and $\tau T^2$.  These three terms are the most dominant
terms in the expression and represent most of the change in
intensity for temperatures of several keV and velocities of several
hundred km/s. If $\Delta I_{\nu}$ were exactly a function only of
$\tau T$, $\tau v$, and $\tau T^2$ in eq. (1), then $\Delta I_{\nu}$
would be exactly a linear function of $a$, $b$, and $c$. In that
case, the measured $\textbf{\emph{a}}_{\mathrm{eff}}$ would be equal
to $\sum \mathrm{C} \textbf{\emph{p}}_{i} $, where $\mathrm{C}$ is
given in eq. (5), and the sum is over the gas properties
$\textbf{\emph{p}}_{i}$ of each element $i$ along the line of sight.

The full SZ intensity shift expression in eq. (1) includes terms
non-linear in $\tau T$, $\tau v$, and $\tau T^2$. However, the
addition of these non-linear terms in the calculation of $\Delta
I_{\nu}$ integrated along a line of sight only introduces a slight
bias between $\textbf{\emph{a}}_{\mathrm{eff}}$ and $\int \mathrm{C}
d\textbf{\emph{p}}_{i}$. In figures 12-14, within the 1-$\sigma$
contours of $\textbf{\emph{a}}_{\mathrm{eff}}$, we indicate the best
fit $\textbf{\emph{a}}_{\mathrm{eff}}$ found by the MCMC, for a
given noise realization, by a star ($\bigstar$).  The best fit
$\textbf{\emph{a}}_{\mathrm{eff}}$ obtained if the clusters are
observed with an ideal instrument without detector noise are
indicated by dots ($\bullet$).  Diamond shapes ($\blacklozenge$)
mark the values of the line-of-sight integrals given by $\int
\mathrm{C} d\textbf{\emph{p}}_{i}$ and calculated using the 3D
cluster simulations.  In figure 12a, the difference between
$a_{\mathrm{eff}}$ from an ideal, no noise instrument and from the
line-of-sight integral is $\Delta a=0.001$. The difference between
$b_{\mathrm{eff}}$ from an ideal instrument and from the
line-of-sight integral is $\Delta b=-0.03$.  Clearly, in the absence
of detector noise, the correspondence between $a_{\mathrm{eff}}$ and
$b_{\mathrm{eff}}$ and the line-of-sight integrals is very close.
Moreover, the difference between $a_{\mathrm{eff}}$ and
$b_{\mathrm{eff}}$ and the line-of-sight integrals given realistic
detector noise is still well within the 1-$\sigma$ errors on
$a_{\mathrm{eff}}$ and $b_{\mathrm{eff}}$ given by the MCMC. Similar
results can be seen in figures 13a and 14a. These simulations
demonstrate an agreement to within 1-$\sigma$ between the
$a_{\mathrm{eff}}$ and $b_{\mathrm{eff}}$ parameters constrained by
SZ measurements and line-of-sight integrals given by $\int
\mathrm{C} d\textbf{\emph{p}}_{i}$.

Figure 15 shows this correspondence more explicitly.  Plotted on the
y-axis are best-fit $\textbf{\emph{a}}_{\mathrm{eff}}$ obtained via
a MCMC using intensities from simulated noise-free SZ images of the
9 keV and 3 keV clusters assuming ACT-like observations. On the
x-axis are plotted $\int \mathrm{C} d\textbf{\emph{p}}_{i}$ for the
corresponding lines of sight.  Four different lines of sight through
both the 9 keV and 3 keV simulated clusters are plotted.  These
lines of sight are 0', 1', 1.5', and 2' from the central pixel of
the simulated SZ images. Filled shapes correspond to the 9 keV
cluster and unfilled shapes to the 3 keV cluster.  For the
$a_{\mathrm{eff}}$ and $b_{\mathrm{eff}}$ parameters, which can be
well constrained, their equivalence to $\int \mathrm{C}
d\textbf{\emph{p}}_{i}$ is extremely close.  This confirms the near
linearity of the SZ intensity shift with respect to the a, b, and c
parameters.  The $c_{\mathrm{eff}}$ parameters demonstrate some
scatter around the $y=x$ line, and this is because the degeneracy in
the $c$ direction prevents a MCMC from settling on the correct
$c_{\mathrm{eff}}$ value.

If we could tightly constrain $a_{\mathrm{eff}}$,
$b_{\mathrm{eff}}$, and $c_{\mathrm{eff}}$ via SZ measurements, we
could solve for the quantities $\tau T_{\tau}$, $\tau v_{\tau}$, and
$\tau (T^{2})_{\tau}$, where the subscript $\tau$ corresponds to
optical-depth-weighted integrals (e.g. $T_{\tau} = \int T d\tau /
\int d\tau$).  From these one can find $T_{y}$, $\tau
T_{\tau}/T_{y}$, and $v_{\tau} T_{y}/T_{\tau}$, following the
algebra in \citet{knox}, where the subscript $y$ corresponds to a
pressure-weighted integral.  Therefore SZ measurements would
constrain the pressure-weighted temperature (arising from the
relativistic corrections), the optical-depth-weighted velocity times
a correction factor and the optical depth times a similar correction
factor which is the ratio of different weighted temperatures.  Since
degeneracies will allow SZ measurements to constrain only
$a_{\mathrm{eff}}$ and $b_{\mathrm{eff}}$, information from an
external source will be needed to constrain the above physically
interesting quantities.

\section{BREAKING PARAMETER DEGENERACY WITH AN X-RAY MEASUREMENT OF $\textbf{$T_{e}$}$}

\subsection{Measured Effective Velocity is approximately $\int v d\tau / \int d\tau$}

Assuming contamination sources can be dealt with effectively, future
SZ observations should be able to constrain two quantities given by
\begin{eqnarray}
a_{\mathrm{eff}} & \approx & \tau (c_{1} T_{\tau} + c_{2}
(T^{2})_{\tau}) + c_{3} \tau v_{\tau}  {\textrm{\hspace{0.3cm} and}}\\
b_{\mathrm{eff}} & \approx & \tau (c_{4} T_{\tau} + c_{5}
(T^{2})_{\tau}) + c_{6} \tau v_{\tau},
\end{eqnarray}
where the $c$'s are elements of the matrix $\mathrm{C}$ defined in
eq. (5). It is conceivable that temperature measurements from an
X-ray survey of clusters could allow the determination of $\tau$ and
$v_{\tau}$.

Formally X-ray observations would need to provide a constraint on
$c_{i} T_{\tau} + c_{j} (T^{2})_{\tau}$, where the $c$'s are known
constants.  However, $T_{\tau}$ and $(T^{2})_{\tau}$ are not
obviously given by X-ray observations.  An X-ray derived temperature
is also not equivalent to $T_{\tau}$. The two may differ by as much
as 1 keV \citep{math}.  If X-ray observations gave $T_{x} =
T_{\tau}$ and it was true that $(T_{\tau})^{2} = (T^{2})_{\tau}$,
then the effective velocity we would get from a MCMC, after adding
the $T_{x}$ prior, would be equal to $v_{\tau}$.  For our 9 keV
simulated cluster, $(T_{\tau})^{2}$ and $(T^{2})_{\tau}$ agree to
within $5\%$ on average, and for our 3 keV simulated cluster, the
two agree to within $12\%$ on average.

To get an estimate of the biases incurred by not having the correct
weighted temperatures, we assume $(T_{\tau})^{2} = (T^{2})_{\tau}$
and add a temperature prior of $T_{\tau}$ to our MCMC.   Adding a
$T_{\tau}$ prior, in the manner we discuss further in \S7.2, we find
for the central pixel of the 9 keV cluster, from simulated,
noise-free, ACT-like SZ images, an effective velocity of 230 km/sec
from the MCMC and an optical-depth-weighted line-of-sight velocity
of 218 km/sec from the three dimensional cluster simulation. For the
central pixel of the 3 keV cluster SZ image, we find an effective
velocity of -10 km/sec from the MCMC and an optical-depth-weighted
line-of-sight velocity of -3 km/sec from the three dimensional
cluster simulation. The bias between the velocity from the MCMC and
$v_{\tau}$ is most likely due to the breakdown of the
$(T_{\tau})^{2} = (T^{2})_{\tau}$ assumption.

To quantify the bias incurred from $T_{x}$ differing from
$T_{\tau}$, we add a $\pm$ 1 keV offset to our $T_{\tau}$ prior.  We
find for the central pixel of the 9 keV cluster SZ image, an
effective velocity of 228 km/sec from the MCMC for both + and - 1
keV offsets. We find for the central pixel of the 3 keV cluster SZ
image, an effective velocity of -45 km/sec from the MCMC for a +
1keV offset and an effective velocity of 6 km/sec for a - 1 keV
offset.  This would suggest a total bias between the measured
effective velocity and $v_{\tau}$ of about 15 km/sec for the 9 keV
cluster and between 10 and 40 km/sec for the 3 keV cluster.

\subsection{MCMC Errors on $v_{\mathrm{eff}}$ and $\tau$ Given an
X-ray $T_{\mathrm{eff}}$ Prior}

To determine realistic errors on $v_{\mathrm{eff}}$ and $\tau$ from
adding a measurement of $T_{\mathrm{eff}}$ to our simulated SZ data,
we again use a MCMC.  We weight each point in our MCMC by the factor
$e^{-((T-T_{\mathrm{eff}})/\Delta T_{\mathrm{eff}})^{2}/2}$ where we
calculate $T_{\mathrm{eff}} = \int T_{e} d\tau/\int d\tau$ from the
3D cluster simulation and $\Delta T_{\mathrm{eff}}$ is the assigned
measurement error on $T_{\mathrm{eff}}$.

For a 1 keV error on $T_{\mathrm{eff}}$, our ACT simulation of the 9
keV cluster gives $\sigma_{v} = 20$ km/sec and $\sigma_{\tau} =
0.002$ for the central pixel.  Our ACT simulation of the 3 keV
cluster gives $\sigma_{v} = 60$ km/sec and $\sigma_{\tau} = 0.004$
for the central pixel.  Assuming a 2 keV error on
$T_{\mathrm{eff}}$, the ACT simulation of the 9 keV cluster gives
$\sigma_{v} = 40$ km/sec and $\sigma_{\tau} = 0.004$, and the ACT
simulation of the 3 keV cluster gives $\sigma_{v} = 100$ km/sec and
$\sigma_{\tau} = 0.005$.  Our Planck simulation of the 9 keV cluster
gives $\sigma_{v} = 500$ km/sec and $\sigma_{\tau} = 0.0024$ for a 1
keV error on $T_{\mathrm{eff}}$ and $\sigma_{v} = 560$ km/sec and
$\sigma_{\tau} = 0.0062$ for a 2 keV error, for the central pixel.
Table \ref{Table2} lists these 1-$\sigma$ errors for convenient
reference.

The errors obtained on $\sigma_{v}$ and $\sigma_{\tau}$ from our
MCMC are smaller than those obtained using a Fisher matrix.
However, the MCMC errors are more accurate than those from a
Fisher matrix since a Fisher matrix approximates the likelihood
surface by an ellipsoidal Gaussian, which can result in overestimated errors
for likelihood surfaces with strong spatial curvature such as these. These
results show that, in the absence of contamination from imperfect
point source and primary microwave background removal, adding
X-ray temperature measurements to the data from upcoming ACT-like
SZ surveys can determine cluster peculiar velocities to within 100
km/sec or less. Large-scale velocity fields obtained from galaxy
clusters out to high redshift could provide an
interesting probe of dark matter and dark energy.

\section{SOURCES OF CONTAMINATION TO THE SZ SIGNAL}

Major sources of possible contamination have been neglected in this
exercise. In particular, it is possible that primary and secondary
microwave fluctuations and point sources (both radio and infrared)
could be problematic.

Primary fluctuations are in some ways both the largest and the
smallest concern. The fluctuation amplitudes are on the order of 100
$\mu K$ and have the exact same spectral behavior as the kinematic
SZ effect, providing a noise source that is an order of magnitude
larger than the signal. However, these fluctuations will be highly
coherent over the extent of the cluster, so the pixel-by-pixel
component separation will naturally measure this extended emission.
At that point, a simple spatial filter can be applied to remove the
primary microwave fluctuations. This spatial filter will have the
effect of removing roughly half of the cluster kinematic SZ signal
\citep{holder}, thereby reducing the signal-to-noise by
approximately this same factor.

Secondary fluctuations will be dominated by kinematic SZ from the
quasi-linear regime (the Vishniac effect; \citet{vishniac}) and the
thermal SZ background. The kinematic SZ fluctuations are expected to
be below the pixel noise, and therefore subdominant, the thermal SZ
background will simply add noise to the component that is separated
as thermal SZ. The rms is expected to be roughly 10 $\mu K$, which
will most likely serve as the dominant source of noise for this
component. However, this is much smaller than the expected thermal
SZ signal from each cluster and should not impact the component
separation process at a noticeable level.

Radio point sources that are uncorrelated with galaxy clusters are
not a concern \citep{knox}, but radio point sources within galaxy
clusters can ``fill in'' the SZ decrement and severely impact
cluster SZ measurements. This is a long-standing concern for low
frequency SZ measurements \citep{moffett89}, and could be a concern
even at frequencies as high as 150 GHz. The spectra of radio sources
up to such high frequencies are not well known, but very rough
estimates can be made of the most likely contamination. Radio
surveys at 1.4 GHz have been done of nearby Abell clusters
\citep{ledlow96} and distant X-ray selected clusters
\citep{stocke99} that find that a typical galaxy cluster has of
order one radio source at 1.4 GHz that would be of order a few mJy
at a cosmological distance. Detailed studies of spectra of bright
radio sources indicate \citep{herbig92} that typical radio sources
have spectra that are falling and steepening with frequency. In
particular, in their sample they found only a handful of sources
that were as bright at 40 GHz as at 1.4 GHz. Most sources with
rising spectra at 1.4 GHz eventually turned over and had lower
fluxes at 40 GHz than at 1.4 GHz, indicating that studies based on
spectral indices at low frequency will not provide accurate
estimates of behavior at high frequencies.  Assuming that cluster
sources have the same spectral behavior and that the steepening at
high frequencies continues, this would indicate that only a few
percent of clusters will have a radio source contributing of order
mJy flux at 150 GHz. In this handful of clusters, radio sources will
be a concern, as this flux would be an order of magnitude larger
than the pixel noise. However, in the majority of clusters the
contamination due to cluster radio galaxies would be comparable to
or smaller than the pixel noise.

Infrared point sources, consisting largely of dusty star forming
galaxies, are likely to be a major source of contamination
\citep{knox, white}. A detailed treatment is beyond the scope of
this paper, but the results of \citet{knox} suggest that hot
clusters (with temperatures above about 6 keV) will have a large
enough SZ signal that the infrared point sources can be estimated
simultaneously with the thermal SZ and kinematic SZ signals,
assuming an independent measure of the gas temperature. In this work
we have found that the three measurements by an ACT-like experiment
provide only two effective constraints, suggesting that there is
redundancy in the measurements and that an additional degree of
freedom (infrared point sources) could be allowed without
significantly degrading the constraints. The surest solution is to
use ALMA to measure the relevant point source fluxes. As a point of
comparison, ALMA could image 100 square degrees (comparable to the
ACT survey area) to a point source sensitivity of 0.1 mJy at 140 GHz
in less than 1 month. If one were to instead focus on the inner 2'
of the largest 100 galaxy clusters, this would take several hours of
observing time. Note that these same observations could be used to
estimate the SZ effects, but the small primary beam of ALMA (due to
the large telescopes) would require careful mosaicing of the cluster
to avoid resolving out much of the cluster flux.

\section{DISCUSSION AND CONCLUSIONS}

Instruments such as ACT, SPT, APEX, and Planck will find thousands
of galaxy clusters in the near future via SZ observations.  In
addition to determining the number density of clusters, which can
put limits on cosmological parameters, these surveys will reveal
information about the gas properties of individual clusters.
Ideally, SZ observations would be made in at least three frequency
bands with one frequency around 300 GHz, one around 150 GHz, and
another either near 90 GHz or better yet near 30 GHz.
Arcminute-resolution observations at those frequencies with
$\simeq 1\mu K$ detector noise would tightly constrain the cluster
gas temperature, line-of-sight velocity, and optical depth in the
absence of excessive point source and primary microwave background
contamination from imperfect subtraction.  Without this set of SZ
observations, parameter degeneracies prevent disentanglement of
these three cluster parameters.

Current limitations in technology and instrument availability will
make it impractical to obtain 30 GHz, arcminute-resolution, $1\mu
K$ sensitivity, SZ observations of the majority of the clusters
that will be found.  SZ surveys that will have 90 GHz channels
will still have parameter degeneracies resulting from detector
noise $\simeq 10\mu K$.  However, we find that upcoming SZ surveys
will be able to tightly constrain two cluster gas parameters which
are linear combinations of $\tau T_{e}$, $\tau v$, and $\tau
T_{e}^{2}$. The constrained parameters are roughly $\tau T_{e}$ and
a single linear combination of the other two terms.
We demonstrated that this is the case for both
individual
isothermal gas regions and for 3D simulated Nbody + hydro clusters.

The SZ intensity shift that microwave photons experience passing
through a cluster is nearly a linear function of $\tau T_{e}$, $\tau
v$, and $\tau T_{e}^{2}$, these being the most dominant terms in the
intensity shift expression.  This near-linearity results in a close
correspondence between the two effective parameters SZ surveys will
constrain and simple line-of-sight integrals of these parameters
through the three dimensional cluster. We illustrated this
correspondence with our three dimensional cluster simulations. This
will greatly simplify data analysis of multi-frequency SZ data: it
will not be necessary (or useful) to model the intensity as a
superposition of elements along the line of sight but instead the SZ
effect can be modeled as a single gas element with a single
effective temperature and velocity.

We have shown that a temperature constraint added to SZ data breaks
the parameter degeneracy between $\tau$, $T_{\mathrm{eff}}$, and
$v_{\mathrm{eff}}$.  Using the above linearity, we show that the
effective velocity constrained by combining SZ with an independent
temperature measure is
approximately the optical-depth-weighted velocity integrated along
the cluster line of sight.  Since X-ray derived temperatures do not
give us precisely the weighted temperature measurements that are
required to
determine $\int v d\tau / \int d\tau$ exactly, we find the measured
effective velocity will be biased away from $\int v d\tau / \int
d\tau$ by about 15 to 40 km/sec, with a smaller bias for hotter,
relaxed clusters.

Errors on $\tau$ and $v_{\mathrm{eff}}$ are calculated via a Markov
chain Monte Carlo method assuming a temperature prior in addition to
SZ data.  We find for ACT-like SZ simulations of our 9 keV cluster,
$\sigma_{v}$ = 20 km/sec and $\sigma_{\tau}$ = 0.002 for a 1 keV
error on $T_{\mathrm{eff}}$, and $\sigma_{v}$ = 40 km/sec and
$\sigma_{\tau}$ = 0.004 for a 2 keV error on $T_{\mathrm{eff}}$. For
our 3 keV simulated cluster, $\sigma_{v}$ = 60 km/sec and
$\sigma_{\tau}$ = 0.004, and $\sigma_{v}$ = 100 km/sec and
$\sigma_{\tau}$ = 0.005 for 1 keV and 2 keV errors on
$T_{\mathrm{eff}}$ respectively.  The Markov chain errors we find on
$v_{\mathrm{eff}}$ and $\tau$ are smaller than those obtained via a
Fisher matrix.  A Fisher matrix overestimates the errors because the
likelihood surface is strongly curved in this parameter
representation, strongly violating the implicit assumption of
ellipsoidal symmetry over the parameter region of interest.  Note
that the errors on velocities will be increased when residual
primary microwave contamination is included, and that bulk flows
within the clusters provide comparable noise in matching observed
peculiar velocities to the true bulk velocity of the cluster
\citep{daisuke2,holder,diaferio}.

If an independent cluster temperature estimate from X-ray
spectroscopy is unavailable, temperature estimates can also be
obtained from either the cluster velocity dispersion \citep{lubin}
or the cluster integrated SZ flux \citep{benson} and scaling
relations.  Since an accuracy of only 2 keV is required on an
additional temperature measurement to obtain very interesting
velocity estimates, the use of these scaling relations could prove
to be a very beneficial tool.

As discussed in \S8, contamination from primary microwave background
fluctuations and point sources add another source of noise that must
be factored into these parameter constraints.  Radio point sources
due to emission from galaxy cluster members themselves and infrared
point sources will both be non-negligible sources of SZ signal
contamination.  Studies of the effect point source contamination
will have on cluster parameter extraction have been carried out by
\citet{knox} and \citet{aghanim2004}. Both studies have found that
the contamination could potentially be serious; however the latter
study considers the effect of point source contamination if no
attempt is made to filter point sources out of the observations or
model them into the parameter extraction routines.  Moreover, even
in the worst case point source contamination scenario, observations
with an instrument such as ALMA will allow straightforward point
source subtraction from SZ images.  Clearly either filtering
techniques or additional ALMA type observations will be needed to
minimize both the point source and primary microwave background
contamination of SZ signals.

Near-future SZ surveys will open the door to a wealth of information
about galaxy clusters.  Determining the number density of galaxy
clusters as a function of redshift is potentially a strong probe of
dark energy's equation of state and variability over time
\citep{haiman, holder2, hu, majumdar1, majumdar2}. However, galaxy
clusters offer more information that can also yield cosmological
information. The kinematic SZ signature of galaxy clusters can
reveal large-scale velocity fields out to high redshift that can
provide an alternative probe of large-scale dark matter and dark
energy \citep{peel}. Cluster optical depth information yields
cluster gas mass estimates, and optical depths are crucial to any of
the tests that have been proposed using the ({\em extremely}
difficult to measure) polarization of scattered microwave photons at
the position of galaxy clusters \citep{kamion}.  The gas parameters
$T_{e}$, $\tau$, and $v$ of individual galaxy clusters are of direct
interest for cluster astrophysics. Arcminute-resolution SZ
observations can begin to probe cluster substructure and offer more
information about cluster gas profiles and internal gas dynamics. In
summary, SZ observations are entering new territory, where large
scale surveys will offer new understandings of galaxy clusters and
cosmology.

\acknowledgements

The authors would like to thank Daisuke Nagai for generously
providing us with his cluster simulations.  NS would also like to
thank Rouven Essig for useful discussions.  This work was supported
by the NSF award AST-0408698 for the ACT project.

\begin{table}
\begin{center}
\begin{tabular}{|c|c|c|c|c|c|c|}
\hline Observing & Detector &$kT_{e}$&$v$&$\tau$&$\sigma_{T_{e}}$&$\sigma_{v}$  \\
Frequencies (GHz)&Noise ($1\mu K$)&(keV)&(km/sec)&&(keV)&(km/sec) \\
\hline \hline
\hline 30, 150, 300&1&10&200&0.012&0.5&25 \\
\hline 90, 150, 300&1&10&200&0.012&1&50 \\
\hline 145, 225, 265&1&10&200&0.012&6&220 \\
\hline 30, 150, 300&1&10&-200&0.012&0.5&50 \\
\hline 90, 150, 300&1&10&-200&0.012&1&100 \\
\hline 145, 225, 265&1&10&-200&0.012&6.5&400 \\
\hline 30, 145, 225, 265&1&10&200&0.012&0.5&25 \\
\hline 90, 145, 225, 265&1&10&200&0.012&1&50 \\
\hline 30, 150, 300&10&10&200&0.012&5&200 \\
\hline 90, 150, 300&10&10&200&0.012&8&250 \\
\hline 145, 225, 265&10&10&200&0.012&9&350 \\
\hline 30, 150, 300&1&7&200&0.009&1&50 \\
\hline 30, 150, 300&1&3&200&0.004&3.5&200 \\
\hline
\end{tabular}
\caption{The 1-$\sigma$ errors on $T_{e}$ and $v$ for different
observing frequencies, detector noise, and gas parameters.
\label{Table1}}
\end{center}
\end{table}

\begin{table}
\begin{center}
\begin{tabular}{|c|c|c|c|c|}
\hline Simulated& Simulated Cluster &Error on Temp.&$\sigma_{\tau}$&$\sigma_{v}$  \\
Experiment&Avg. Temp. (keV)&Prior (keV)&&(km/sec) \\
\hline \hline
\hline ACT-like&9&1&0.002&20 \\
\hline ACT-like&3&1&0.004&60 \\
\hline ACT-like&9&2&0.004&40 \\
\hline ACT-like&3&2&0.005&100 \\
\hline Planck-like&9&1&0.002&500 \\
\hline Planck-like&9&2&0.006&560 \\
\hline
\end{tabular}
\caption{The 1-$\sigma$ errors on $\tau$ and $v$ for ACT-like and
Planck-like experiments using 9 keV and 3 keV simulated clusters
with varying errors on the temperature prior $T_{\mathrm{eff}}$.
\label{Table2}}
\end{center}
\end{table}

\begin{center}
\begin{figure}
$\begin{array}{c@{\hspace{0.1in}}c@{\hspace{0.1in}}c}
\epsscale{.31}\plotone{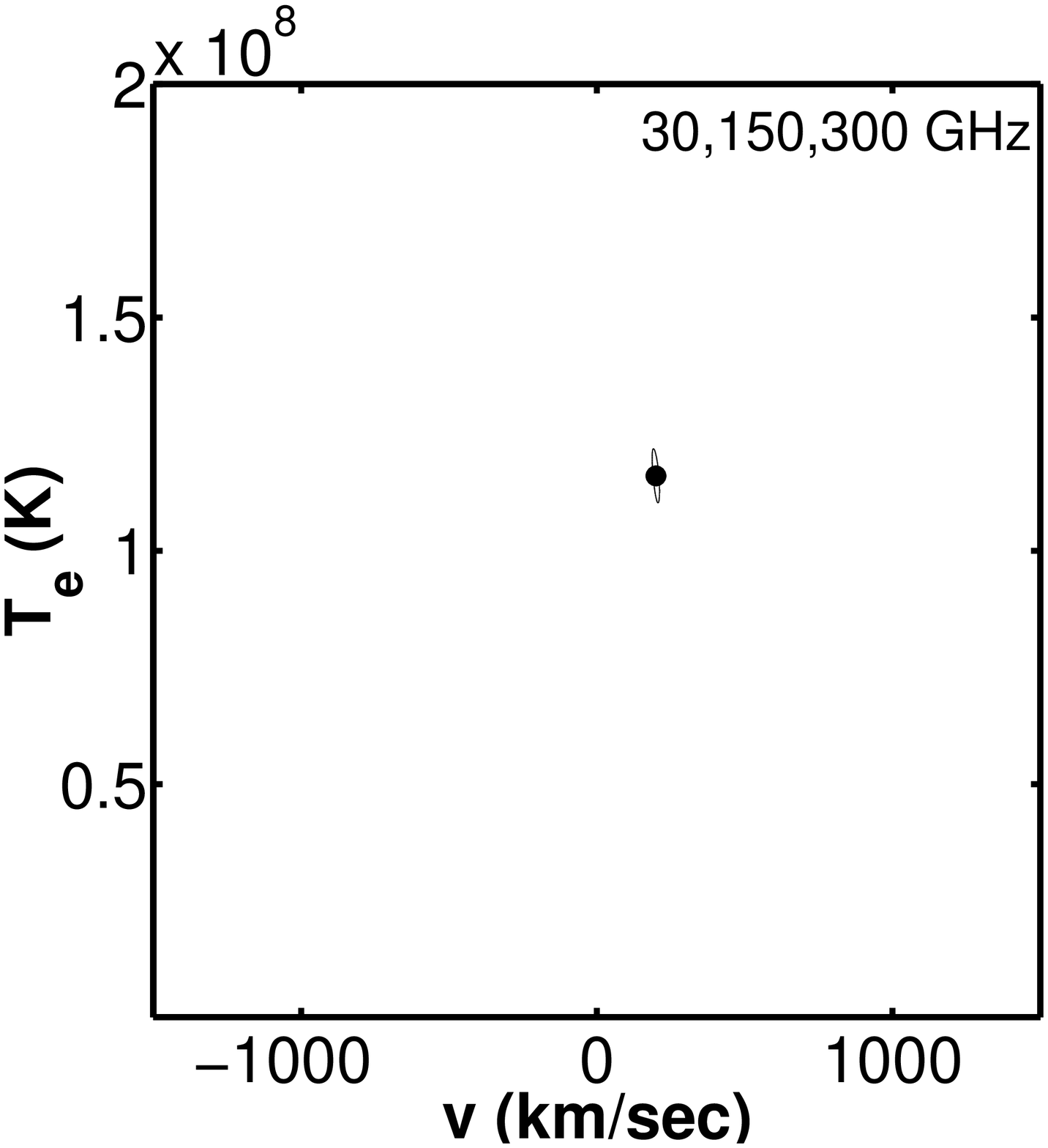}&\plotone{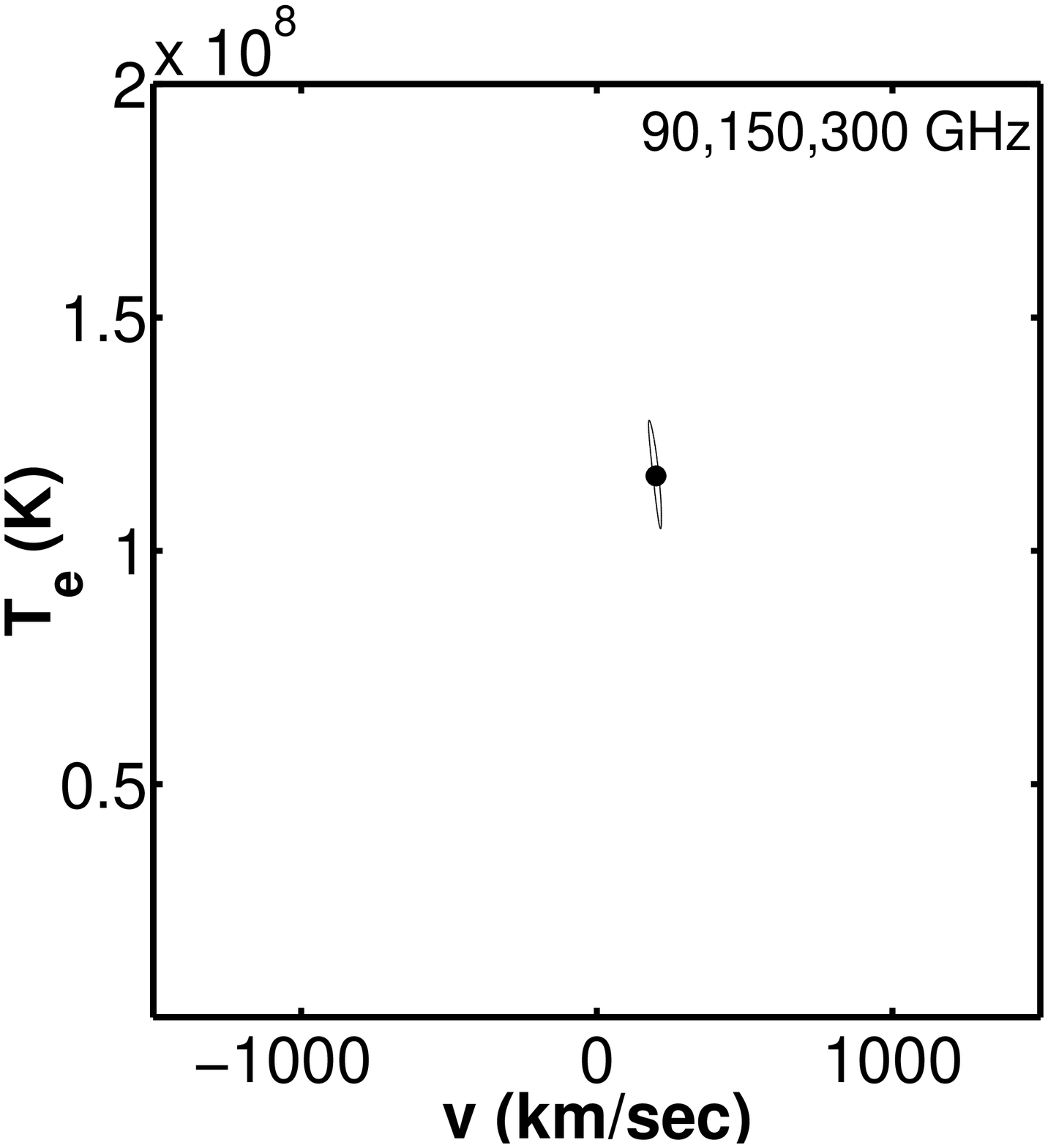}&\plotone{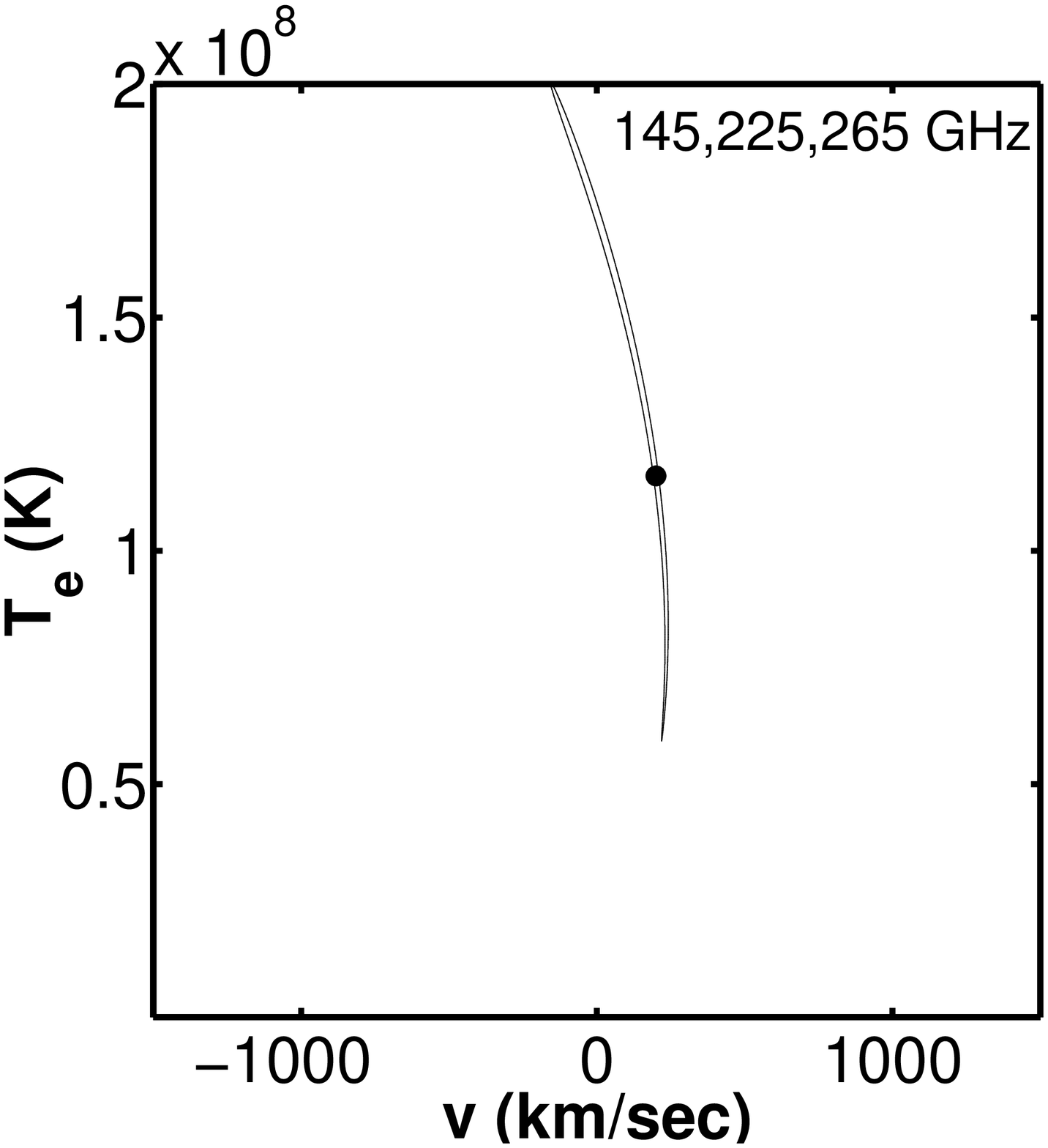}\cr
\mbox{\hspace{0.26in}(a)}&\mbox{\hspace{0.26in}(b)}&\mbox{\hspace{0.26in}(c)}
\end{array}$ \caption{1-$\sigma$ likelihood contours for a
simulated region of gas of $T_{e}$ = 10 keV ($1.2\times 10^{8}$ K),
$v$ = 200 km/sec, and $\tau$ = 0.012 obtained by calculating the SZ
effect for three different observing frequency sets assuming $1\mu
K$ detector noise.  The frequency sets are 1a) 30, 150, 300 GHz, 1b)
90, 150, 300 GHz, and 1c) 145, 225, 265 GHz. The SZ intensity shifts
from the gas region are calculated using the formula in
\citet{itoh2}, and the ratios of the intensity shifts at different
frequencies in each set are computed to eliminate the dependence on
optical depth.  A dot ($\bullet$) marks the input fiducial gas
region.\label{f1a}\label{f1b}\label{f1c}}
\end{figure}
\end{center}

\begin{figure}
\begin{center}
$\begin{array}{c@{\hspace{0.1in}}c@{\hspace{0.1in}}c}
\epsscale{.31}\plotone{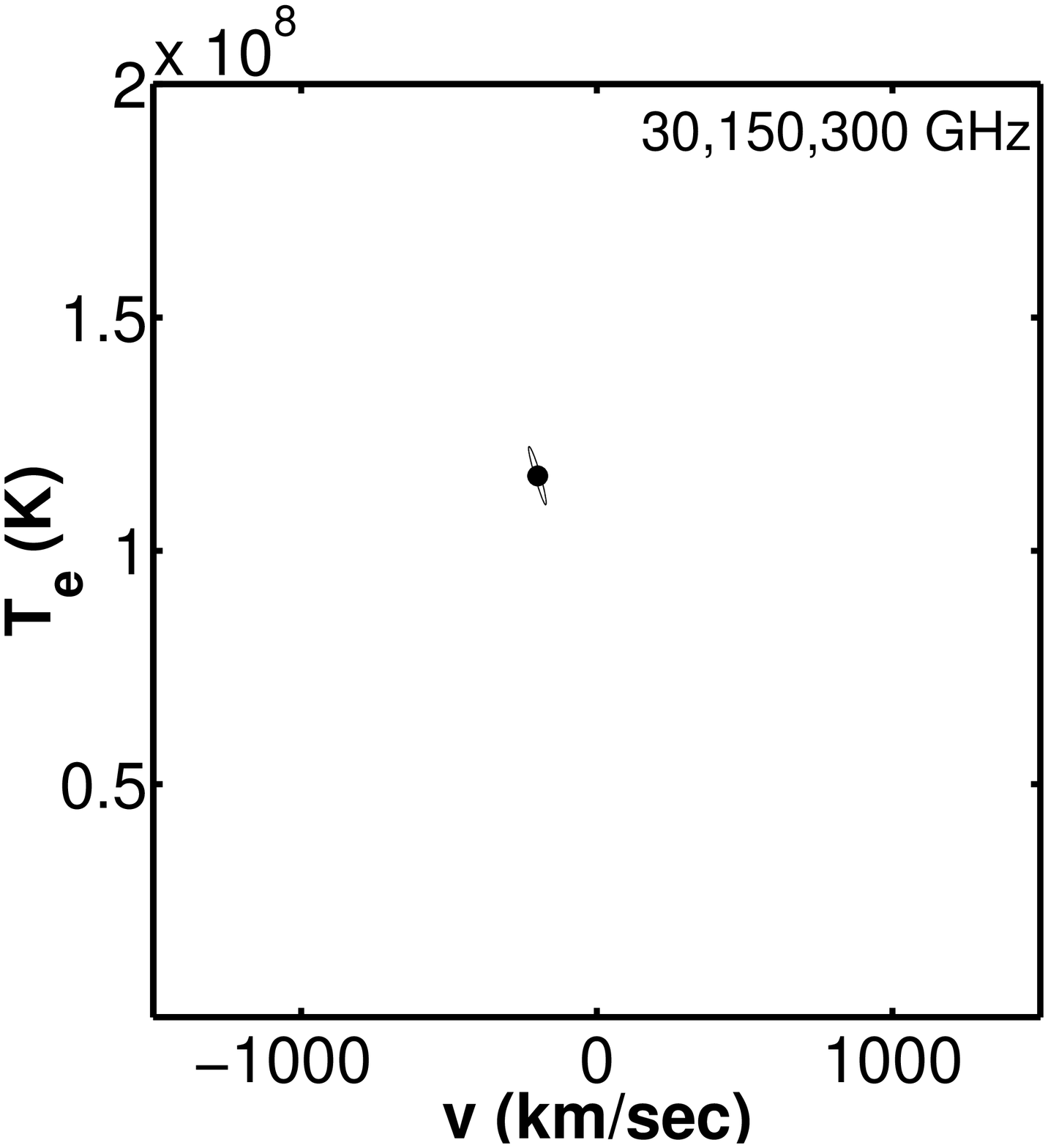}&\plotone{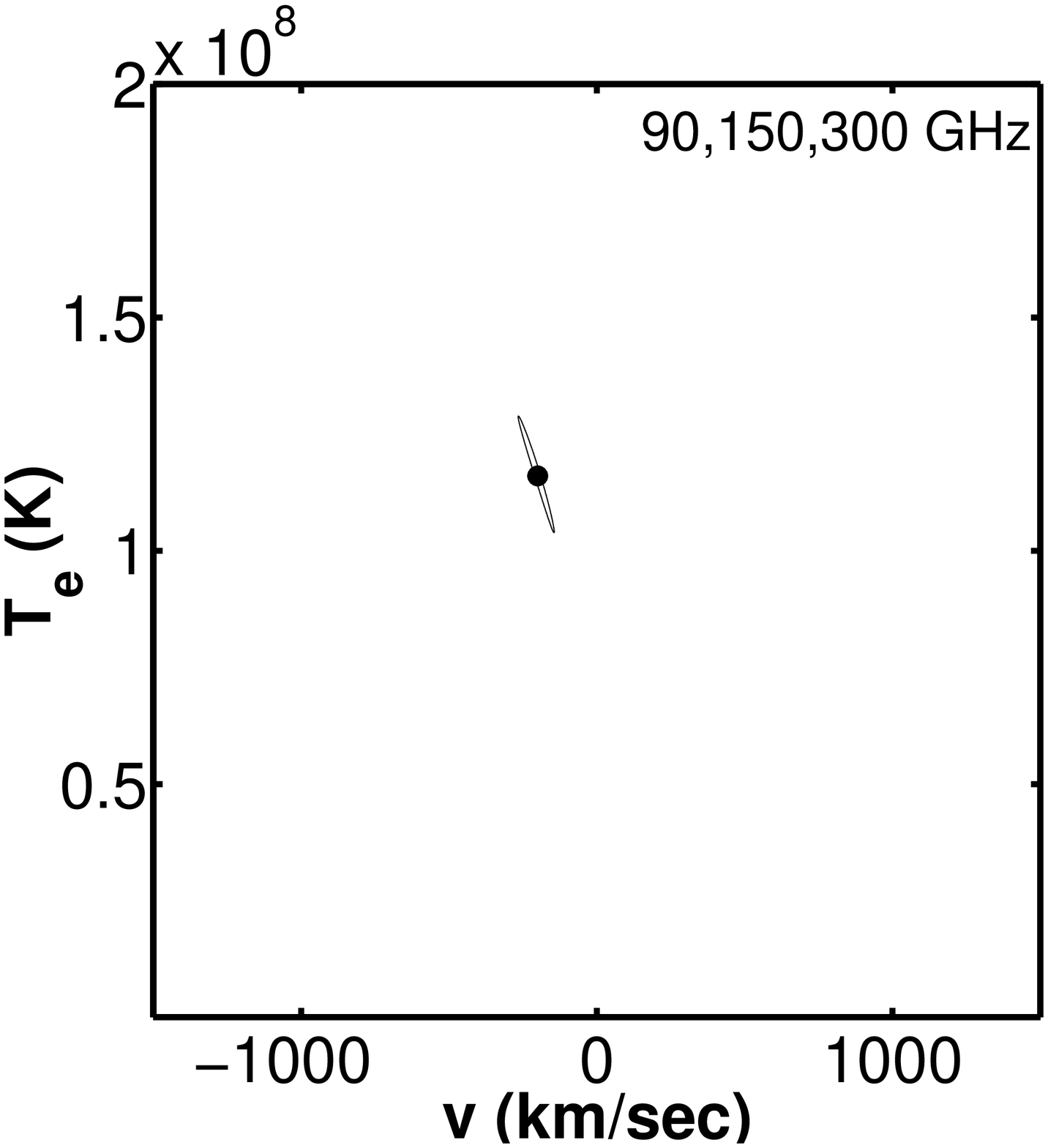}&\plotone{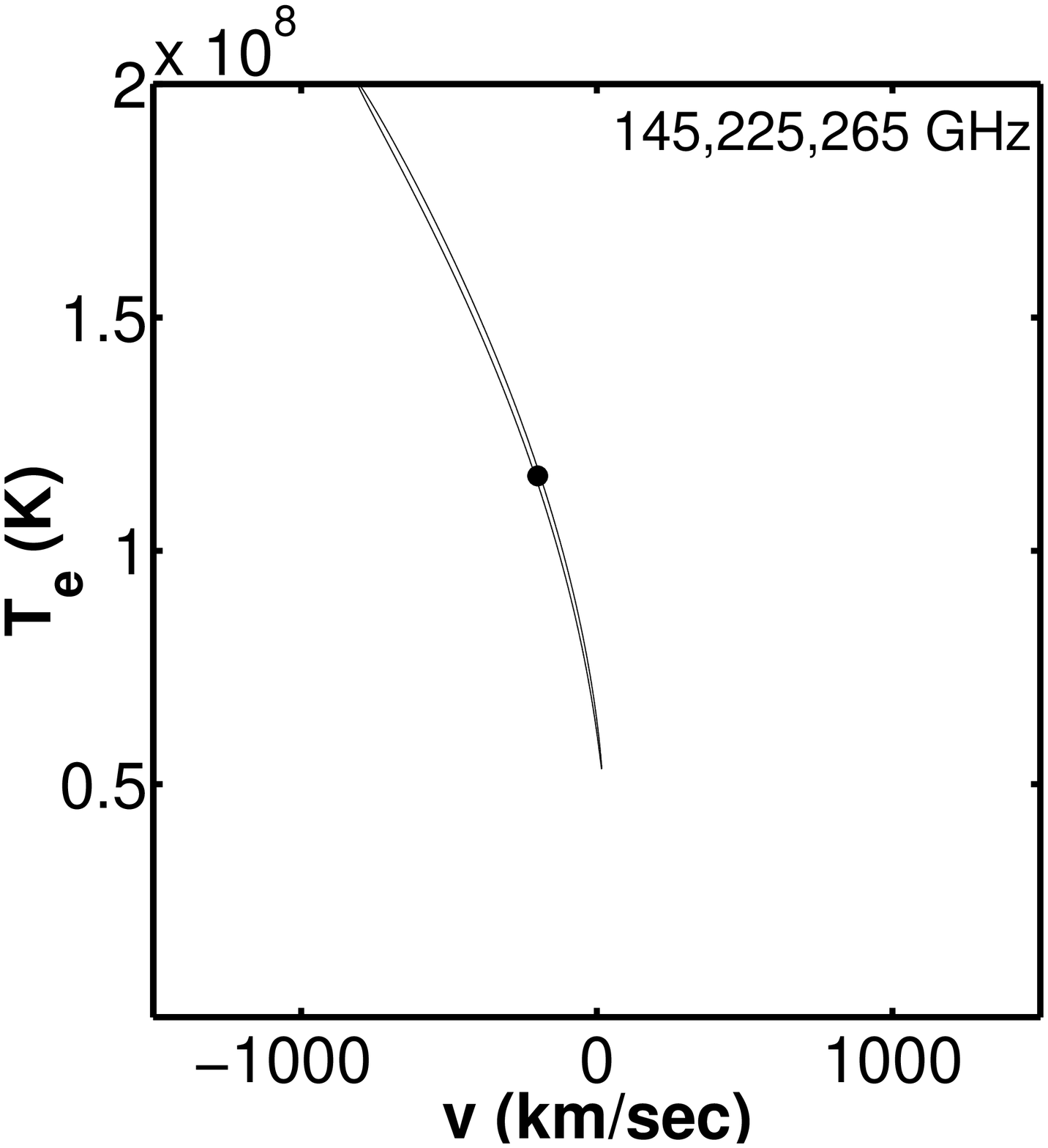}\cr
\mbox{\hspace{0.26in}(a)}&\mbox{\hspace{0.26in}(b)}&\mbox{\hspace{0.26in}(c)}
\end{array}$ \caption{Same as Fig. 1 except for $v$ = -200
km/sec.}
\end{center}
\end{figure}

\begin{figure}
\begin{center}
$\begin{array}{c@{\hspace{0.1in}}c@{\hspace{0.1in}}c}
\epsscale{.31}\plotone{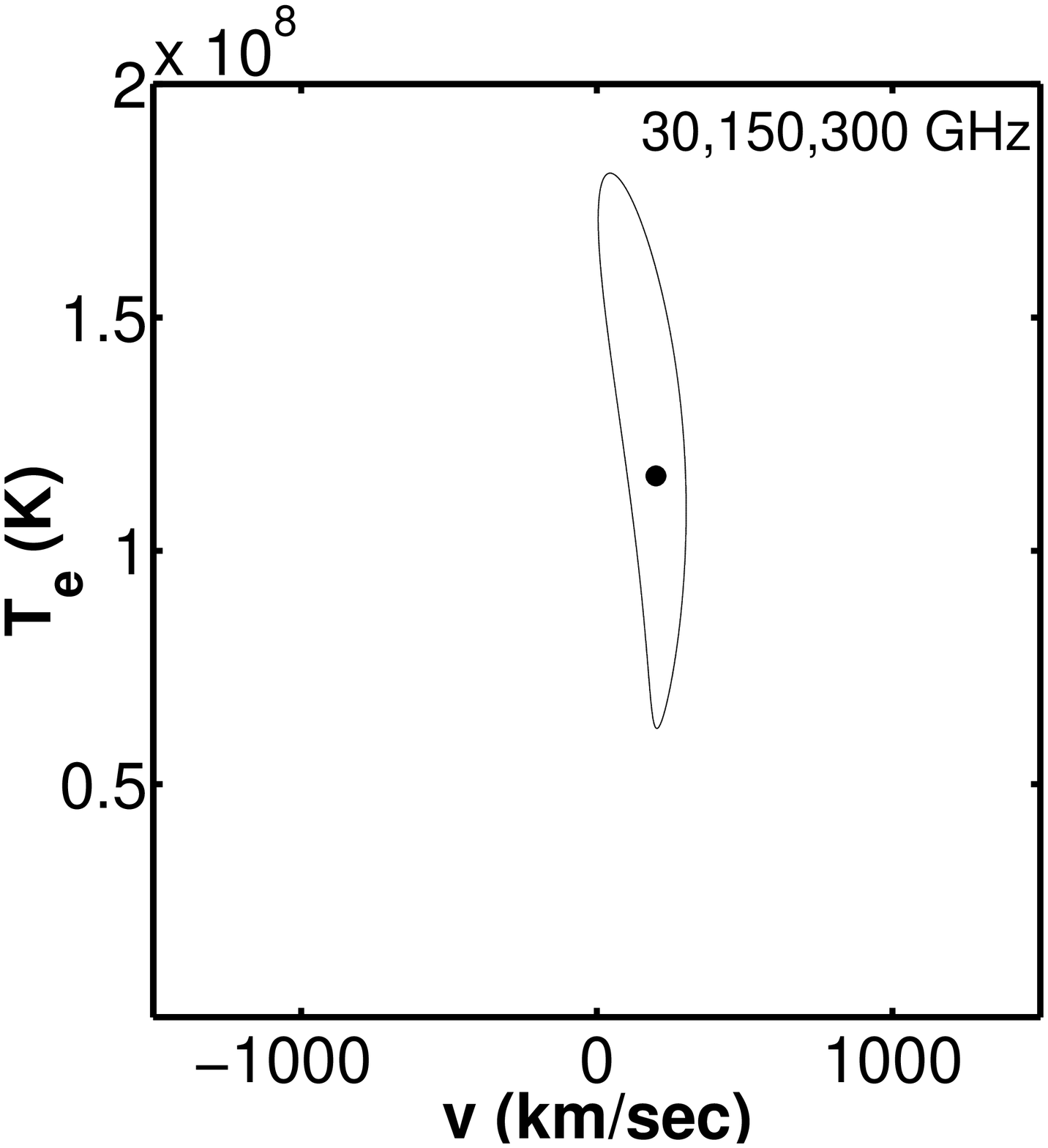}&\plotone{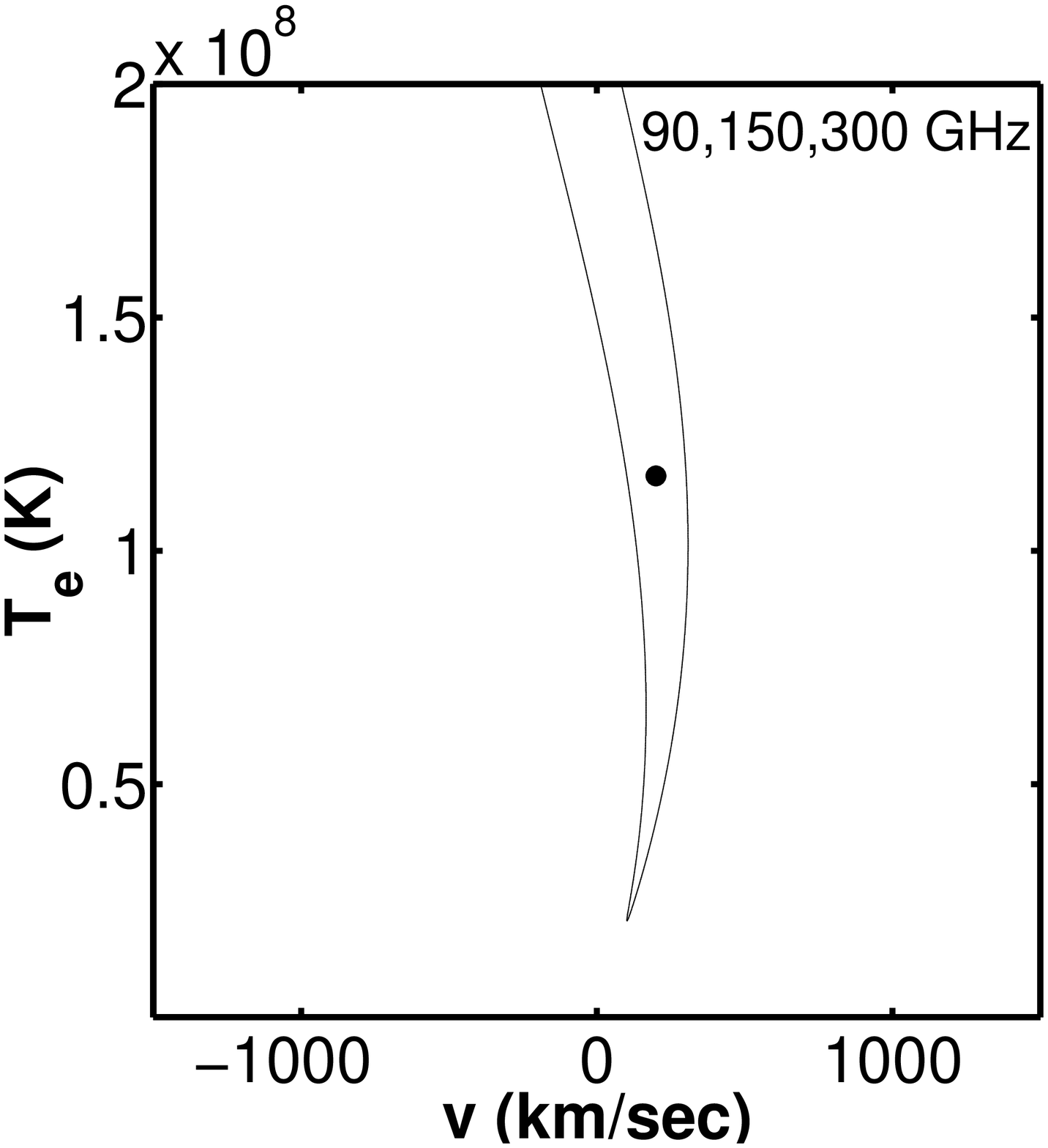}&\plotone{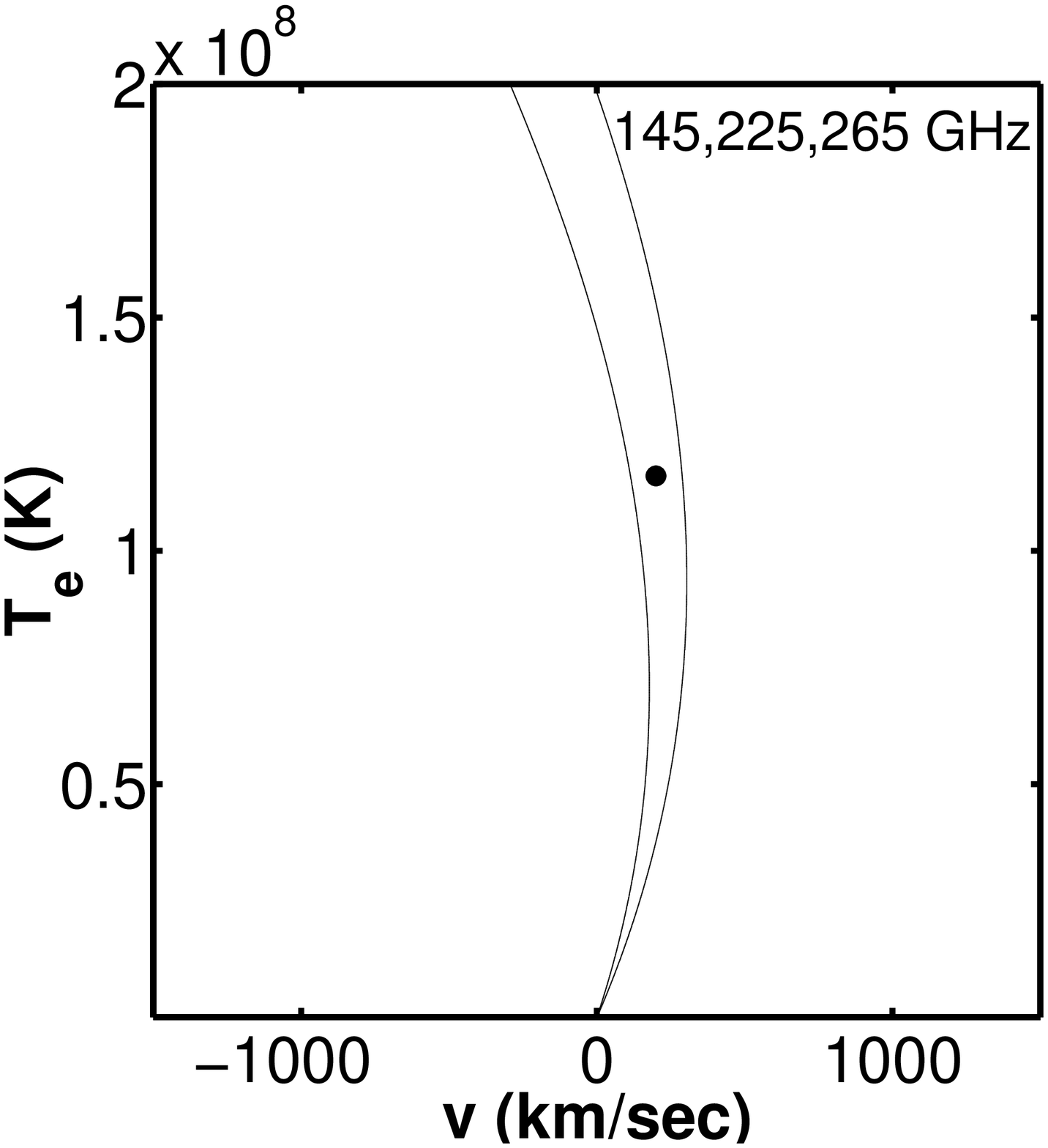}\cr
\mbox{\hspace{0.26in}(a)}&\mbox{\hspace{0.26in}(b)}&\mbox{\hspace{0.26in}(c)}
\end{array}$  \caption{Same as Fig. 1 except for detector noise of
$10 \mu K$.}
\end{center}
\end{figure}

\begin{figure}
\begin{center}
$\begin{array}{c@{\hspace{0.1in}}c}
\epsscale{.31}\plotone{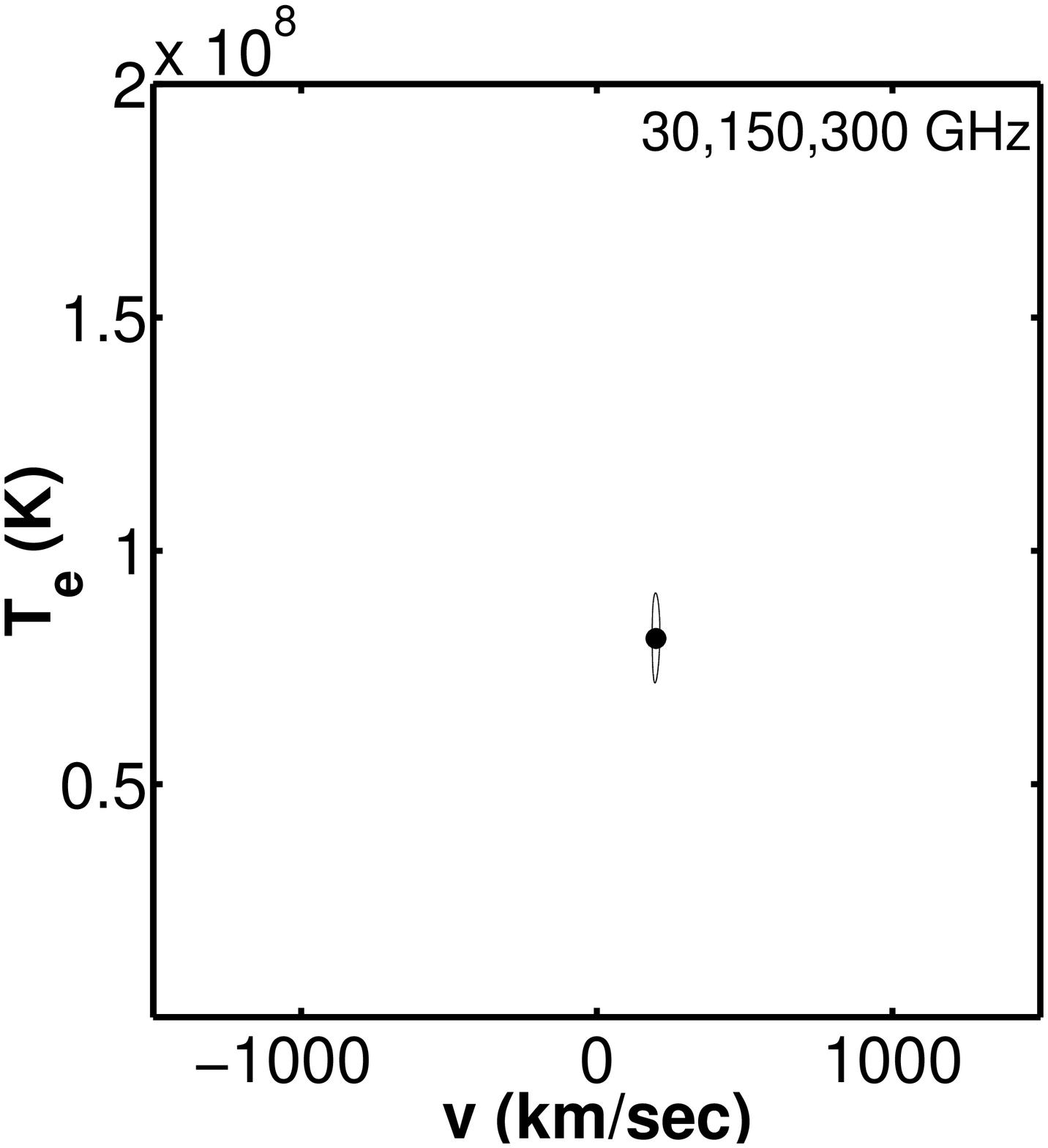}&\plotone{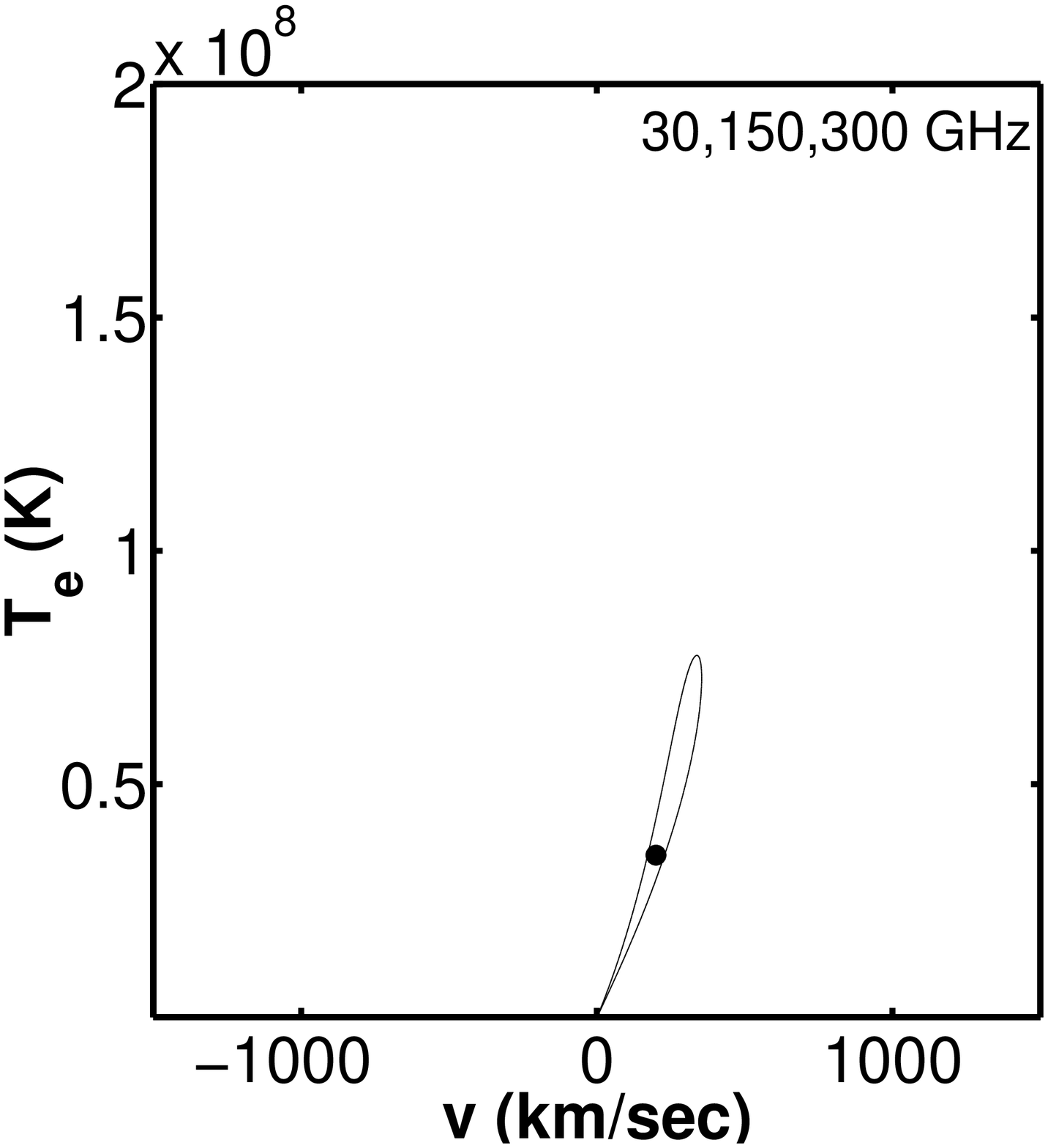}\cr
\mbox{\hspace{0.26in}(a)}&\mbox{\hspace{0.26in}(b)}
\end{array}$ \caption{Same as Fig. 1a except for gas $T_{e}$ and
$\tau$ values: 4a) $T_{e}$ = 7 keV, $\tau$ = 0.009; 4b) $T_{e}$ = 3
keV, $\tau$ = 0.004.}
\end{center}
\end{figure}

\begin{figure}
\begin{center}
$\begin{array}{c@{\hspace{0.11in}}c@{\hspace{0.11in}}c}
\epsscale{.3}\plotone{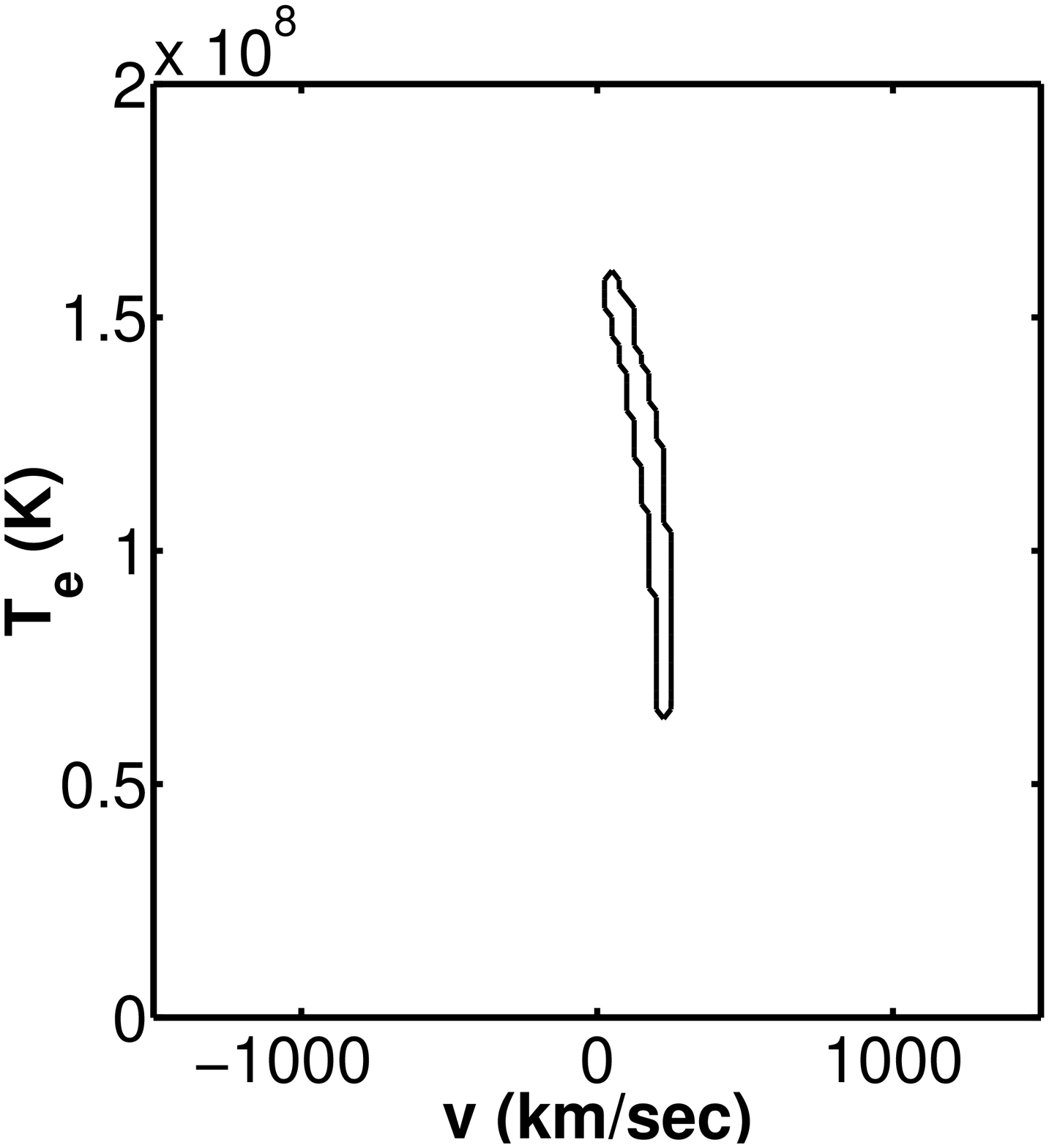}&\epsscale{.311}\plotone{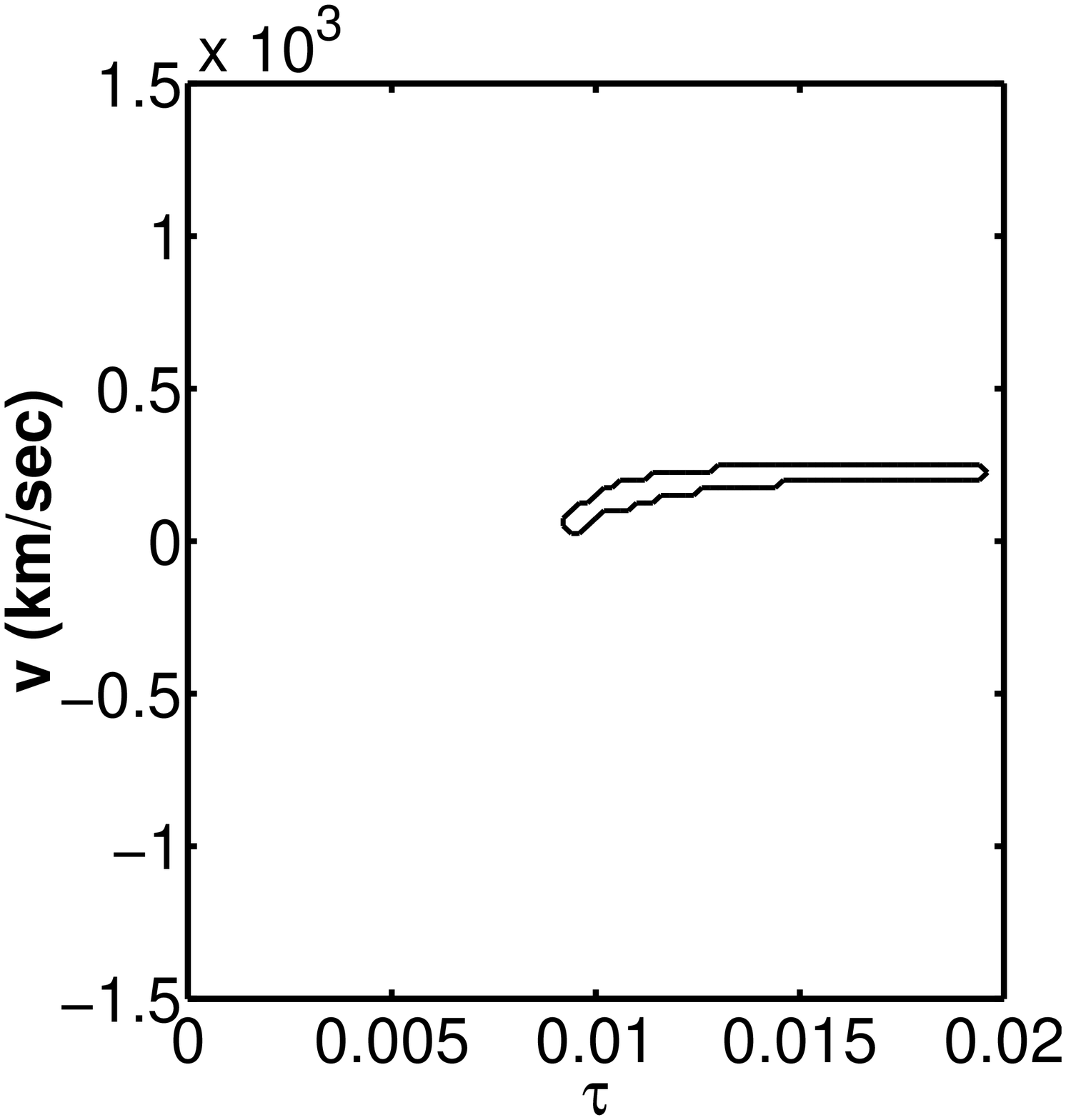}&\epsscale{.31}\plotone{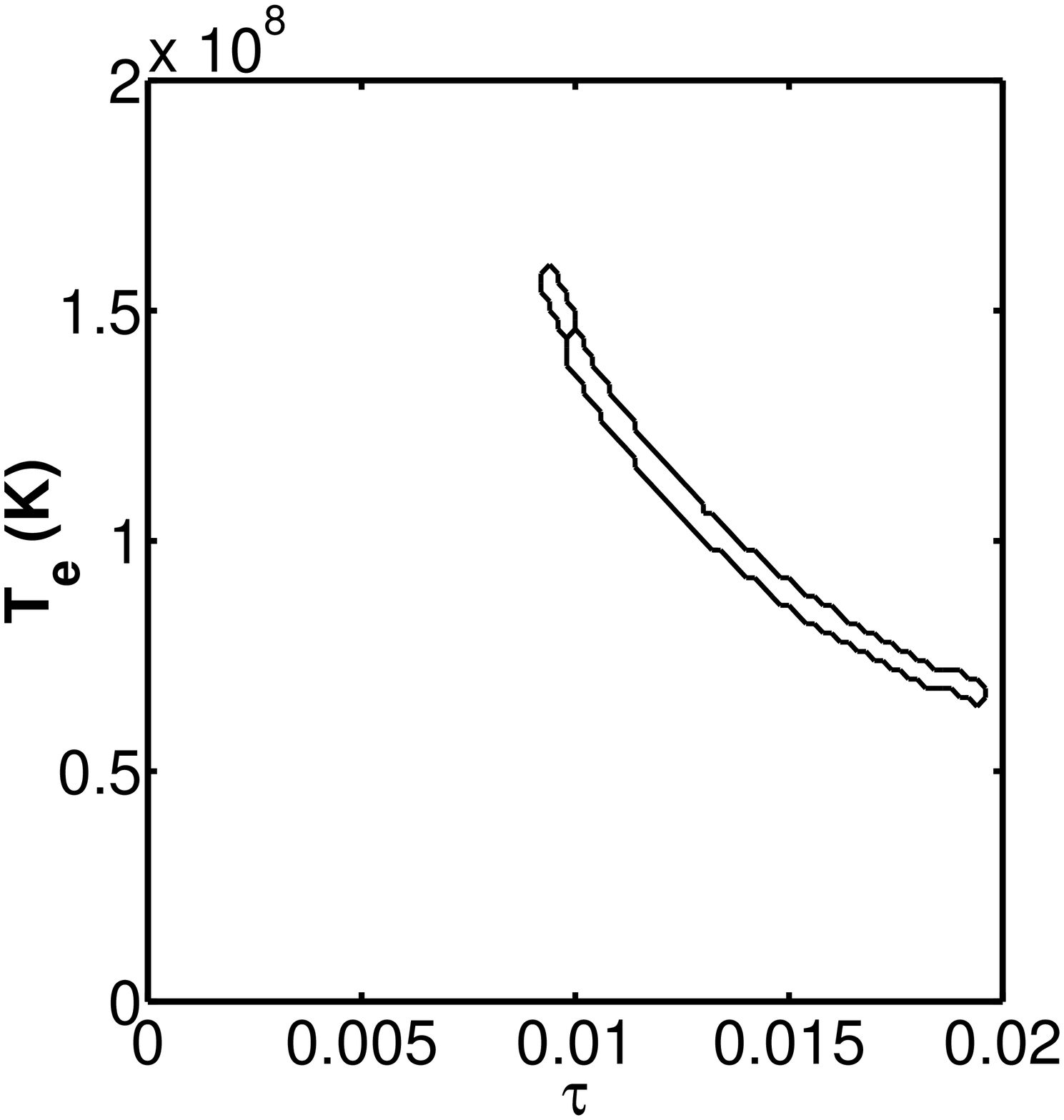}\cr
\mbox{\hspace{0.26in}(a)}&\mbox{\hspace{0.26in}(b)}&\mbox{\hspace{0.26in}(c)}
\end{array}$ \caption{Projected 1-$\sigma$ likelihood contours for
the ($T_{e}$, $v$, $\tau$) parameter space from SZ intensity
shifts at 145, 225, and 265 GHz for a simulated gas region of
$T_{e}$ = 10 keV ($1.2\times 10^{8}$ K), $v$ = 200 km/sec, and
$\tau$ = 0.012 assuming $1\mu K$ detector noise.}
\end{center}
\end{figure}

\begin{figure}
\begin{center}
$\begin{array}{c@{\hspace{0.1in}}c@{\hspace{0.1in}}c}
\epsscale{.306}\plotone{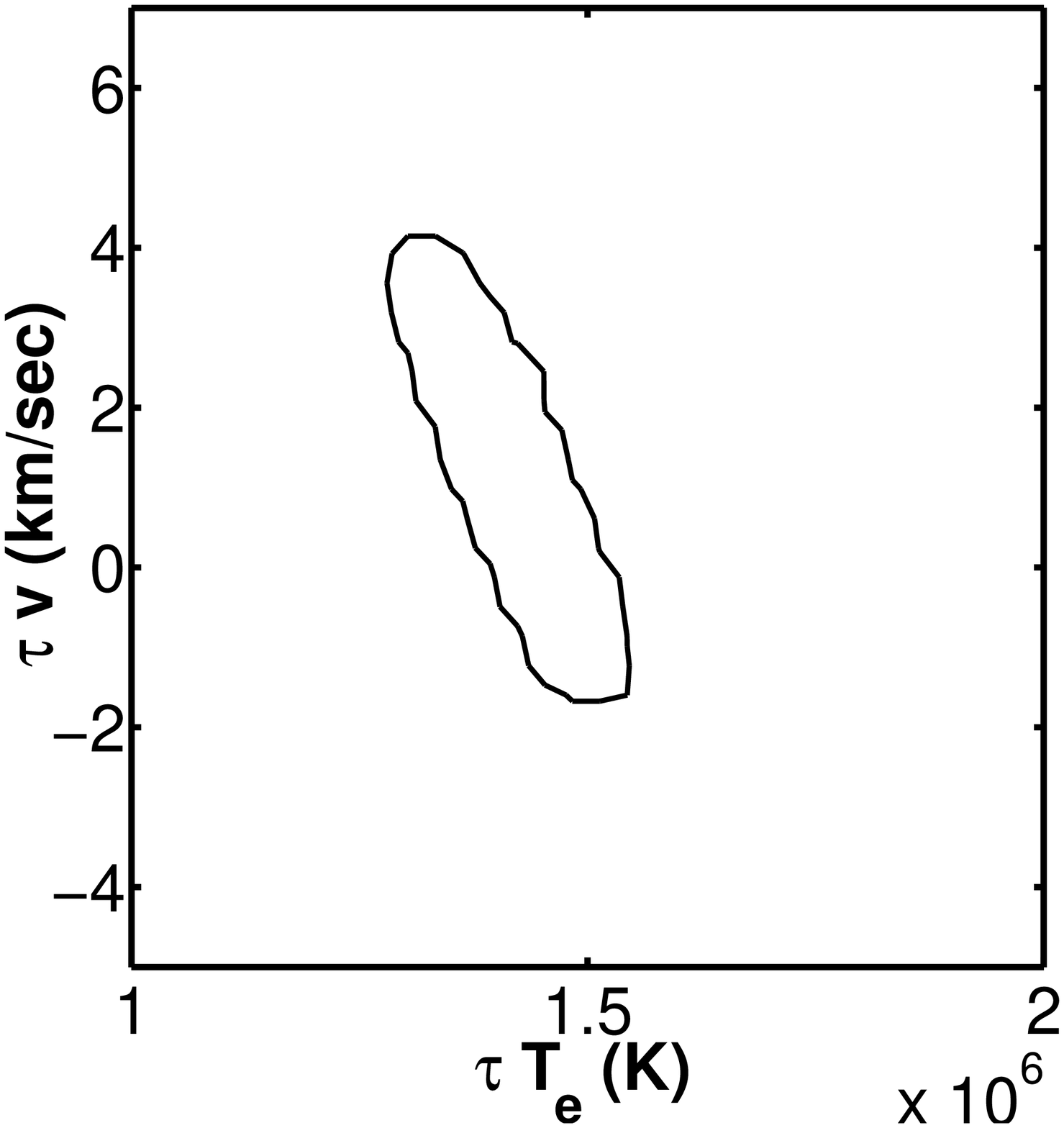}&\epsscale{.31}\plotone{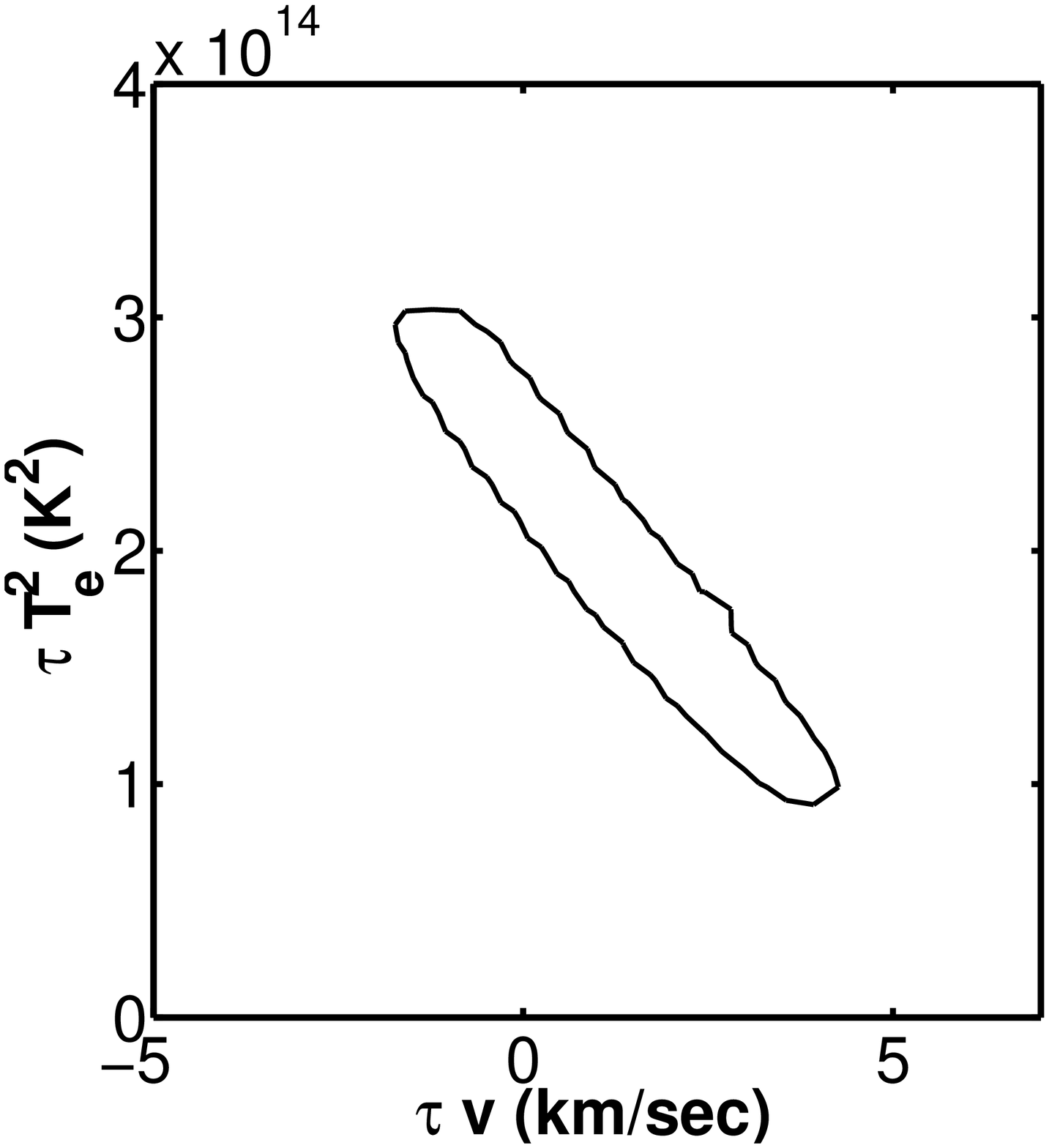}&\epsscale{.31}\plotone{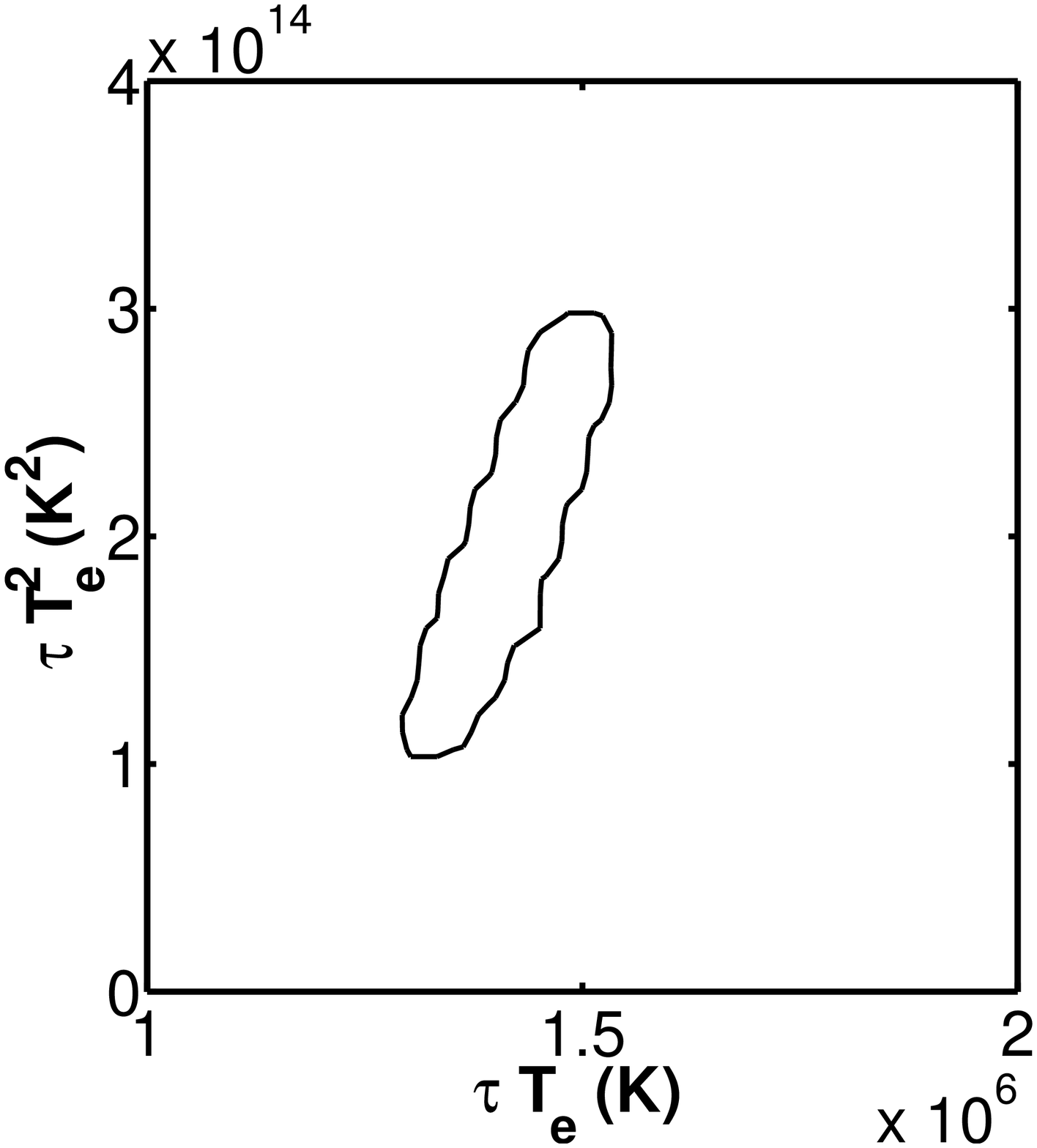}\cr
\mbox{\hspace{0.26in}(a)}&\mbox{\hspace{0.26in}(b)}&\mbox{\hspace{0.26in}(c)}
\end{array}$
\caption{Same as Fig. 5 except for the parameter space ($\tau
T_{e}$, $\tau v$, $\tau T_{e}^{2}$), corresponding to the physical
parameters expected to be important.}
\end{center}
\end{figure}

\begin{figure}
\begin{center}
$\begin{array}{c@{\hspace{0.1in}}c@{\hspace{0.1in}}c}
\epsscale{.31}\plotone{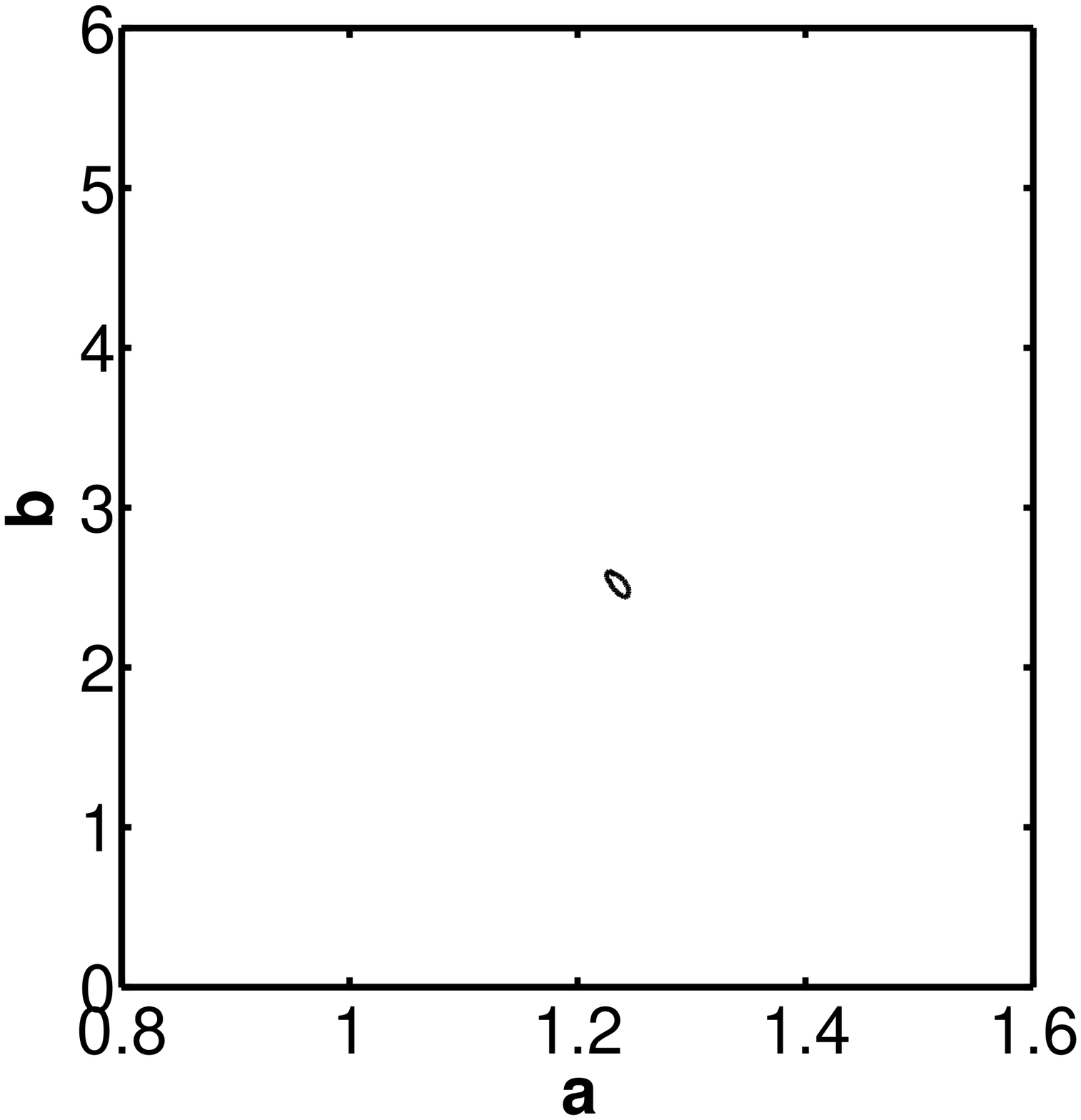}&\epsscale{.304}\plotone{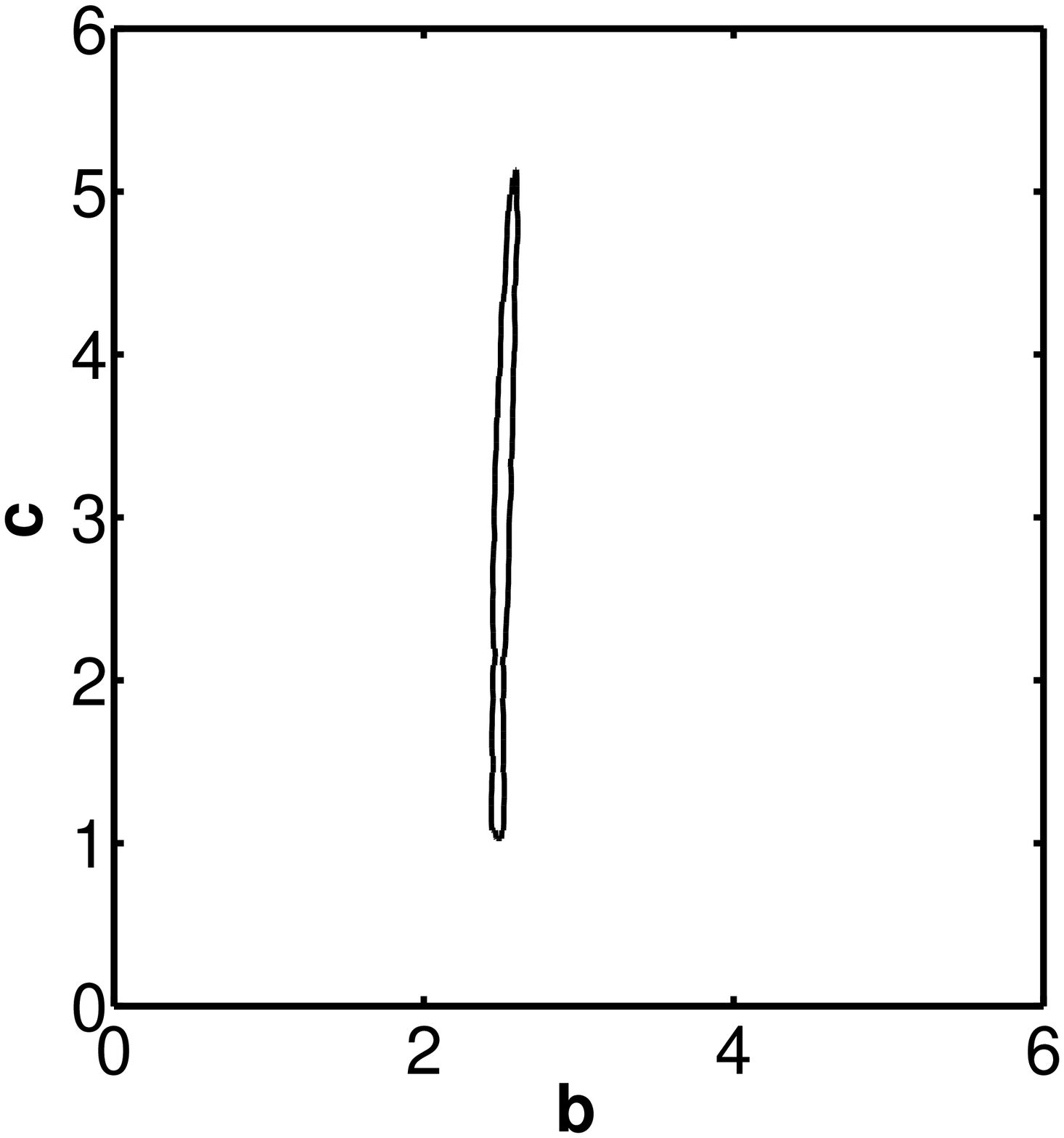}&\epsscale{.308}\plotone{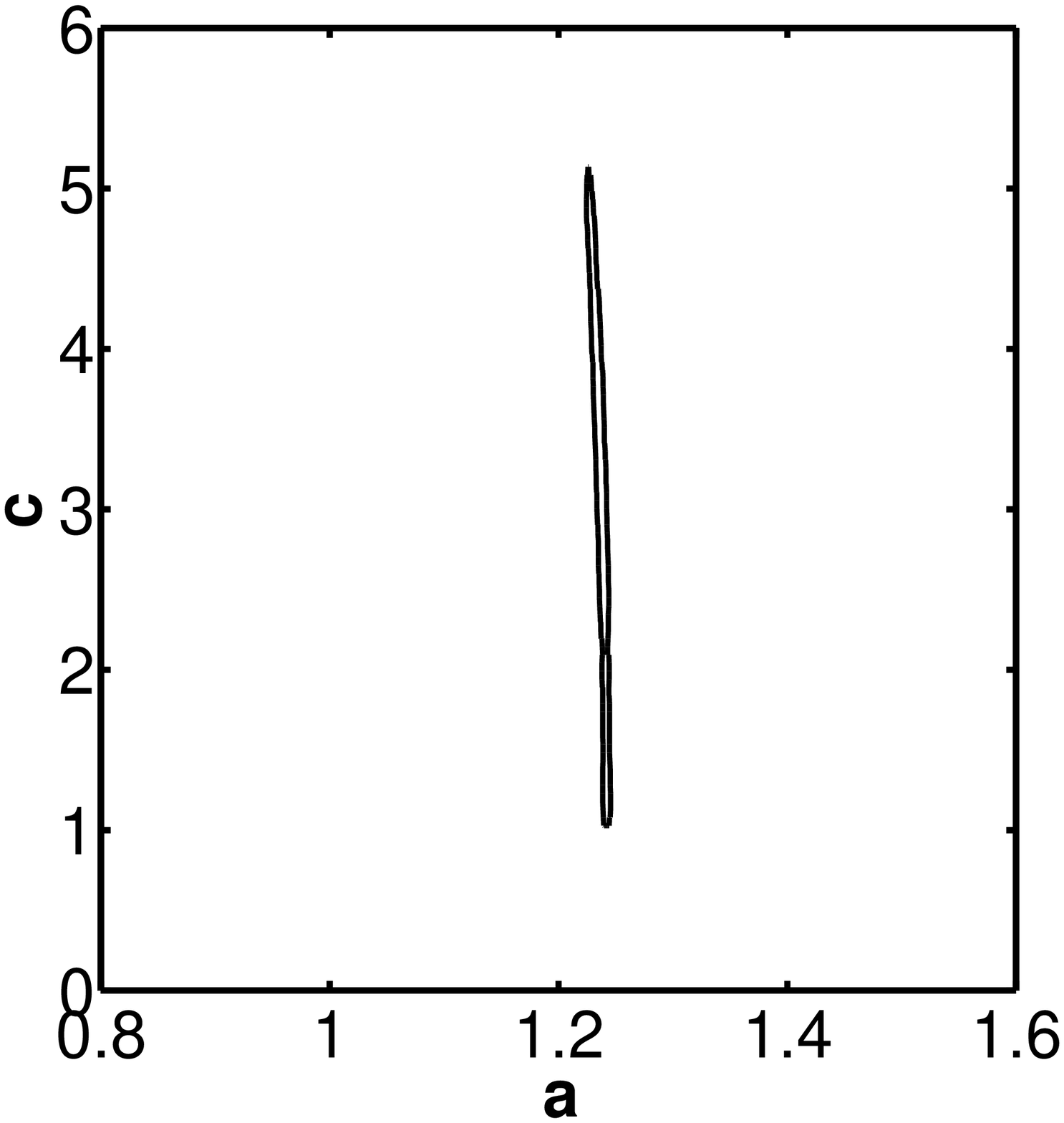}\cr
\mbox{\hspace{0.15in}(a)}&\mbox{\hspace{0.15in}(b)}&\mbox{\hspace{0.15in}(c)}
\end{array}$
\caption{Same as Fig. 5 except for the parameter space ($a$, $b$,
$c$), the combinations of the physical parameters that are best
constrained by SZ observations.}
\end{center}
\end{figure}

\begin{figure}
\begin{center}
$\begin{array}{c@{\hspace{0.005in}}c@{\hspace{0.005in}}c@{\hspace{0.005in}}c}
\epsscale{.047}\plotone{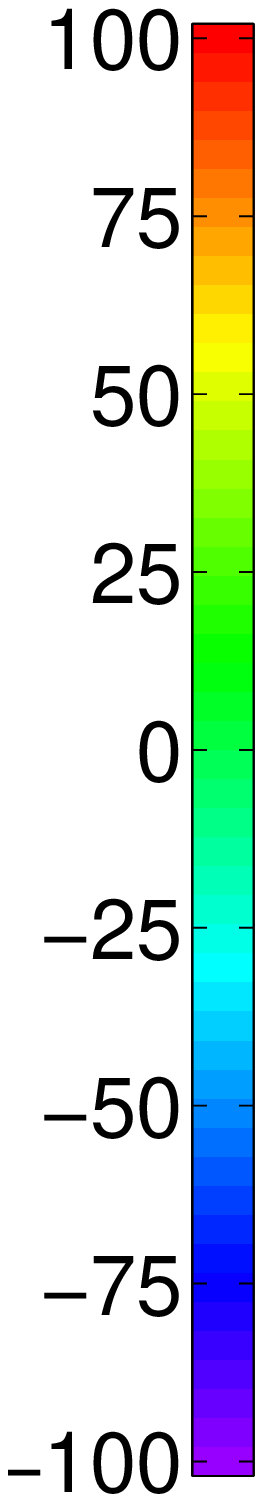}&\epsscale{.3049}\plotone{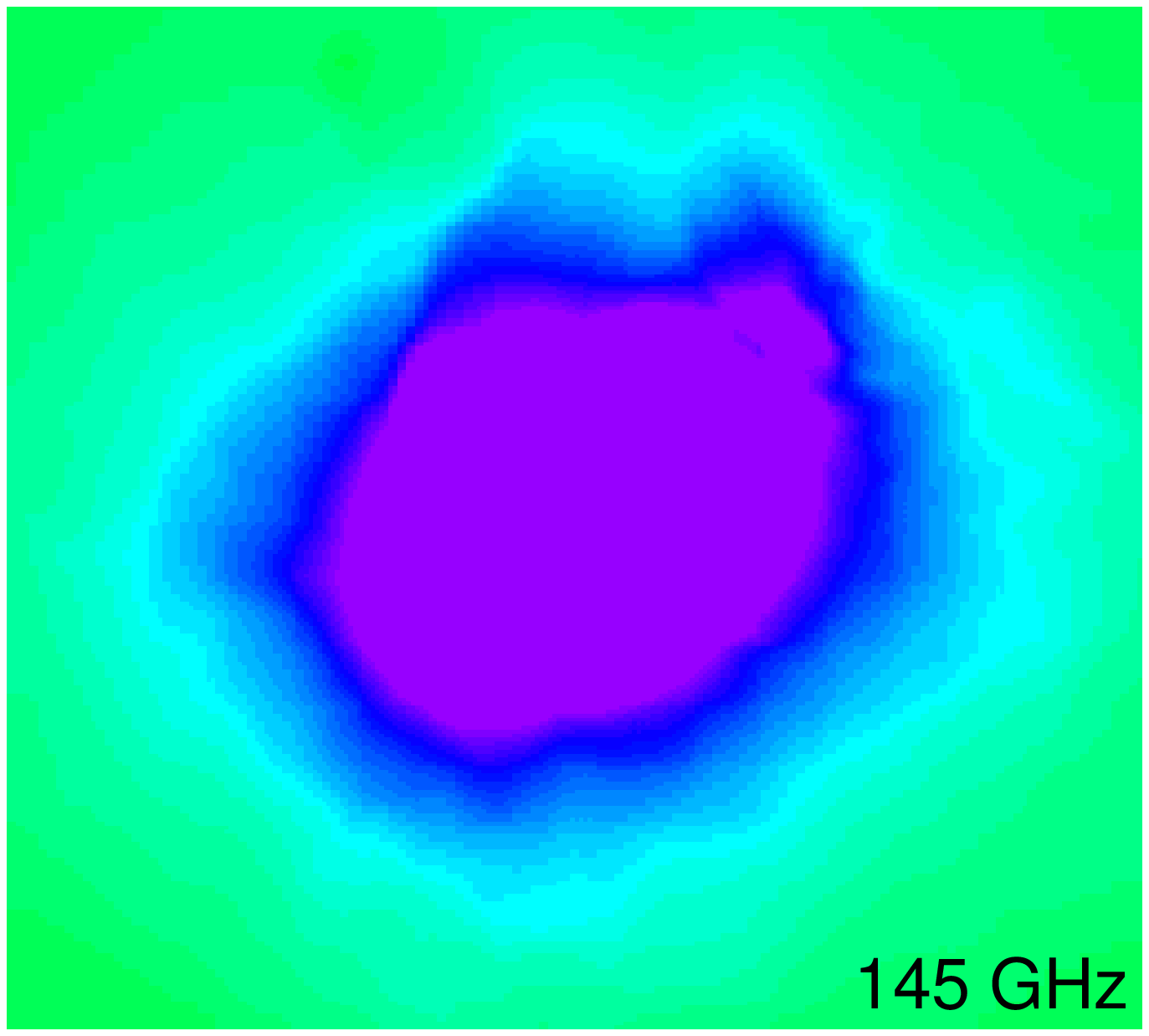}&\epsscale{.306}\plotone{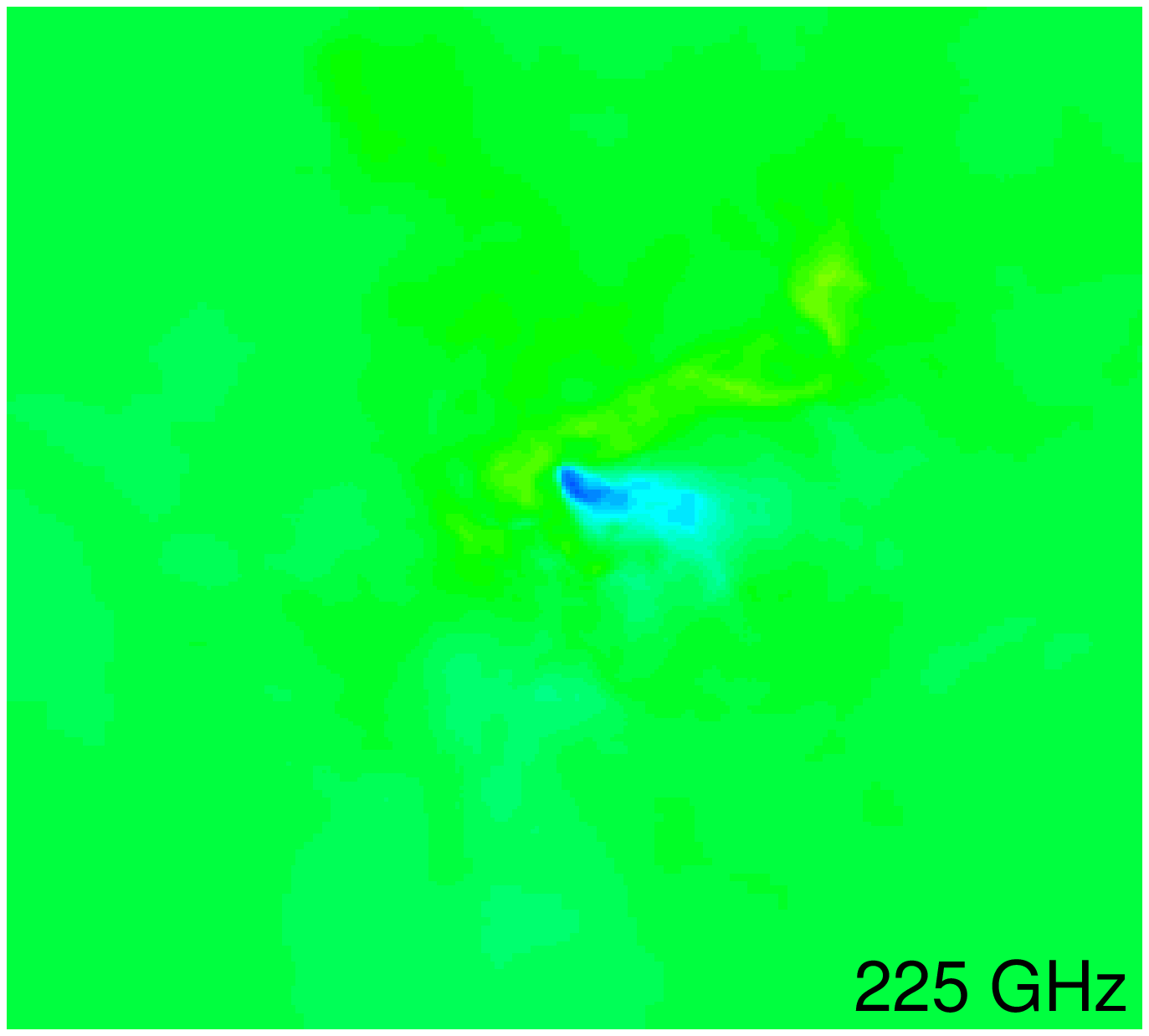}&\epsscale{.308}\plotone{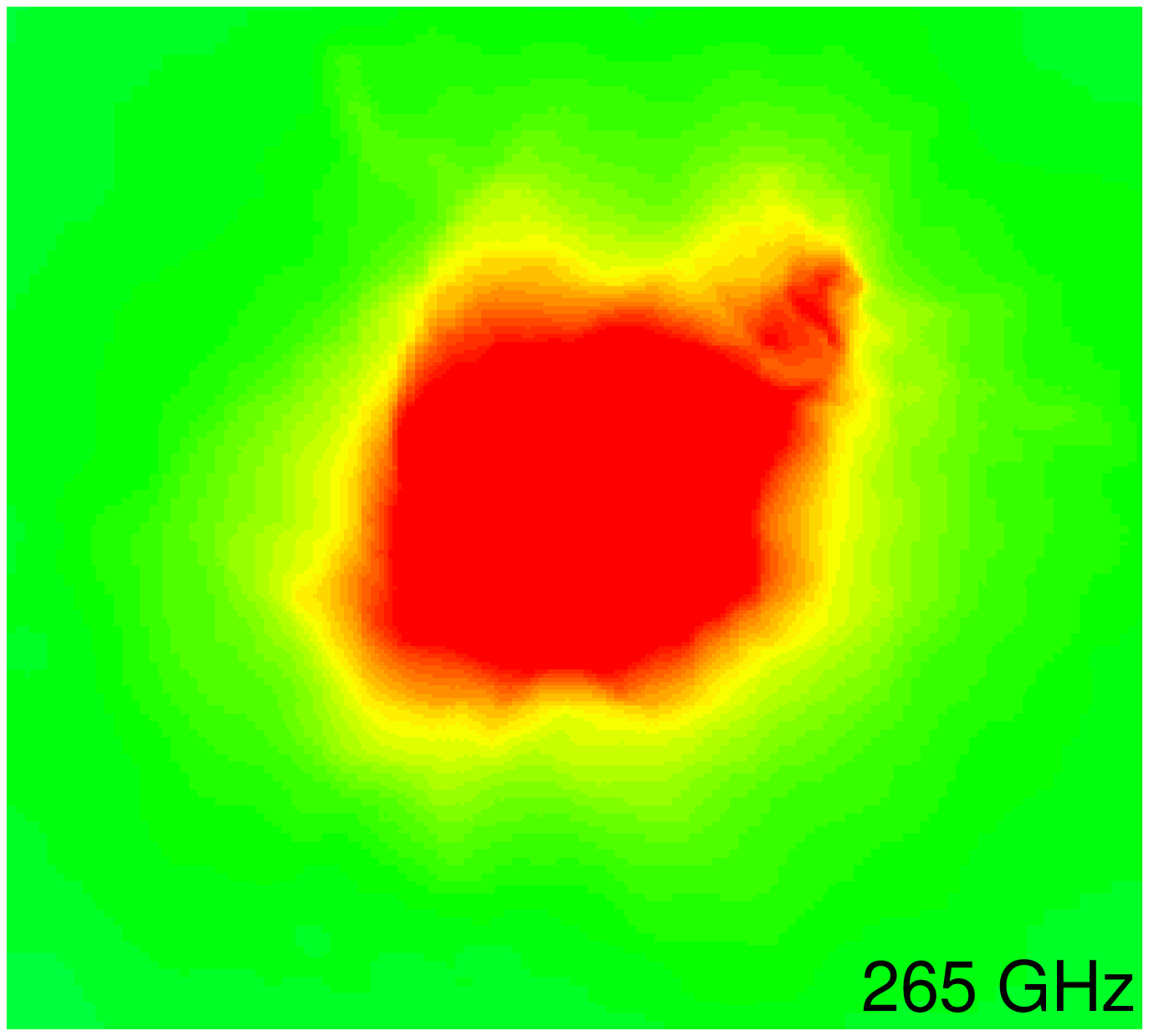}\cr
\mbox{}&\mbox{(a)}&\mbox{(b)}&\mbox{(c)}
\end{array}$
\caption{SZ simulations of a simulated Nbody+hydro cluster before
smoothing and adding detector noise. The cluster is about $10^{15}
M_{\odot}$, has an average gas temperature of about 9 keV, and is at
z=0.43. Figures 8a, 8b, and 8c are of the 145, 225, and 265 GHz
bands respectively. Each figure is about 6' x 6' with a pixel size
of 0.02' x 0.02'. The images are converted to temperature
differences from the mean microwave background temperature.  The
color scale is from -100$\mu K$ to 100$\mu K$. Primary microwave
background fluctuations and point source contamination are not
included.}
\end{center}
\end{figure}

\begin{figure}
\begin{center}
$\begin{array}{c@{\hspace{0.005in}}c@{\hspace{0.005in}}c@{\hspace{0.005in}}c}
\epsscale{.047}\plotone{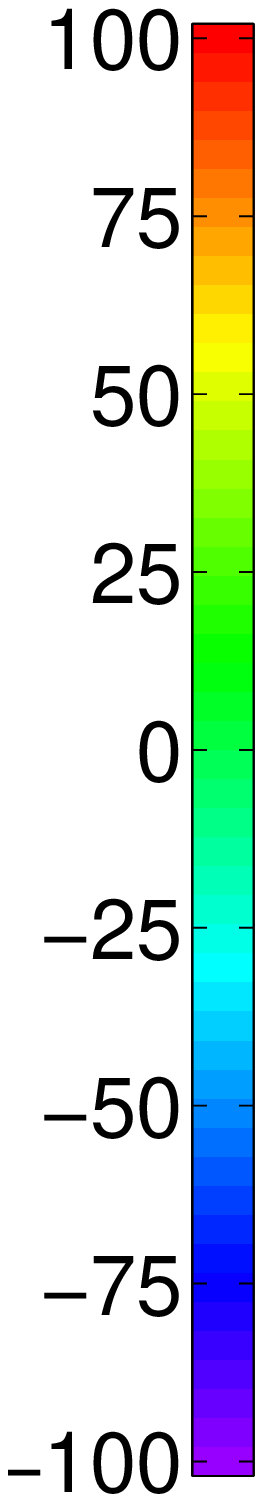}&\epsscale{.309}\plotone{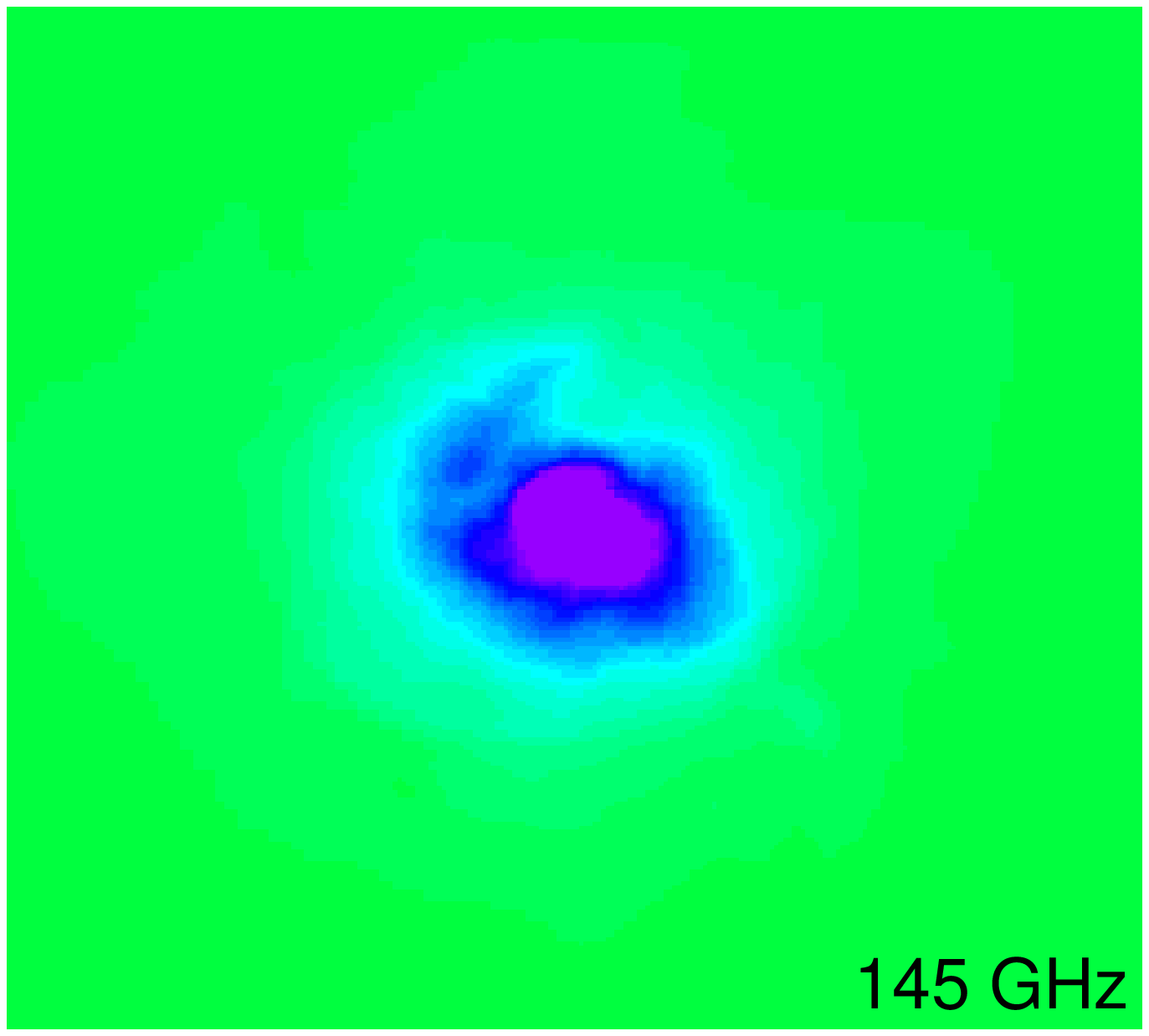}&\epsscale{.309}\plotone{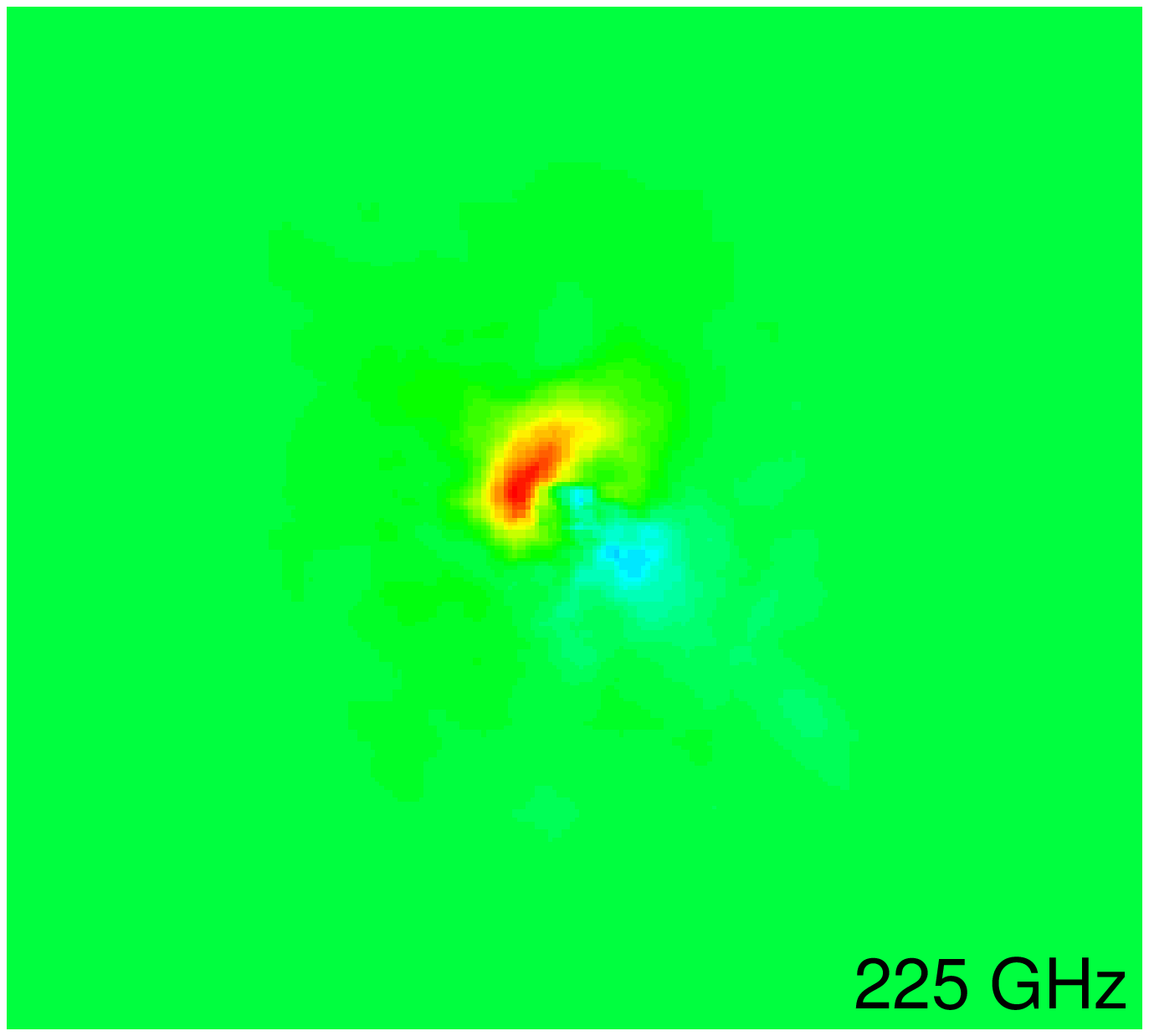}&\epsscale{.31105}\plotone{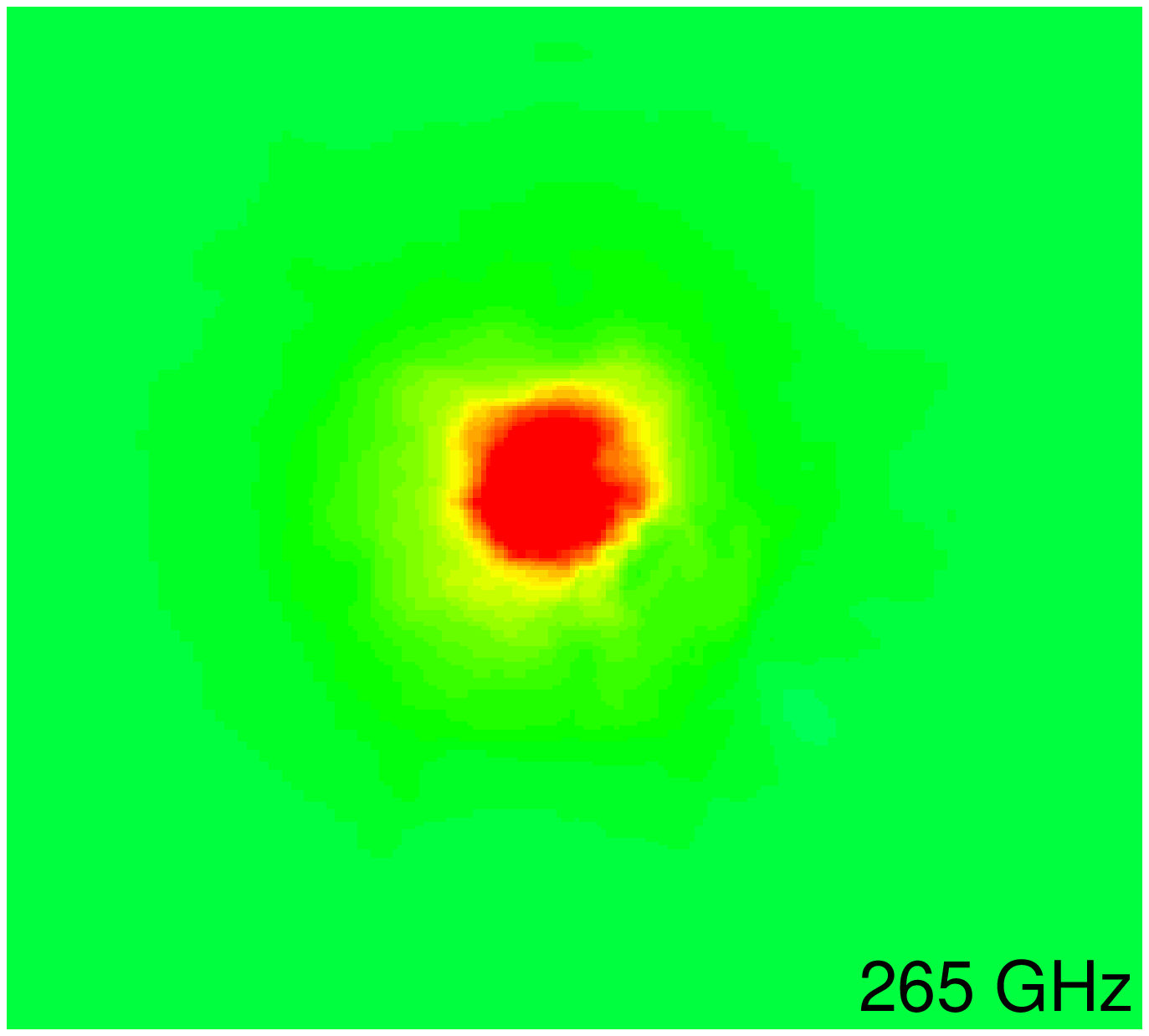}\cr
\mbox{}&\mbox{(a)}&\mbox{(b)}&\mbox{(c)}
\end{array}$
\caption{Same as Fig. 8 except the cluster is about $2 \times
10^{14} M_{\odot}$ and has an average gas temperature of about 3
keV.}
\end{center}
\end{figure}

\begin{figure}
\begin{center}
$\begin{array}{c@{\hspace{0.005in}}c@{\hspace{0.005in}}c@{\hspace{0.005in}}c}
\epsscale{.047}\plotone{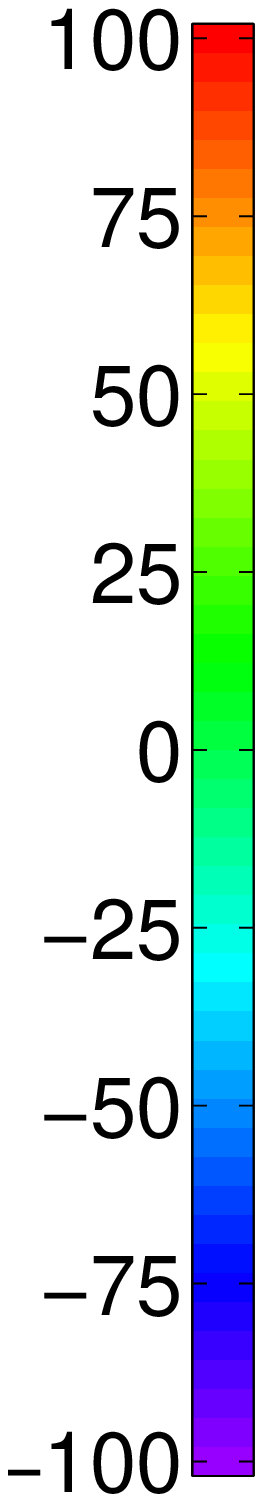}&\epsscale{.309}\plotone{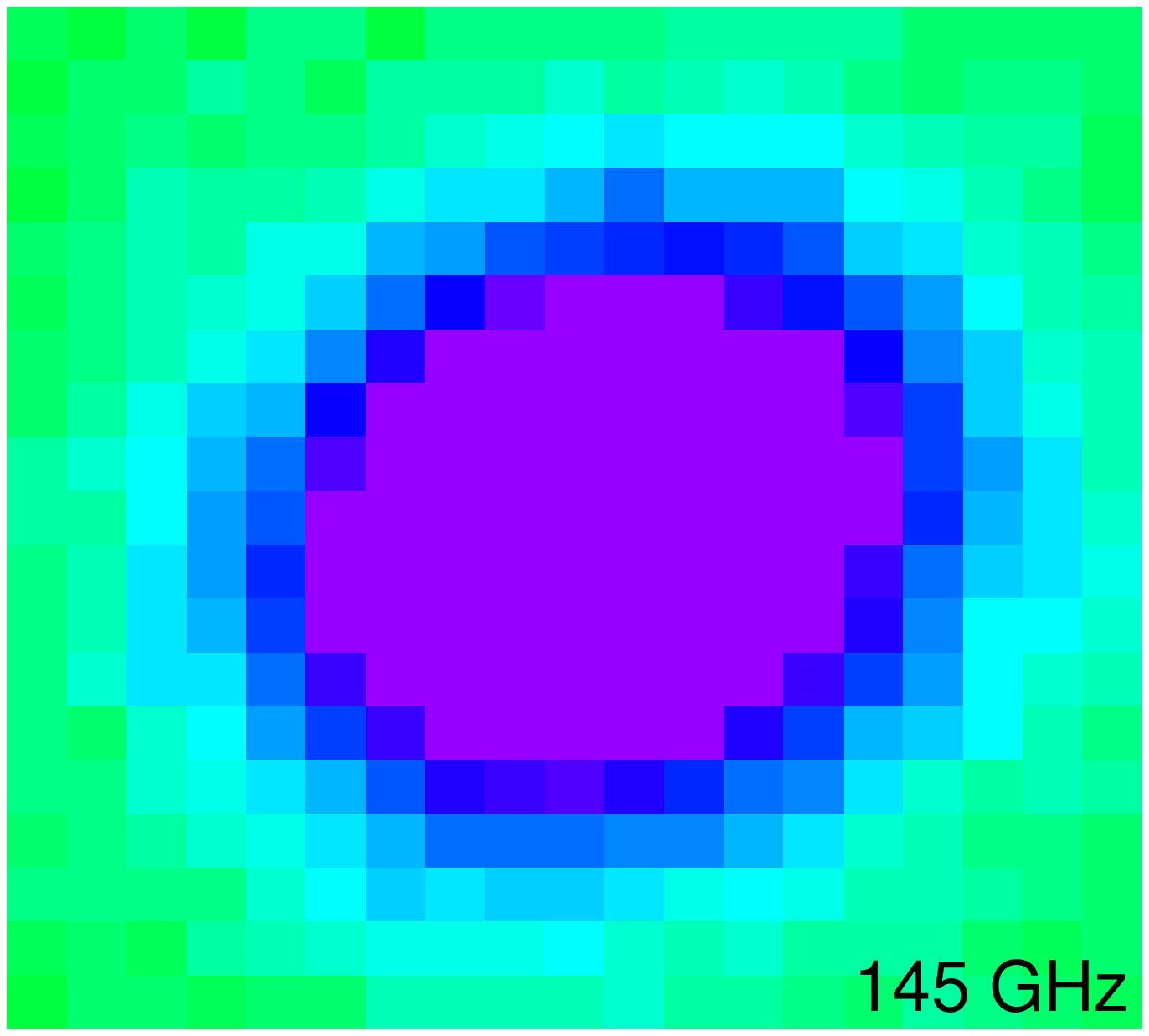}&\epsscale{.31}\plotone{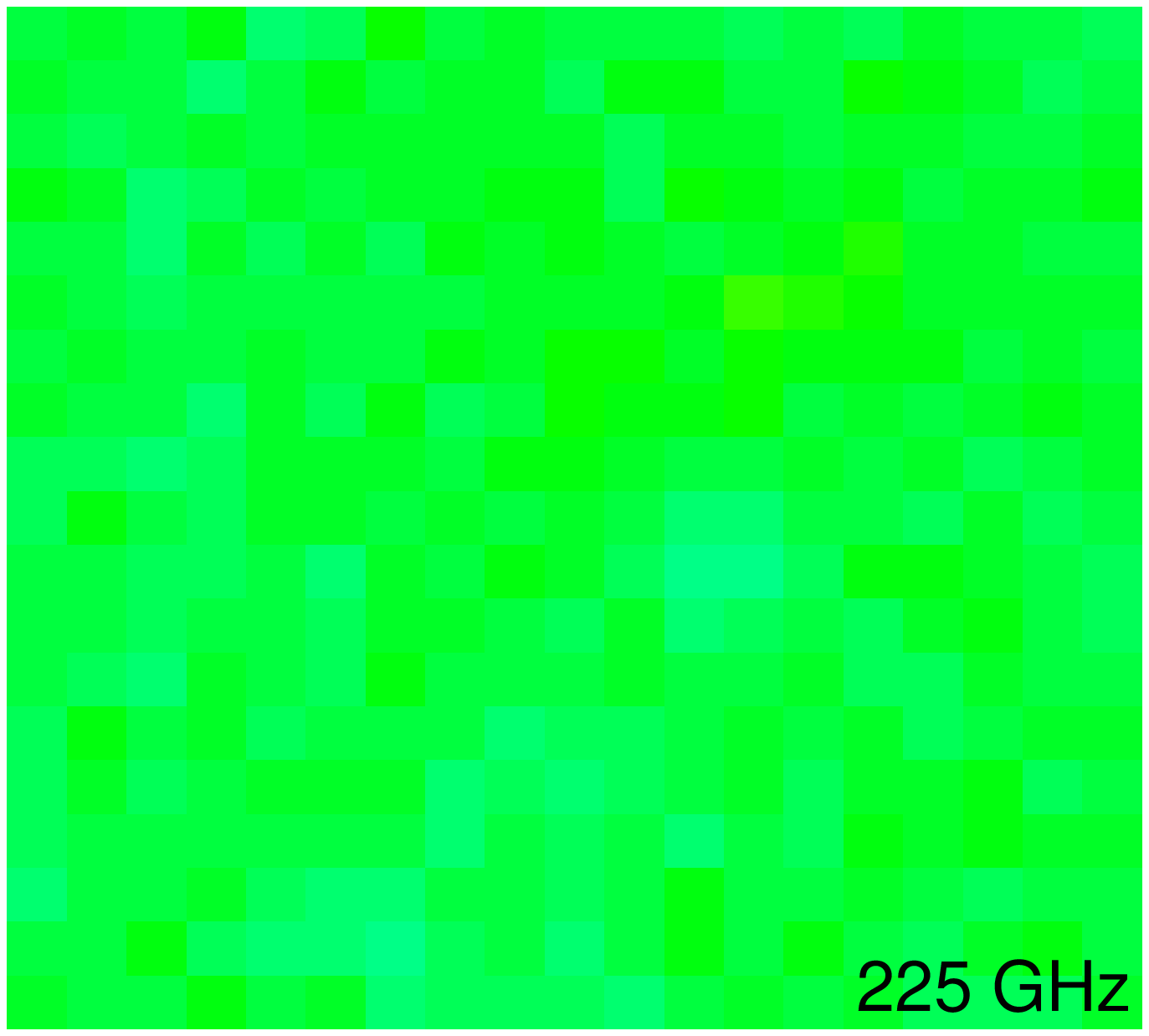}&\epsscale{.309}\plotone{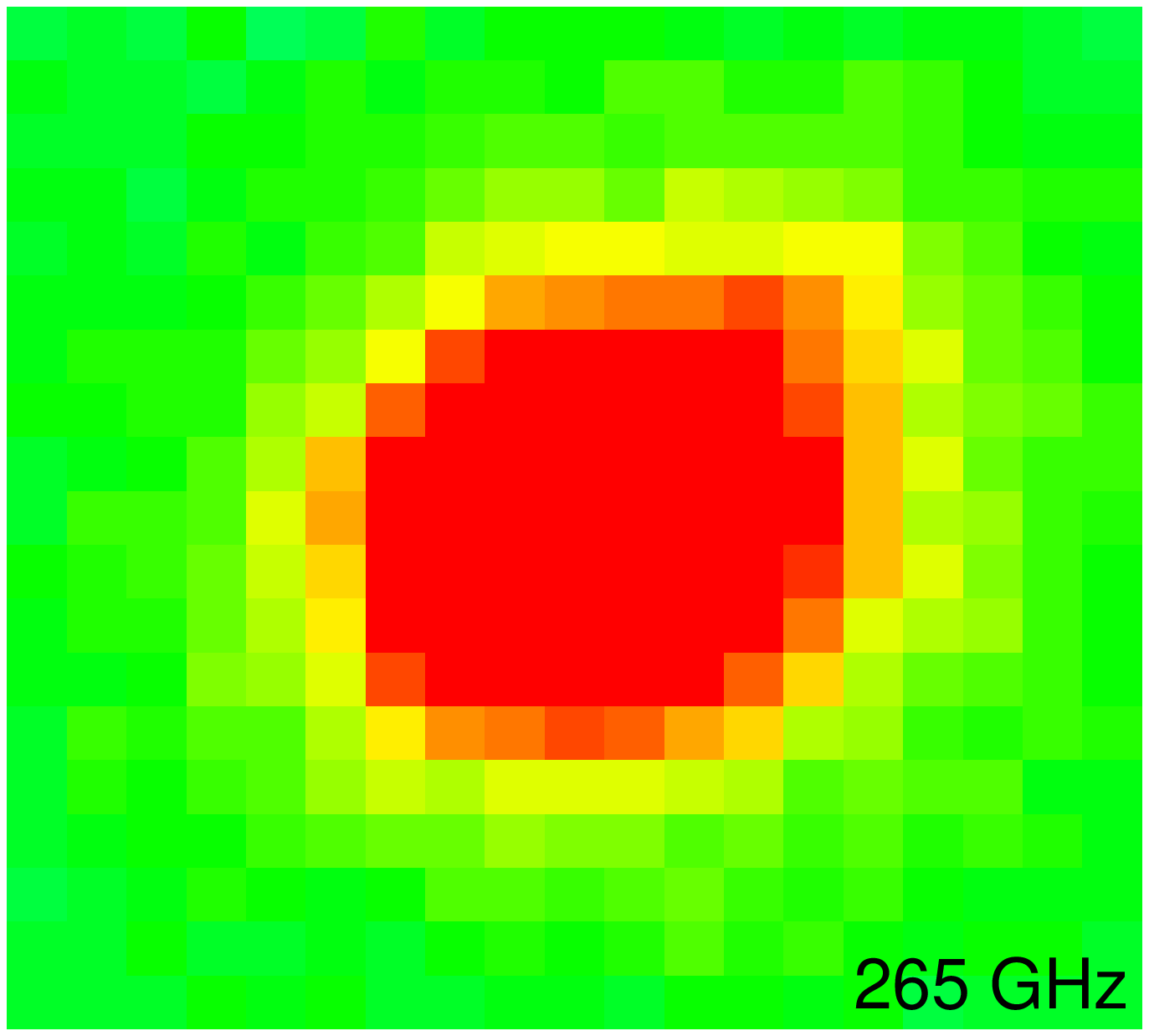}\cr
\mbox{}&\mbox{(a)}&\mbox{(b)}&\mbox{(c)}
\end{array}$
\caption{ACT-like SZ simulations of the 9 keV simulated cluster
shown in Fig. 8 with 1' resolution and $3\mu K$ gaussian random
instrument noise in each 0.3' x 0.3' pixel.  The remaining figure
specifications are the same as in Fig. 8.}
\end{center}
\end{figure}

\begin{figure}
\begin{center}
$\begin{array}{c@{\hspace{0.005in}}c@{\hspace{0.005in}}c@{\hspace{0.005in}}c}
\epsscale{.047}\plotone{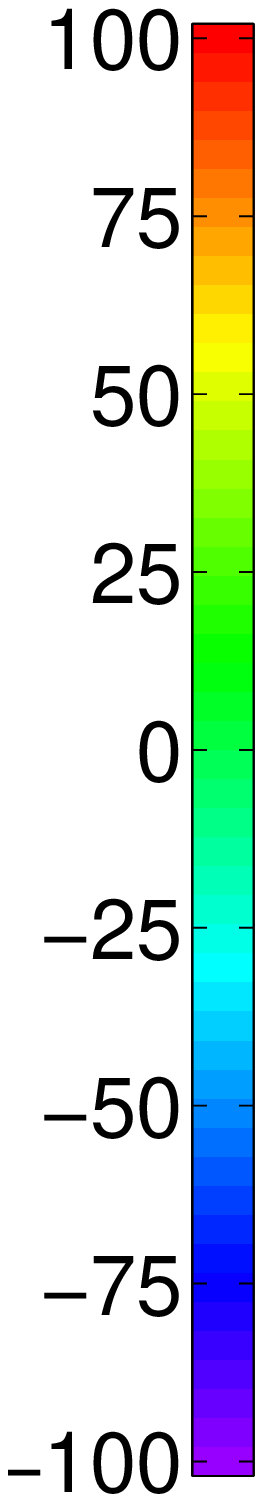}&\epsscale{.309}\plotone{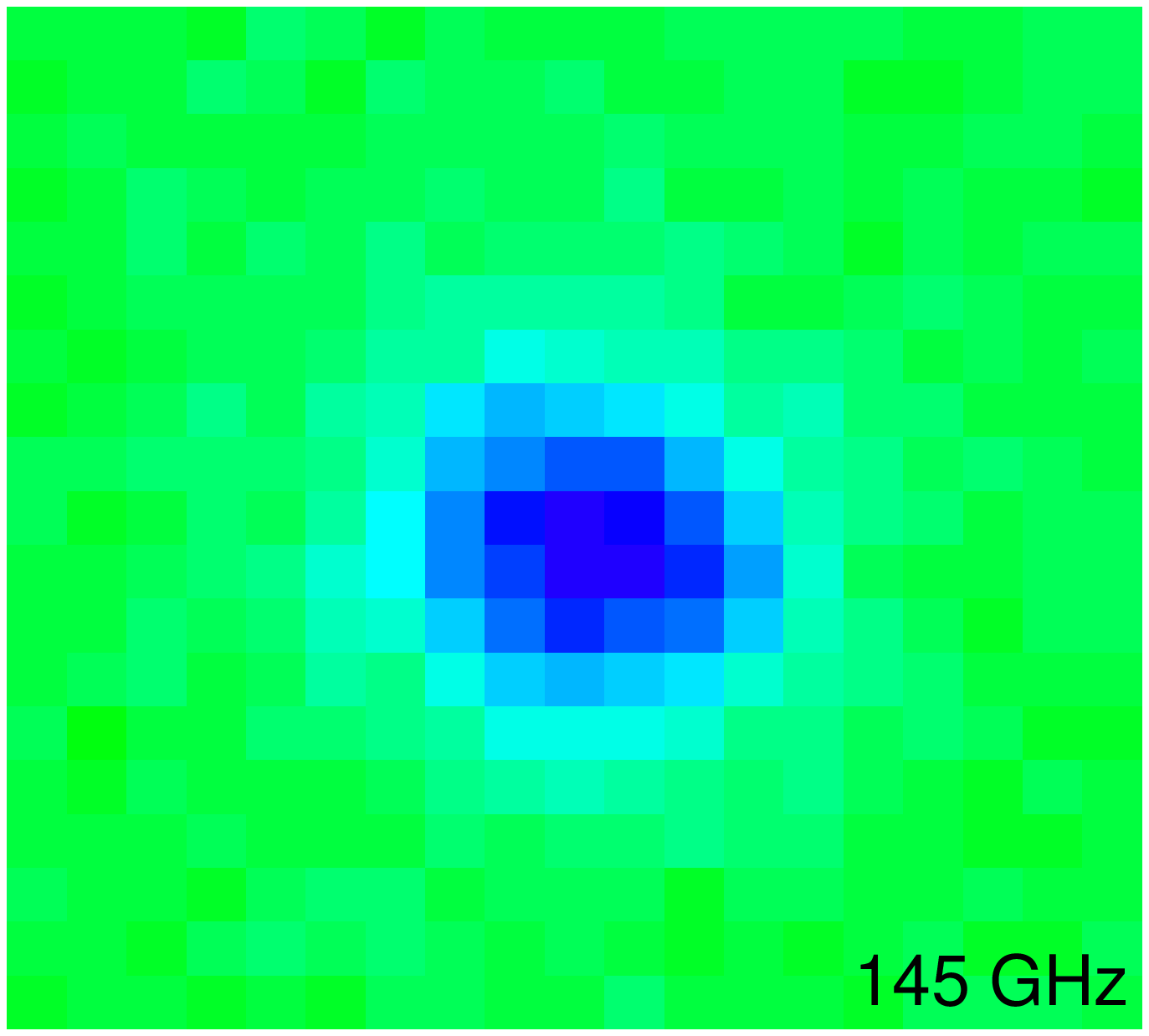}&\epsscale{.308}\plotone{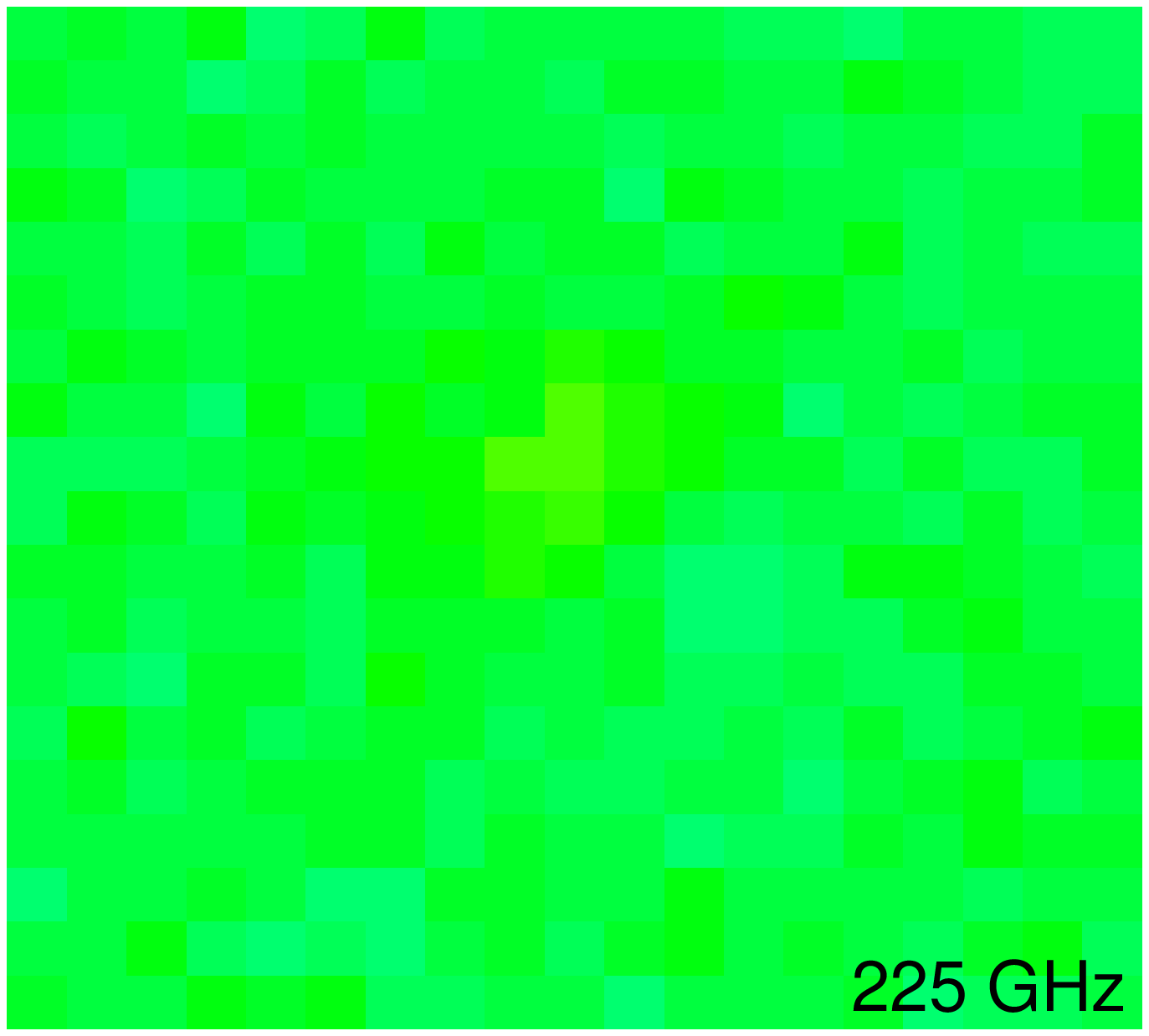}&\epsscale{.31}\plotone{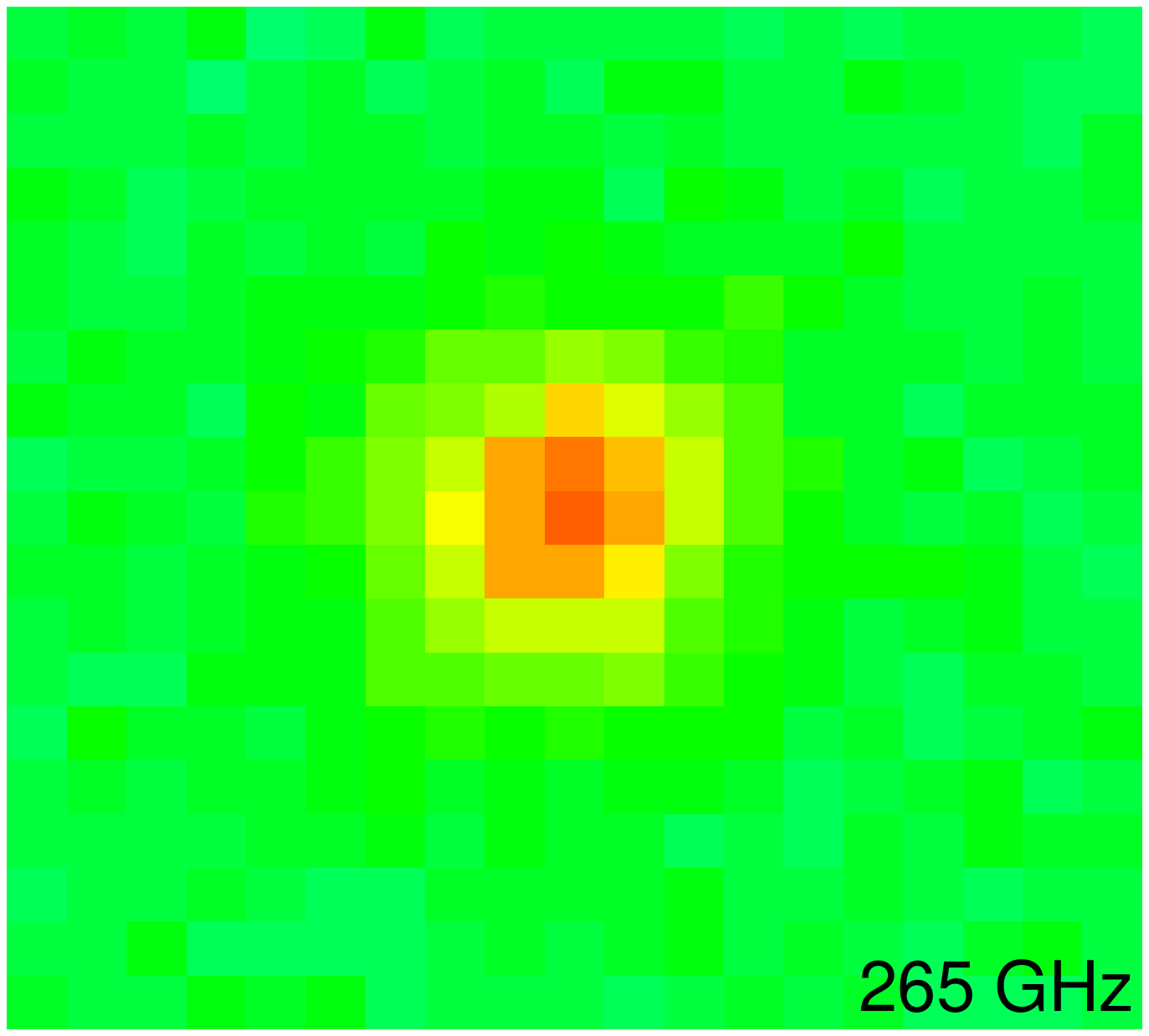}\cr
\mbox{}&\mbox{(a)}&\mbox{(b)}&\mbox{(c)}
\end{array}$
\caption{Same as Fig. 10 except using the 3 keV simulated cluster
shown in Fig. 9.}
\end{center}
\end{figure}

\begin{figure}
\begin{center}
$\begin{array}{c@{\hspace{0.1in}}c@{\hspace{0.1in}}c}
\epsscale{.31}\plotone{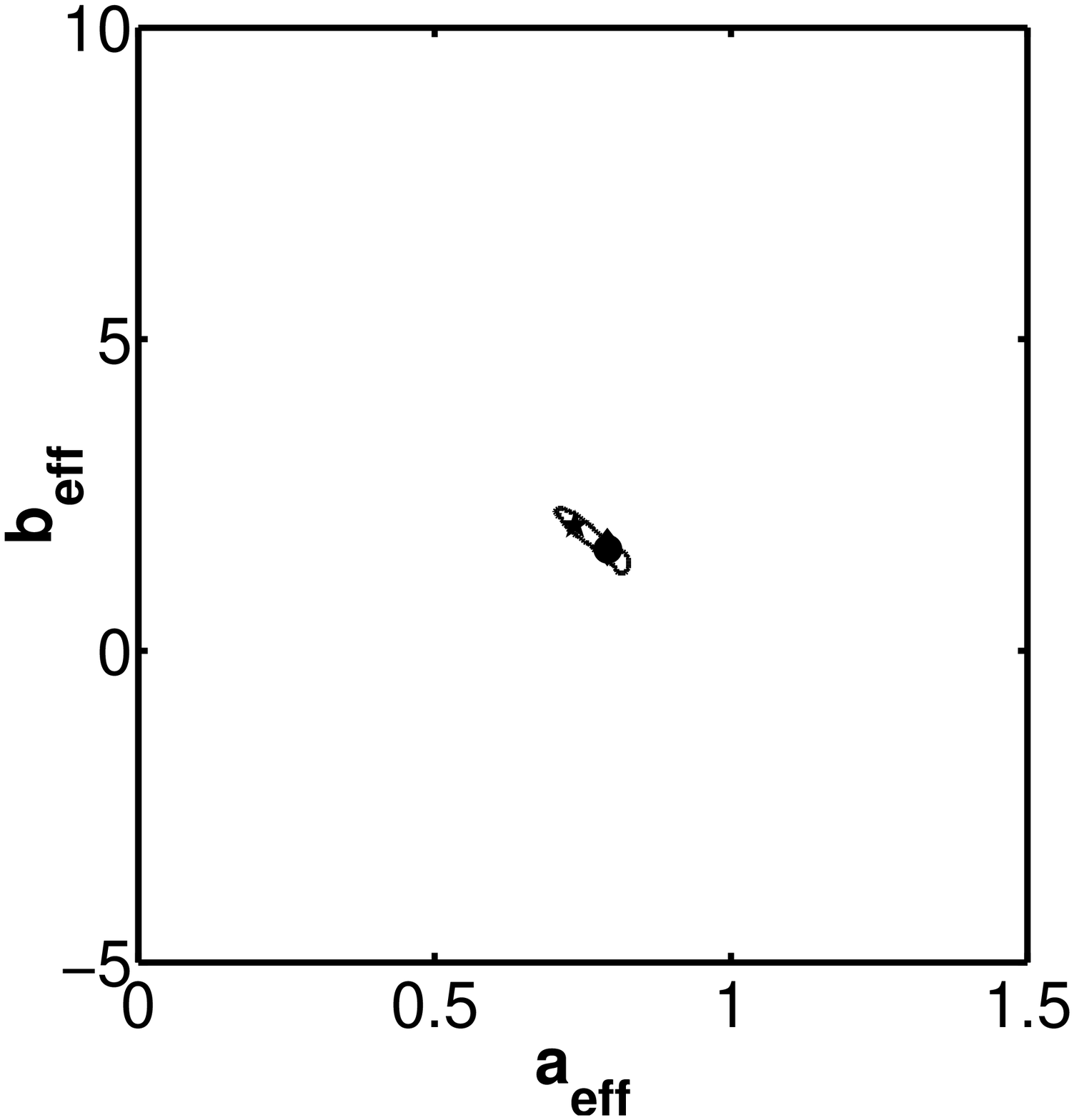}&\epsscale{.31}\plotone{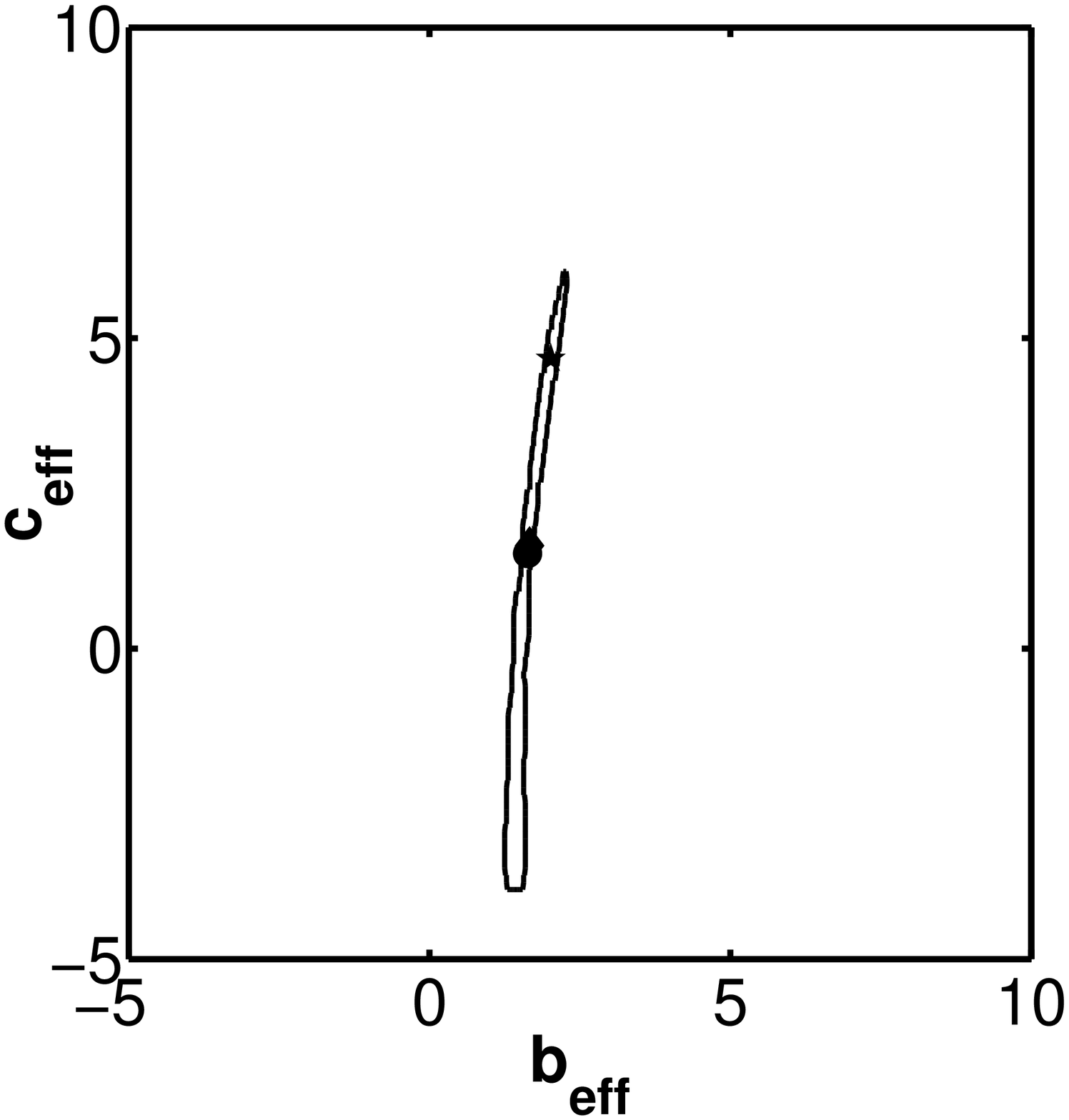}&\epsscale{.312}\plotone{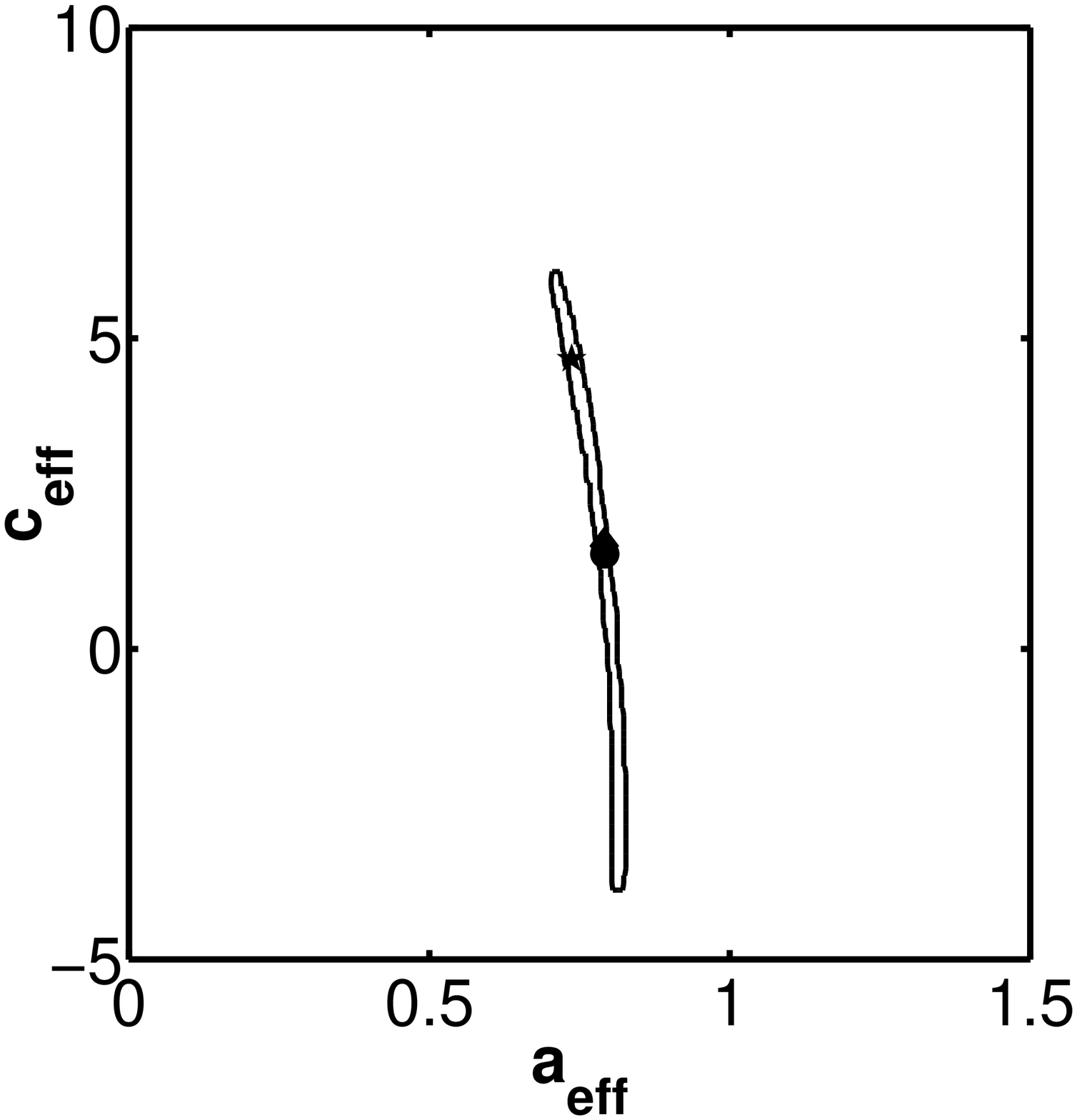}\cr
\mbox{\hspace{0.19in}(a)}&\mbox{\hspace{0.2in}(b)}&\mbox{\hspace{0.2in}(c)}
\end{array}$
\caption{The projected 1-$\sigma$ contours of the likelihood surface
for the ($a_{\mathrm{eff}}$, $b_{\mathrm{eff}}$, $c_{\mathrm{eff}}$)
parameter space from the central pixel of simulated ACT-like SZ
images of a simulated 9 keV Nbody+hydro cluster. The likelihood
contours are generated using a Markov chain and are for a given
noise realization.  The star ($\bigstar$) indicates the best fit
$\textbf{\emph{a}}_{\mathrm{eff}}$ values from the Markov chain. The
dot ($\bullet$) indicates the best fit
$\textbf{\emph{a}}_{\mathrm{eff}}$ values obtained with an ideal
instrument without detector noise. A diamond shape ($\blacklozenge$)
marks the values of line-of-sight integrals through the three
dimensional cluster.}
\end{center}
\end{figure}

\begin{figure}
\begin{center}
$\begin{array}{c@{\hspace{0.1in}}c@{\hspace{0.1in}}c}
\epsscale{.312}\plotone{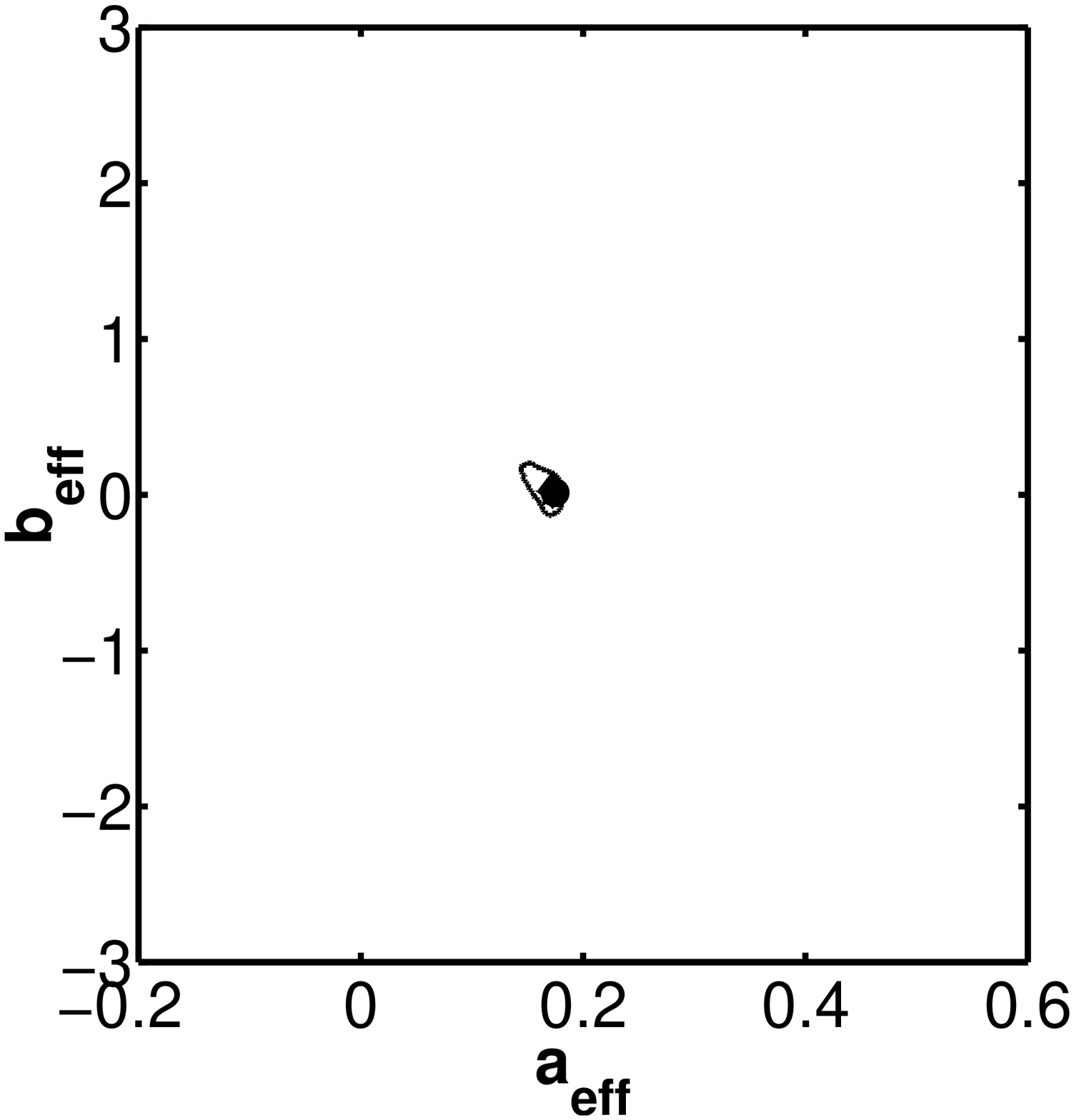}&\epsscale{.30}\plotone{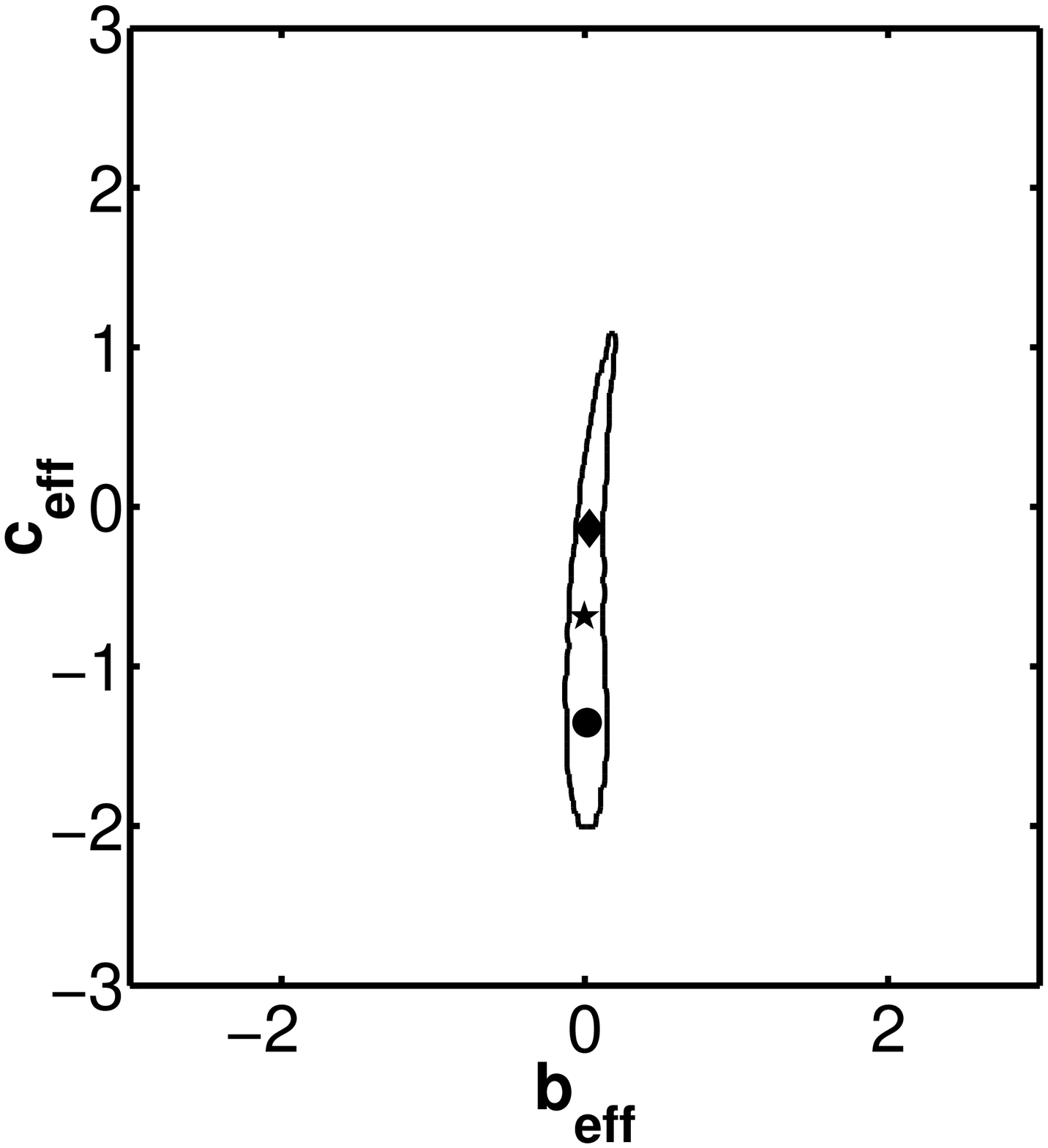}&\epsscale{.321}\plotone{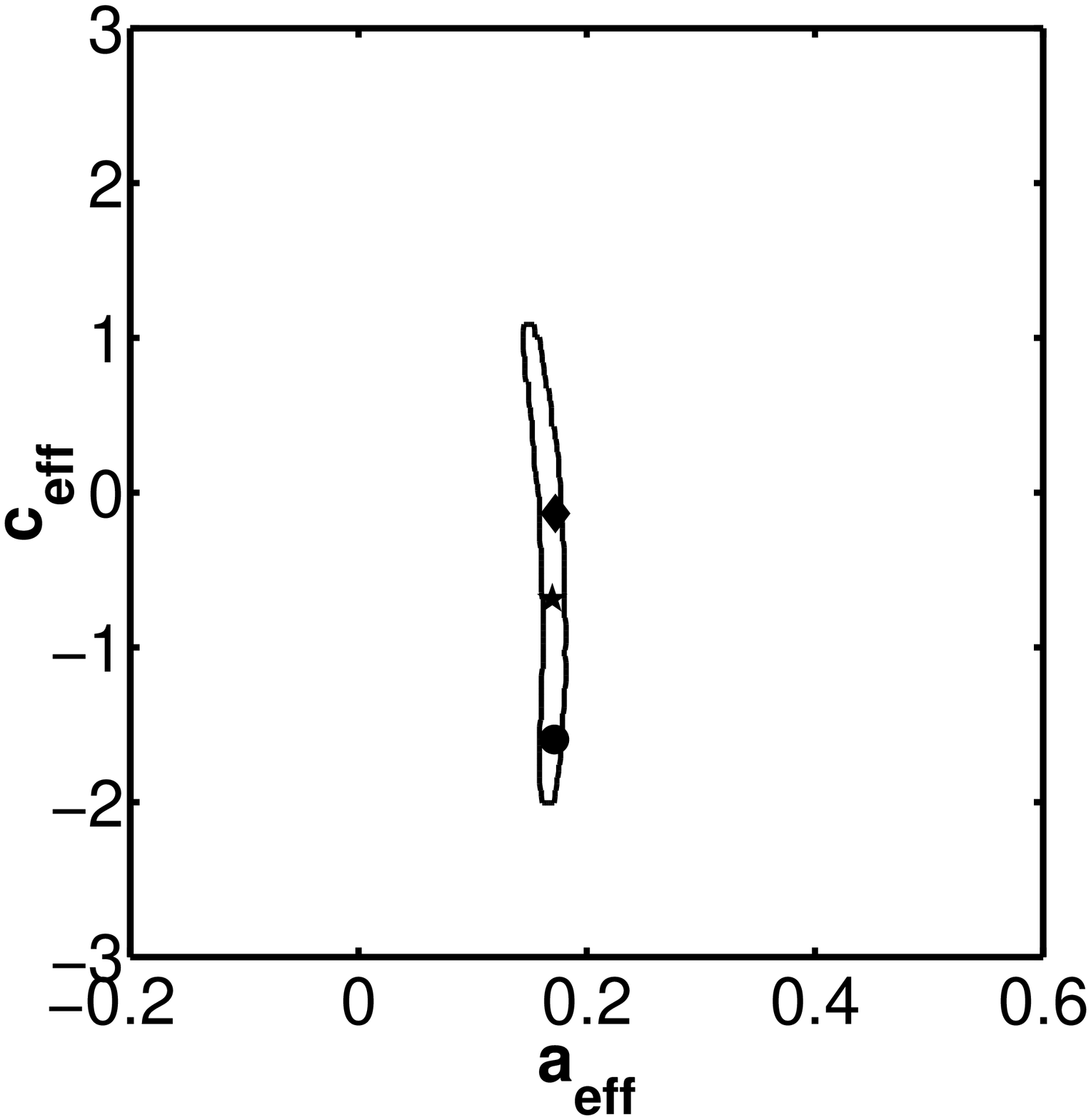}\cr
\mbox{\hspace{0.2in}(a)}&\mbox{\hspace{0.22in}(b)}&\mbox{\hspace{0.23in}(c)}
\end{array}$
\caption{The projected 1-$\sigma$ contours of the likelihood surface
for the ($a_{\mathrm{eff}}$, $b_{\mathrm{eff}}$, $c_{\mathrm{eff}}$)
parameter space from the central pixel of simulated ACT-like SZ
images of a simulated 3 keV Nbody+hydro cluster. The symbols within
the contours are as for Fig. 12; note the axes are different
scales.}
\end{center}
\end{figure}

\begin{figure}
\begin{center}
$\begin{array}{c@{\hspace{0.1in}}c@{\hspace{0.1in}}c}
\epsscale{.30305}\plotone{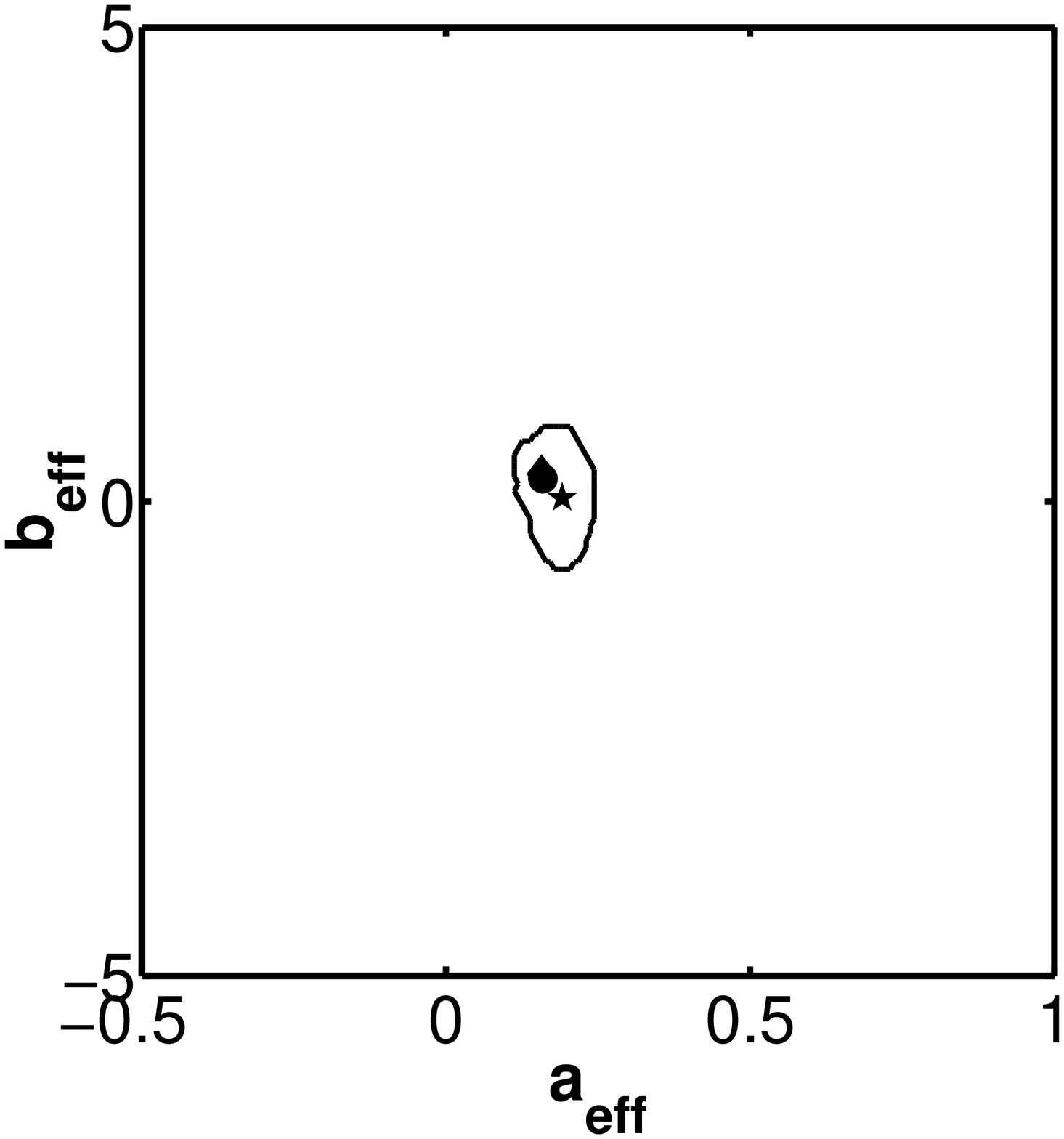}&\epsscale{.31}\plotone{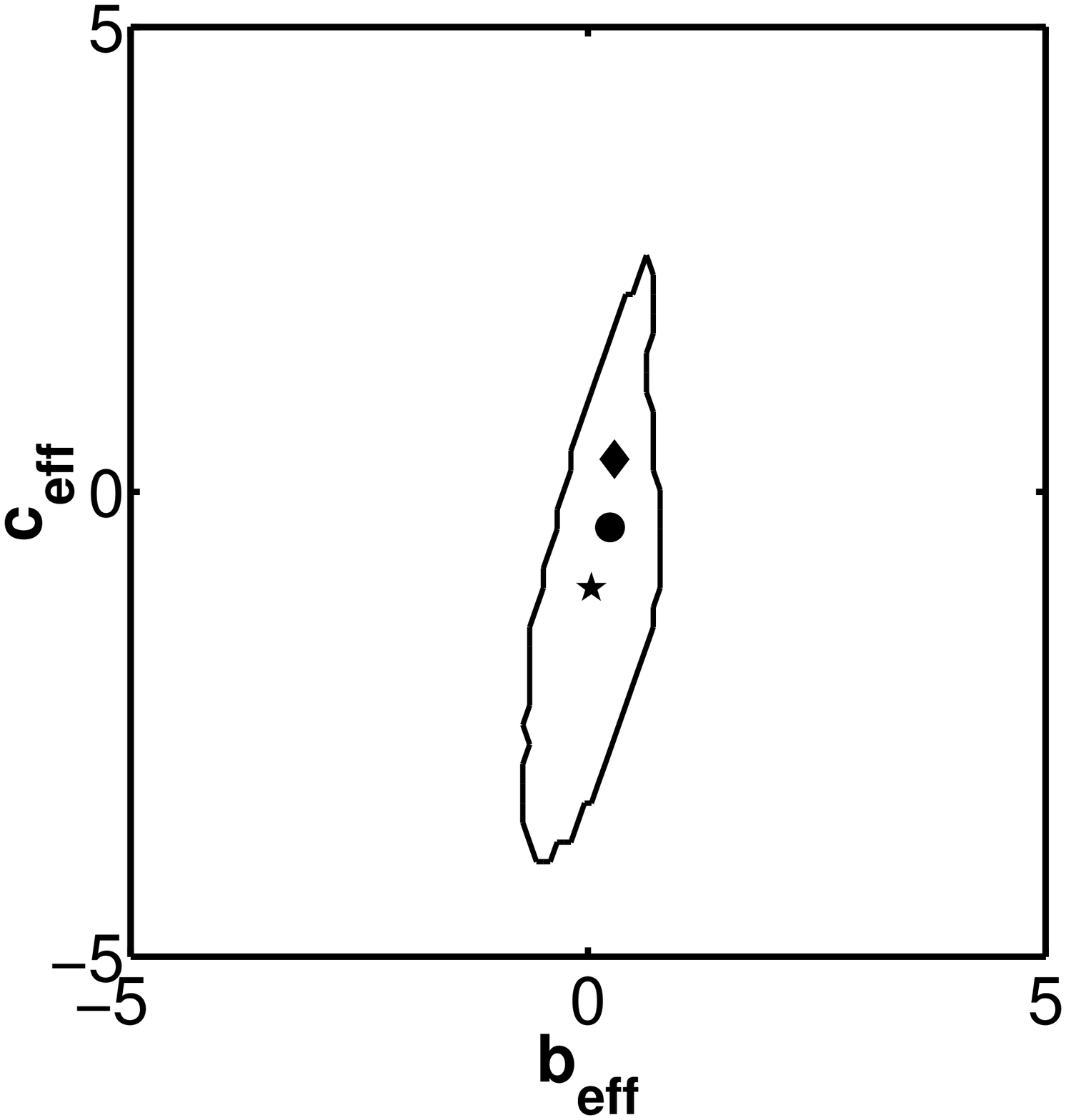}&\epsscale{.309}\plotone{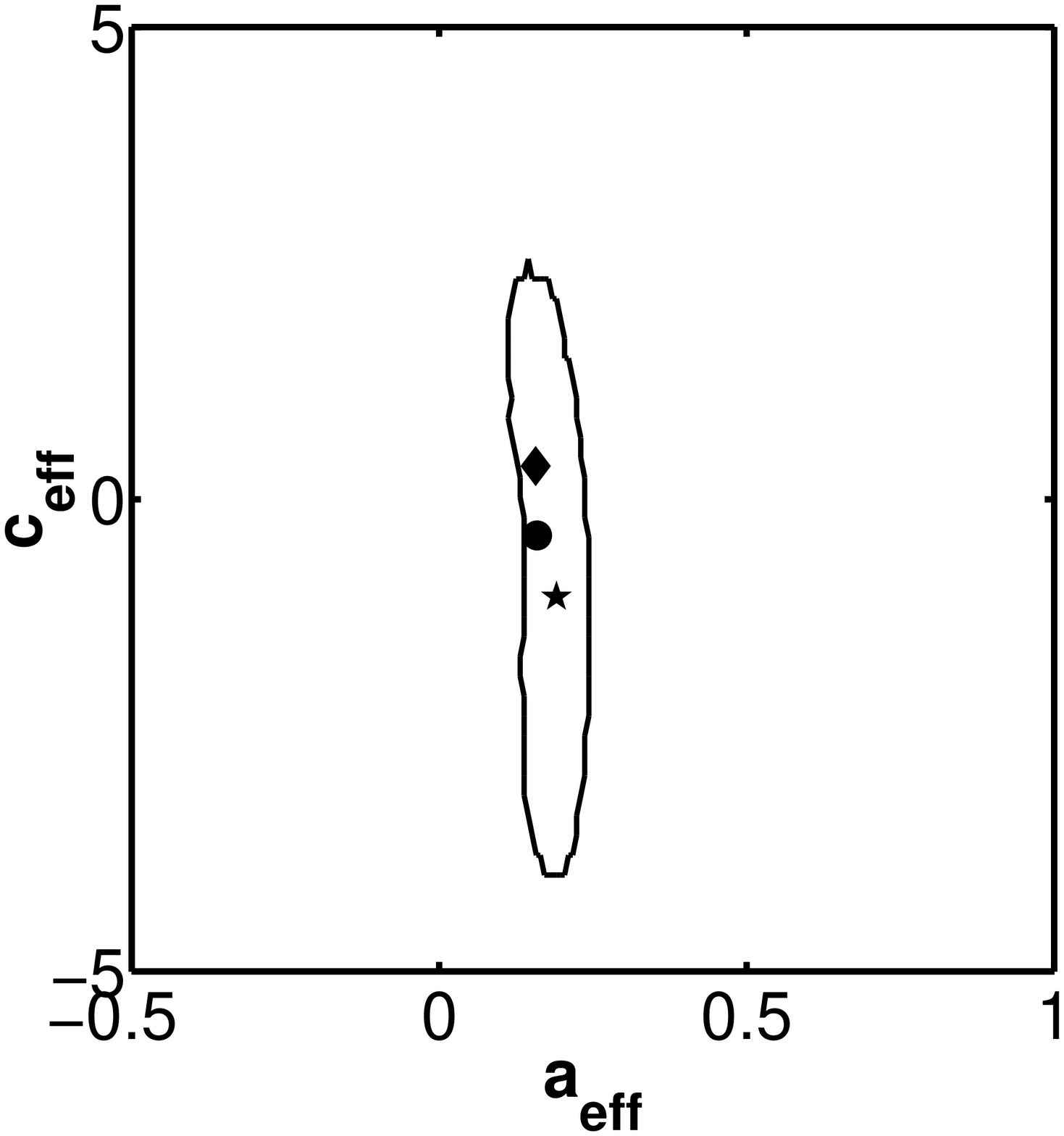}\cr
\mbox{\hspace{0.2in}(a)}&\mbox{\hspace{0.21in}(b)}&\mbox{\hspace{0.21in}(c)}
\end{array}$
\caption{The projected 1-$\sigma$ contours of the likelihood surface
for the ($a_{\mathrm{eff}}$, $b_{\mathrm{eff}}$, $c_{\mathrm{eff}}$)
parameter space from the central pixel of simulated Planck-like SZ
images of a simulated 9 keV Nbody+hydro cluster. The symbols within
the contours are as for Fig. 12; note the axes are different
scales.}
\end{center}
\end{figure}

\begin{figure}
\begin{center}
\epsscale{1.0}\plotone{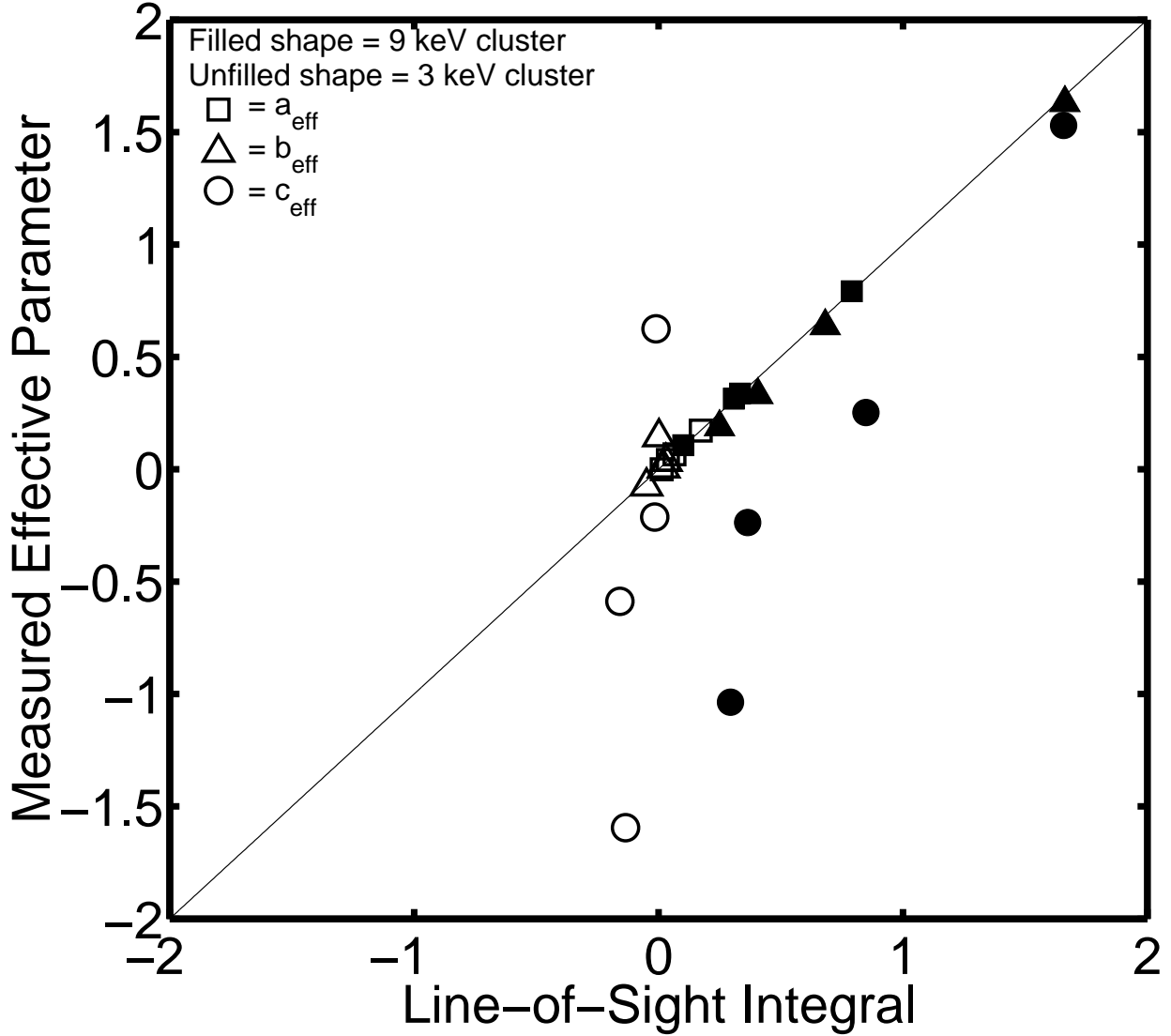} \caption{Best fit
$\textbf{\emph{a}}_{\mathrm{eff}}$ from simulated, noise-free,
ACT-like SZ images found using a Markov chain versus $\int
\mathrm{C} d\textbf{\emph{p}}$ integrated along the cluster line of
sight. The results for four different lines of sight through both
the 9 keV and 3 keV simulated clusters are plotted. The lines of
sight are 0', 1', 1.5', and 2' from the central pixel of the
simulated SZ images. $d\textbf{\emph{p}}=(T d\tau, v d\tau, T^{2}
d\tau)$, and $\mathrm{C}$ is a matrix of constants introduced in eq.
(5).  $\square$ = $a_{\mathrm{eff}}$, $\triangle$ =
$b_{\mathrm{eff}}$, and {\small{$\bigcirc$}} = $c_{\mathrm{eff}}$,
and filled (unfilled) shapes correspond to the 9 keV (3 keV)
cluster.  Typical error bars for $1\mu K$ detector noise per 1' beam
are $\pm$ 0.04, $\pm$ 0.3, and $\pm$ 3 for $a_{\mathrm{eff}}$,
$b_{\mathrm{eff}}$, and $c_{\mathrm{eff}}$ respectively.}
\end{center}
\end{figure}

\end{document}